\documentclass[usenatbib]{mn2e}
\usepackage{amssymb,amsmath,graphicx,natbib}
\begin{document}
\title{Genesis of the dusty Universe: modeling submillimetre source counts}
\author[A.~Rahmati and P.P.~van~der~Werf]
  {A.~Rahmati$^1$\thanks{rahmati@strw.leidenuniv.nl} and
  P.P.~van der Werf$^{1}$ \\
  $^1$Sterrewacht Leiden, Leiden University, P.O. Box 9513, 2300 RA Leiden, 
  The Netherlands}

\maketitle

\begin{abstract}
  We model the evolution of infrared galaxies using a phenomenological approach to match the observed source counts at different infrared wavelengths.
  In order to do that, we introduce a new algorithm for reproducing source counts which is based on direct integration of probability distributions rather than using Monte-Carlo sampling. 
  We also construct a simple model for the evolution of the luminosity function and the colour distribution of infrared galaxies which utilizes a minimum number of free parameters; we
  analyze how each of these parameters is constrained by observational data. The
  model is based on pure luminosity evolution, adopts the Dale \& Helou Spectral
  Energy Distribution (SED) templates, allowing for evolution in the infrared
  luminosity-colour relation, but does not consider Active Galactic Nuclei as a
  separately evolving population. We find that the $850\,\mu$m source counts and
  redshift distribution depend strongly on the shape of the luminosity evolution
  function, but only weakly on the details of the SEDs. Based on this
  observation, we derive the best-fit evolutionary model using the
  $850\,\mu$m counts and redshift distribution as constraints.  Because of the
  strong negative $K$-correction combined with the strong luminosity evolution,
  the fit has considerable sensitivity to the sub-$L_*$ population at high
  redshift, and our best-fit shows a flattening of the faint end of the
  luminosity function towards high redshifts. Furthermore, our best-fit model
  requires a colour evolution which implies the typical dust temperatures of
  objects with the same luminosities to decrease with redshift. We then compare
  our best-fit model to observed source counts at shorter and longer wavelengths which indicates our model reproduces the $70\,\mu$m and $1100\,\mu$m source counts remarkably well, but under-produces the counts at intermediate wavelengths.  Analysis reveals that the discrepancy arises at low redshifts, indicating that revision of the adopted SED library towards lower dust temperatures (at a fixed infrared luminosity) is required. This modification is equivalent to a population of cold galaxies existing at low redshifts, as also indicated by recent Herschel results, which are underrepresented in IRAS sample. We show that the modified model successfully reproduces the source counts in a wide range of IR and submm wavelengths.
\end{abstract}


\section{Introduction}

Although dust is an unimportant component in the mass budget of galaxies,
  its presence radically alters the emergent spectrum of star forming galaxies.
  Since stars are born in dusty clouds, most of the energy of young stars is
  absorbed by dust particles, which are heated by the absorption process and
  radiate their energy away by thermal emission at infrared (IR) and
  submillimetre (submm) wavelengths. As a result, the infrared radiation of star
  forming galaxies is a useful measure of the massive star formation rate.
Mainly due to the atmospheric opacity, the thermal radiation from dust could not
be systematically studied until the launch of IRAS in the mid-1980s, which
  provided the first comprehensive view of the nearby dusty Universe.  Ten
years later COBE discovered the Cosmic Infrared
Background (CIB) \citep{Puget96,Fixsen98} and it turned out the observed power
of the CIB is comparable to what can be deduced from the optical Universe. This was in contrast to the
observations of local galaxies which suggest only one third of the energy output
of galaxies is in IR bands \citep{Lagache05}.  Moreover, the relatively flat
slope of the CIB at long wavelengths indicated a population of dusty galaxies
which are distributed over a wide range of redshifts \citep{Gispert00}.  Thanks
to various large surveys performed with different satellites and ground based
observatories, this background radiation is now partly resolved into point
sources at different wavelengths. While at shorter wavelengths the emission is
mainly coming from local and low redshift galaxies, longer wavelengths
contain information about larger distances.

The shape of emission spectrum of warm dust particles resembles a modified
blackbody spectrum with a peak varying with the typical dust temperature which is observed to be
around $T\sim 30-40$K in galaxies; therefore the far-IR (FIR) and submm spectrum of a
typical dusty galaxy consists of a peak at rest-frame wavelengths around
$\lambda \sim 100-200 \mu m$ which drops on both sides (see Figure~\ref{fig:SED}). While the presence of other emitters like polycyclic aromatic hydrocarbons (PAHS) and Active Galactic Nuclei (AGNs) complicates
the shape of spectra at shorter wavelengths, at submm wavelengths the spectra
behave like modified blackbodies and their amplitudes drop steeply. In
fact, because of this steep falloff in the Rayleigh-Jeans tail of the Spectral
Energy Distribution (SED), the observed flux density at a fixed submm
wavelength can rise by moving the SED in redshift space (the so called
$K$-correction), which counteracts the cosmological dimming. It turns out
that the interplay between these two processes enables us to observe galaxies
at submm wavelengths out to very high redshifts (see also the discussion in
Section~\ref{sec:others} and Figure~\ref{fig:short-long-fluxes}).


After the first observations of SCUBA at $850\,\mu$m confirmed the importance of
submm galaxies at high redshifts \citep{Smail97}, there were many subsequent
surveys using different instruments which explored different cosmological fields
\citep{Smail02,Webb03,Borys03,Greve04,Coppin06,Bertoldi07,Austermann10,Scott10,Vieira10}
and also some surveys which used gravitational lensing techniques to
extend the observable submm Universe to sub-mJy fluxes
\citep{Smail97,Chapman02,Knudsen08, Johansson11}. 

While low angular resolution makes individual identifications and
  spectroscopy a daunting task, the surface density of sources as a function of
  brightness (i.e., the source counts) can be readily analysed and contains
  significant information about the population properties and their evolution.
One can assume simple smooth parametrized models for the evolution of
dusty galaxies and relate their low redshift observed properties (e.g., the
total IR luminosity function, which we simply call the luminosity function, or
in short LF hereafter) to their source counts \citep{Blain93,Guiderdoni97,Blain99,Chary01,Rowan-Robinson01,Dole03,Lagache04,Lewis05,LeBorgne09,Valiante09}.
Such backward evolution models usually combine SED templates and the low
  redshift properties of IR galaxies which are allowed to change with redshift
  based on an assumed parametric form, together with Monte-Carlo techniques to
  reproduce the observed source counts. These models are therefore purely
  phenomenological, and suppress the underlying
  physics. Their power lies exclusively in providing a parametrized description
  of statistical properties of the galaxy population under study, and the
  evolution of these properties. The results of such modeling provides a
  description of the constraints that must be satisfied (in a statistical sense)
  and could be explored further by more physically motivated models such as hydrodynamical simulations
  embedded in a $\Lambda$CDM cosmology, with subgrid prescriptions for star
  formation and feedback (e.g. \citet{Schaye10}). It is important to realize that backward
  evolution models are constrained only over limited observational parameter
  space (e.g., at particular observing wavelengths, flux levels, redshift
  intervals, etc.), and are ignorant of the laws of physics. Therefore it is dangerous to
  use them outside of their established validity ranges; in other words, these
  models have descriptive power but little explanatory power. As
  a result, one of the most instructive uses of these models is in analyzing
  where they fail to reproduce the observations, and how they can be modified to
  correct for these failures which implies that the underlying
  assumptions must be revised. This is the approach that we use in the
  present paper.
  
  Since nowadays many and various observational constraints exist (e.g. source counts at
  various IR and submm wavelengths, colour distributions, redshift
  distribution), backward evolution models can reach considerable levels of
  sophistication.  However, early backward evolution models used only a few SED templates
  (or sometimes even only one SED) to represent the whole population of dusty
  galaxies. This approach neglects the fact that dusty galaxies are not
  identical and span a variety of dust temperatures and hence SED shapes. As a
  result, such models cannot probe the possible evolution in the SED properties
  of submm galaxies, for which increasing observational support has been
  found during the last few years
  \citep{Chapman05,Pope06,Chapin09,Symeonidis09,Seymour10,Hwang10}. 
  
In addition, existing backward evolution models typically use only
``luminosity'' and/or ``density'' evolution which respectively evolves the
characteristic luminosity (i.e., $L_\star$) and the amplitude of the LF without
changing its shape. In other words, they do not consider any evolution in the
shape of the luminosity function. Finally, the intrinsically slow Monte-Carlo
methods used in many of the models make the search of
parameter space for the best-fit model laborious and tedious.

In this paper, we use a new and fast algorithm which is different from
conventional Monte-Carlo based algorithms, for calculating the source counts
for a given backward evolution model.  We also use a complete library of
IR SED templates which form a representative sample of observed galaxies,
at least at low redshifts. This will enable us to address questions about the
evolution of colour distribution of dusty galaxies during the history of the Universe.
We also consider a new evolution form for the LF which allows us to constrain
evolution in the shape of the LF in addition to quantifying its
amplitude changes.

The structure of the paper is as follows: in \S\ref{sec:ingredients} we
present different ingredients of our parametric colour-luminosity function
(CLF) evolution model and introduce a new algorithm for the fast calculation
of the source counts for a given CLF. Then, after choosing our observational
constraints for $850\,\mu$m objects in \S\ref{sec:observed} we try to find a model consistent with those constraints by studying the sensitivity of the produced source counts to different parameters in \S\ref{sec:modelprop}.
After finding a $850\,\mu$m-constrained best-fit model, we test its performance in producing source counts at other wavelengths in \S\ref{sec:others} which leads us to a model capable of producing source counts in a wide range of IR and submm wavelengths. Finally, after discussing the astrophysical implications of our best-fit model in \S\ref{sec:discussion}, we end the paper with concluding remarks. Throughout this paper, we use the
standard $\Lambda$CDM cosmology with the parameters
$\Omega_{\rm{m}}=0.3$, $\Omega_{\rm{\Lambda}}=0.7$ and $h=0.75$.

\section{Model ingredients}
\label{sec:ingredients}
Although luminosity is the main parameter to differentiate between different
galaxies, it is an integrated property of SED and cannot distinguish between
different SED shapes corresponding to different physical processes taking
place in different objects with similar outgoing integrated energy fluxes.
Using colour indicators can resolve this degeneracy. 
\cite{Dale01} demonstrated that the $R(60,100)$ which is defined as
$\log(S_{60\,\mu \rm{m}}/S_{100 \,\mu \rm{m}})$, is the best single parameter
characterization of IRAS galaxies. 
Moreover, it was shown that IRAS galaxies exhibit a slowly varying
correlation between $R(60,100)$ and luminosity such that objects with
larger characteristic luminosities have warmer characteristic colours 
\citep{Dale01,Chapman05,Chapin09}. Based on these facts, we construct a model
for the Colour-Luminosity Function (hereafter CLF). This model consists of the
observed CLF in the local Universe and a parametric evolution function. Then,
by adopting a suitable set of SED models, we calculate the source count of
IR objects at different wavelengths and compare the results with observations. 
\subsection{CLF at $z=0$}
\label{sec:CLF0}
The local volume density of IRAS galaxies, $\Phi_0(L,C)$, can be parametrized as
a function of total IR luminosity, $L$, and $R(60,100)$ colour, $C$,
where the total infrared luminosity is calculated by integrating over the SED
from $3$ to $1100\,\mu$m \citep{Dale01}. Furthermore, it is possible to express
the local colour-luminosity distribution as the product of the local luminosity
function, $\Phi_0(L)$, and the local conditional probability of a galaxy
having the colour $C$ given the luminosity $L$, $P_0(C|L)$,  
\begin{equation}
\label{eq:CLF-product}
\Phi_0(L,C) = \Phi_0(L)P_0(C|L).
\end{equation}

Since the IRAS galaxies represent an almost complete sample of IR galaxies
out to the redshift $\sim 0.1$ \citep{Saunders90}, we use the luminosity
and colour functions found by analyzing a flux limited sample of
$S_{60\,\mu {\rm{m}}}> 1.2{\rm{Jy}}$ IRAS galaxies \citep{Chapman03, Chapin09}.
\citet{Chapman03} and \citet{Chapin09} analyze this sample which covers
most of the sky, using an accessible volume technique for finding the
LF and fit a dual power law function to the observed luminosity distribution.
The parametric form of luminosity function based on \citet{Chapin09} is given by
\begin{equation}
\label{eq:LF0}
\Phi_0(L) = \rho_*(\frac{L}{L_*})^{1-\alpha}(1+\frac{L}{L_*})^{-\beta},
\end{equation}
where ${L}_*=5.14 \times 10^{10} {L}_{\odot}$ is the characteristic knee
luminosity, $\rho_*=1.22\times 10^{-14} {\rm{Mpc}}^{-3}L_{\odot}^{-1}$ is the
number density normalization of the function at $L_*$,  and $\alpha=2.59$
and $\beta=2.65$ characterize the power-laws at the faint ($L < L_*$) and
bright ($L > L_*$) ends, respectively.

\citet{Chapman03} and \citet{Chapin09} also found a Gaussian representation for the colour distribution of IRAS galaxies
\begin{equation}
\label{eq:CF0}
P_0(C\mid L) = \frac{1}{\sqrt{2\pi} \sigma_{\rm{c}}}\exp[-\frac{1}{2}\times (\frac{C-C_0}{\sigma_{\rm{c}}})^2],
\end{equation}      
where $C_0$ can be represented by a dual power law function
\begin{equation}
\label{eq:c0}
C_0 = C_* - \delta \log(1+\frac{L'}{L})+\gamma \log(1+\frac{L}{L'}),
\end{equation}
with $C_*=-0.48$, $\delta=-0.06$, $\gamma=0.21$ and
$L'=3.2 \times 10^{9} L_{\odot}$, and the distribution width,
$\sigma_{\rm{c}}$, is expressed as
\begin{equation}
\label{eq:sigma_c}
\sigma_{\rm{c}} = \sigma_{\rm{f}}(1-2^{-L'/L})+ \sigma_{\rm{b}}(1-2^{-L/L'}),
\end{equation}
where $\sigma_{\rm{f}}=0.2$ and $\sigma_{\rm{b}}=0.128$ \citep{Chapin09}.
\subsection{CLF evolution}
\label{sec:CLFz}
The CLF introduced in the previous section, contains information about the
distribution of IR galaxies only in the nearby Universe. For modeling the IR
Universe at higher redshifts, its evolution must be modeled.
The necessity of the CLF evolution with redshift is shown directly by
the fact that the observed power of the the CIB is comparable to what
can be deduced from the optical cosmic background cannot be
explained by our understanding of the local Universe which
indicates the infrared output of galaxies is only one third of their
optical output \citep{Lagache05}. Furthermore, several studies have
shown for a fixed total IR luminosity the typical temperature of
infrared sources is lower at higher redshifts
\citep{Chapman05,Pope06,Chapin09,Symeonidis09,Seymour10,Hwang10,Amblard10},
which advocates an IR colour evolution. Therefore, we need to adopt a
reasonable form of evolution in the luminosity and colour distributions to
reproduce correctly the CLF evolution of IR galaxies with redshift.

There are three different ways to evolve the local luminosity function of
IR galaxies: {\bf{(i)}} changing $\rho_*$ with redshift (i.e.,
density evolution); {\bf{(ii)}} changing $L_*$ with redshift (i.e.,
luminosity evolution) and {\bf{(iii)}} changing the bright and faint end
slopes (i.e., $\alpha$ and $\beta$) with redshift. While the density and
luminosity evolution change the abundances of all sources independent of their
luminosities, they leave the shape of the LF unchanged.
Such models are called {\it translational\/} models since they amount to only a
translation of the LF in parameter space. Going beyond translational models,
a variation of the shape of LF with redshift
(i.e., varying $\alpha$ and $\beta$ slopes in equation \ref{eq:LF0})
can be used to change the relative contributions of bright and faint sources
at different redshifts.
    
As discussed by \citet{Blain99}, luminosity and density evolution affect
  the CIB predicted by the models similarly, but luminosity evolution has a much
  stronger effect on the source counts. Combining source counts and integrated
  background can therefore distinguish between luminosity and density
  evolution. The result of this analysis shows luminosity evolution must
  strongly dominate over density evolution, since pure density evolution
  consistent with the observed $850\,\mu$m source counts would overpredict the
  integrated background by a factor 50 to 100 \citep{Blain99}. We therefore
  assume negligible density evolution.
However, as we will show in Section 4, luminosity evolution is not
sufficient to reproduce the correct source count and redshift distribution of
submm galaxies, which forces us to drop the assumption of a purely translational
  LF evolution model. Moreover, there is no reason to assume that the slopes of
LF at faint and bright ends remain the same at all redshifts. Therefore, 
we allow them to change in our model and introduce a redshift dependent
LF which can be written as
\begin{equation}
\label{eq:LF-lum_evol}
\Phi(L,z) = \rho_*(\frac{L}{g(z) L_*})^{1-\alpha_z}(1+\frac{L}{g(z) L_*})^{-\beta_z},
\end{equation}
which is similar to equation (\ref{eq:LF0}) but now $\alpha_z$ and
$\beta_z$ are changing with redshift and $L_*$ is multiplied by $g(z)$,
which we refer to as the luminosity evolution function.

Since the average total-IR-luminosity density and its associated 
star formation density closely follow the luminosity evolution function,
we choose a form of luminosity evolution which is similar to the observed evolution of
the cosmic star formation history (see \citet{Hopkins06}); in other words, we assume that 
the luminosity evolution is a function which is increasing at
low redshifts and after reaching its maximum turns into a decreasing 
function of look-back time at high redshifts: 
\begin{equation}
\label{eq:g_z}
g(z) =\left\{\begin{array}{lll}(1+z)^{n}&\mbox{ if } z\leq z_{a}  \\
(1+z_a)^{n}&\mbox{ if } z_{a}<z\leq z_b \\
(1+z_a)^{n}(1+z-z_b)^{m}&\mbox{ if } z_{b}<z \end{array}\right.
\end{equation}
where $n$, $m$, $z_a$ and $z_b$ are constants. As we will show in Section~\ref{sec:lum_evol}, the source count is not strongly sensitive to the model properties at high redshifts. In other words, the sensitivity of the source count calculation to the difference between $z_a$ and $z_b$ is much less than its high sensitivity to
the value of $z_{a}$ itself. Moreover, the model outputs are not highly
sensitive to the slope of $g(z)$ at high redshifts if it remains negative 
(i.e., $m<0$). Therefore, we can simplify the model by assigning suitable 
fixed values to $z_{b}- z_{a}$ and $m$, without any significant change in its
flexibility. We choose $z_{b}- z_{a} = 1$ and $m = -1$, knowing that any
different choice for those values can be compensated by a very small change
in $z_{a}$ or/and $n$. we will discus this issue in more detail in
Section~\ref{sec:lum_evol}.

We also adopt linear forms for changing $\alpha_z$ and $\beta_z$ with redshift 
\begin{equation}
\label{eq:alpha_z}
\alpha_z =2.59+a_{\alpha}z,
\end{equation}
\begin{equation}
\label{eq:beta_z}
\beta_z =2.65+a_{\beta}z,
\end{equation}
where $a_{\alpha}$ and $a_{\beta}$ are constants. We apply the slope evolution only for $z < \frac{z_a + z_b}{2}$ noting that the observed source count is not very sensitive to the exact properties of our model at very high redshifts
and extending the evolution of LF slopes to even higher redshifts does not
change our results.

As mentioned earlier, some studies have shown that high redshift IR galaxies
have lower dust temperatures than low redshift galaxies, at a fixed total
IR luminosity. In our model, in order to evolve the colour distribution
accordingly, we adopt a relation similar to \citet{Valiante09} to shift the
centre of the colour distribution for a given luminosity towards colours
associated with lower dust temperatures (i.e., smaller $R(60,100)$ values)
at higher redshifts (see Figure~\ref{fig:frac_evol}). In this way, the colour distribution of
IR galaxies can be written as
\begin{equation}
\label{eq:CFz}
P(C\mid L) = \frac{1}{\sqrt{2\pi} \sigma_{\rm{c}}}\exp[-\frac{1}{2}\times (\frac{C-C'_0}{\sigma_{\rm{c}}})^2],
\end{equation}      
where $\sigma_{\rm{c}}$ is given by equation (\ref{eq:sigma_c}) and $C'_0$ is
\begin{equation}
\label{eq:c0z}
C'_0 = C_* - \delta \log(1+\frac{L'(1+z)^w}{L})+\gamma \log(1+\frac{L}{L'(1+z)^w}),
\end{equation}
where $w$ is a constant. Although for colour evolution we adopt a similar
formalism to \citet{Valiante09}, unlike them we allow $w$ to vary as a free
parameter which enables us to constrain the colour evolution as well.

Based on the above formalism, the general CLF of IR galaxies can be written as
\begin{equation}
\label{eq:CLFz}
\Phi(L,C,z) = \Phi(L,z)P(C|L),
\end{equation}
where $\Phi(L,z)$ and $P(C|L)$ are defined by equations
(\ref{eq:LF-lum_evol})-(\ref{eq:c0z}).
\subsection{SED Model}
\label{sec:SED}
\begin{figure}
\centerline{\hbox{\includegraphics[width=0.5\textwidth]
             {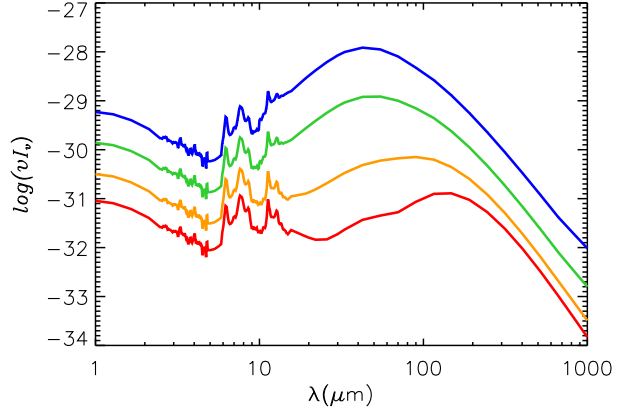}}}
\caption{Some SED samples from the template library of
\citet{Dale02} which we use in our model.
The blue, green, orange and red lines (from top to bottom) are respectively
associated with colours $C =  0.16$, $0.04$, $-0.27$ and $-0.54$.
The normalization is arbitrary.}
\label{fig:SED}
\end{figure}
Several studies have shown that AGNs do not dominate the FIR energy output of
the Universe \citep{Swinbank04,Alexander05,Chapman05,Lutz05,Valiante07,Pope08,Menendez09,
  Fadda10, Jauzac10}.  Therefore, in order to avoid unnecessary complexity
  in our model we adopt a single population of galaxies, modeled with only one
  family of SEDs representing star forming galaxies. 
  
  This choice is justified for long wavelengths, for instance at $850\,\mu$m, by considering the fact that at those wavelengths the AGN contribution to the observed flux is negligible. In other words, to be able to change the $850\,\mu$m fluxes and source counts, AGN should be the dominant contributor to the SED at rest-frame wavelengths longer than $\sim 200\,\mu$m which is highly unlikely.
  
 On the other hand, excluding AGN contribution could be potentially important at shorter wavelengths: if we adopt a simple assumption where the AGN continuum is well represented by a simple torus model \citep{Efstathiou95,Valiante09}, then a strong enough AGN could create a bump, on top of the starburst SED, at rest-frame wavelength ranges $\sim 10-50\,\mu$m (see Figure 2 in \citet{Efstathiou95} and Figure 9 in \citet{Valiante09}). This feature, together with increasing AGN luminosity with redshift could modify the observed source counts at observed wavelengths shorter than $200\,\mu$m but still is unlikely to affect submm counts. 
 
 However, \citet{Mullaney11} recently used the deepest available Herschel survey and showed that for a sample of X-ray selected AGNs up to $z\sim3$, the observed $100\,\mu$m and $160\,\mu$m fluxes are not contaminated by AGN. Consequently, even if all the galaxies which contribute to the FIR and submm source counts host AGN, their observed fluxes is driven by star formation activity at wavelengths around $100\,\mu$m or longer.

In order to get accurate SED templates for star-forming galaxies, we use the
\cite{Dale01} and \citet{Dale02} SED models which are produced by a semi-empirical method
to represent spectral energy distribution of star-forming galaxies in the
IR region of spectrum. \cite{Dale01} add up emission profiles of
different dust families (i.e. large grains, very small grains and polycyclic
aromatic hydrocarbons) which are exposed to a range of radiation field
strengths in a parametrized form to generate the SED of different star-forming
systems. In this way, it is possible to produce the SED corresponding to each
$R(60,100)$ colour and scale it to the desired total infrared luminosity.
In our model, we use spectral templates taken from the \cite{Dale02} catalog
which provides 64 normalized SEDs with different $R(60,100)$ colours, ranging
from $-0.54$ to $0.21$. Some SED examples taken from this template set
are illustrated in Figure~\ref{fig:SED}.       
\subsection{The algorithm}
\label{sec:algorithm}
Given the distribution of objects in the Universe, and their
associated SEDs, it is possible to calculate the number of sources which
have observed fluxes above some detection threshold, $S_{\rm{th}}$,
at a given wavelength $\lambda = \lambda_{\rm{obs}}$:
\begin{equation}
\label{eq:N-general}
N (>S_{\rm{th}})  = \int\int\int Q\times \Phi(L,C,z) \frac{dV}{dz}dz dL dC,
\end{equation}
where $V$ is the volume and $Q$ is the probability that a source
with luminosity $L$, colour $C$ and redshift $z$ has an observed flux density
greater than $S_{th}$ at that wavelength (i.e., its detection probability).
At each point of redshift-colour-luminosity space, $Q$ is either $1$ or $0$
for any particular galaxy, but it is useful to think of $Q$ as the average
probability of detection for galaxies in each cell of that space. 

In order to calculate $Q$, we first note that for each source with a given
luminosity, $L$, and colour, $C$, there is a redshift, $z_{\rm{max}}$,
at which the observed flux is equal to the detection threshold
\begin{equation}
\label{eq:z_max}
S(L,C,\lambda_0,z_{\rm{max}}) = S_{\rm{th}}= \frac{(1+z_{\rm{max}})L_{\nu}(L,C,\lambda_0)}{4\pi D_L^2(z_{\rm{max}})},
\end{equation}
where $D_L$ is the luminosity distance and $L_{\nu}$ is the rest-frame
luminosity density ($\rm{W\,Hz}^{-1}$) of the object with total luminosity
$L$ and colour $C$ at wavelength $\lambda_0 = \lambda_{\rm{obs}}/(z+1)$.

Now consider a cell defined by redshift interval $\Delta z = z_2-z_1$
(where $z_1 < z_2$), luminosity interval $\Delta L = L_2 - L_1$ and
colour interval $\Delta C = C_2-C_1$. Assuming negligible colour and luminosity evolution between $z_1$ and $z_2$, the average detection probability in
the cell is equal to the fraction of detectable objects in that cell which is: 
\begin{equation}
\label{eq:Q}
Q =\left\{\begin{array}{lll}1&\mbox{ if } z_{\rm{max}}>z_2\\
\frac{D^3(z_{\rm{max}}) - D^3(z_1)}{D^3(z_2) - D^3(z_1)}&\mbox{ if }z_1\leq z_{\rm{max}}\leq z_2  \\
0&\mbox{ if } z_{\rm{max}} < z_1\end{array}\right. 
\end{equation}
where $D(z)$ is the proper distance equivalent to redshift $z$.
For writing equation \ref{eq:Q}, we assumed galaxies to be distributed uniformly
in space between redshifts $z_1$ and $z_2$.
Consequently, based on equation \ref{eq:N-general}, the number of objects
which are contributing to the source count in each cell of the
redshift-colour-luminosity space is    
\begin{equation}
\label{eq:dN-interval}
\Delta N (>S_{\rm{th}})  = Q\times \bar{\Phi}(L,C,z)\Delta L \Delta C \Delta V ,
\end{equation}
where $\Delta V$ is the volume corresponding to the redshift interval
$\Delta z$ and $\bar{\Phi}(L,C,z)$ is the average CLF in the cell,
which for small enough $\Delta L$, $\Delta C$ and $\Delta z$ can be written as
\begin{equation}
\label{eq:CLF-aver}
\bar{\Phi}(L,C,z) = \Phi(\bar{L},\bar{C},\bar{z}).
\end{equation}
Then, the total source count is obtained by summing over the contribution
from all cells (see Appendix A).

As a result of using $Q$ and its special form as expressed in
equation \ref{eq:Q}, our algorithm computes the source count for continuously
distributed sources in the Universe, independent of the size of $\Delta z$
which is used in equation \ref{eq:N-general}. Thanks to this property, the
computational cost required for the source count calculation is reduced
significantly, and the small size of $\Delta z$ is only necessary for
accurate calculation of $K$-corrected SEDs and CLFs at different redshifts
and not to guarantee a uniform distribution of sources (see also Appendix A).\\

Now that we have all the tools ready, we can test the capability of our model
in reproducing the observed properties of $850\,\mu$m sources and find which
particular choices of model parameters are implied. In the following sections,
we first discuss the observational constraints that we want to reproduce with
our model and then we will proceed with finding the best-fit model which can
reproduce those properties. 

\begin{figure}
\centerline{\hbox{\includegraphics[width=0.5\textwidth]
             {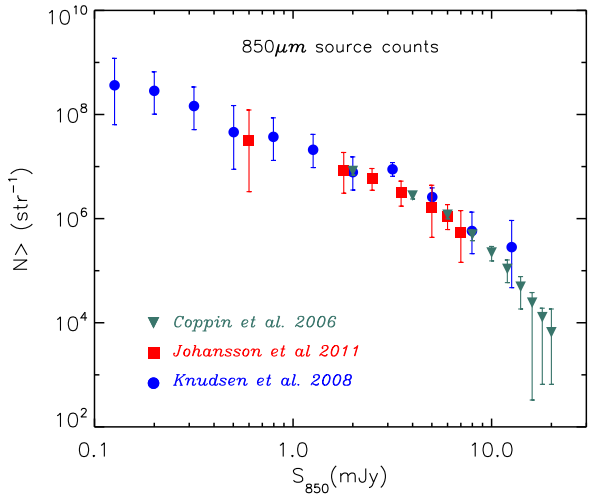}}}
\caption{A compilation of some observed $850\,\mu$m source counts.
Blue circles, green triangles and red squares are respectively data points
taken from \citet{Knudsen08}, \citet{Coppin06} and \citet{Johansson11}.}
\label{fig:850data}
\end{figure}
\section{$850\,\mu$m Observational Constraints}
\label{sec:observed}
\subsection{Observed $850\,\mu$m source count}
Among several existing extragalactic submm surveys which provide source counts
at $850\,\mu$m \citep{Coppin06,Weiss09,Austermann10}, the SCUBA Half-Degree
Extragalactic Survey (SHADES) \citep{Coppin06} is the largest one which has the
most complete and unbiased sample. However, this survey and other surveys which
use JCMT and the same blank field method are restricted by the JCMT confusion
limit of $\sim2$mJy at $850\,\mu$m and cannot probe the source counts of the
fainter population. Using a complementary method, the lensing
technique has been used to probe $850\,\mu$m source counts to flux thresholds as low as
$0.1$mJy \citep{Smail02,Knudsen08,Johansson11}. As we will show later in this section, the sensitivity of bright and faint submm source counts to the evolution of dusty galaxies is completely
different and it is essential to incorporate a large dynamic range to constrain
possible evolutionary scenarios. Therefore, we use the best available
observational information at both faint and bright tails of $850\,\mu$m source
count, by combining all of the SHADES data points with those of
\citet{Knudsen08} at flux thresholds $<2$mJy to assemble our reference source
count which is also in agreement with other observations (see Figure~\ref{fig:850data}).

\subsection{Redshift distribution of bright $850\,\mu$m sources}
Although the observed number counts at $850\,\mu$m, especially at faint fluxes,
are sensitive to the redshift distribution of infrared galaxies,
they do not constrain it directly. As we will discuss later, different models
with different redshift distributions can reproduce the $850\,\mu$m source
counts with the same accuracy.  Therefore, it is important to impose an
additional constraint on redshift distribution of those objects.

Unfortunately, studying the redshift distribution of submm galaxies is extremely
difficult and there are only a few works on spectroscopically confirmed
redshift distribution of bright submm galaxies with observed fluxes grater than
$\sim 4-5$ mJy \citep{Chapman05,Wardlow11}. Those studies show a redshift distribution peaking around $z\sim 2$ (see the right panel in Figure~\ref{fig:850-best}). We use this redshift distribution as one of our observed constraints and force our model to reproduce a redshift distribution
which peaks at the same redshift.

\section{Finding $850\,\mu$m best-fit Model}
\label{sec:modelprop}
In order to be able to extract meaningful trends and differentiate between them,
it is important to keep our phenomenological model as simple as possible. It is
therefore desirable to find the minimum number of free parameters without which
the model cannot produce an acceptable result.  Furthermore, the role of
different parameters are not identical: while some parameters are not strongly
constrained, small changes in others can change the result significantly. It
is also important to look at degeneracies between various parameters; some
parameters are not independent and varying one may be compensated by varying
the others. In this section we investigate our model
to understand those issues before presenting our best-fit model. 
\begin{figure*}
\centerline{\hbox{\includegraphics[width=0.45\textwidth]
             {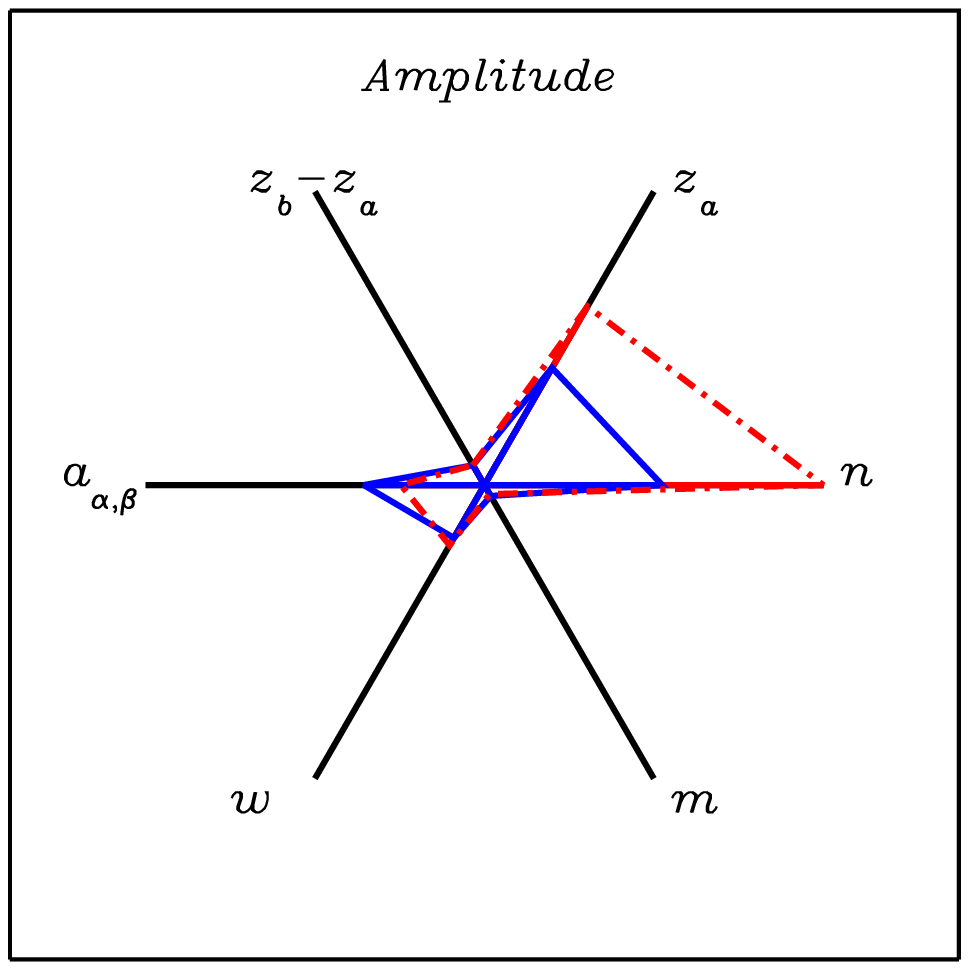}}
            \hbox{\includegraphics[width=0.45\textwidth]
             {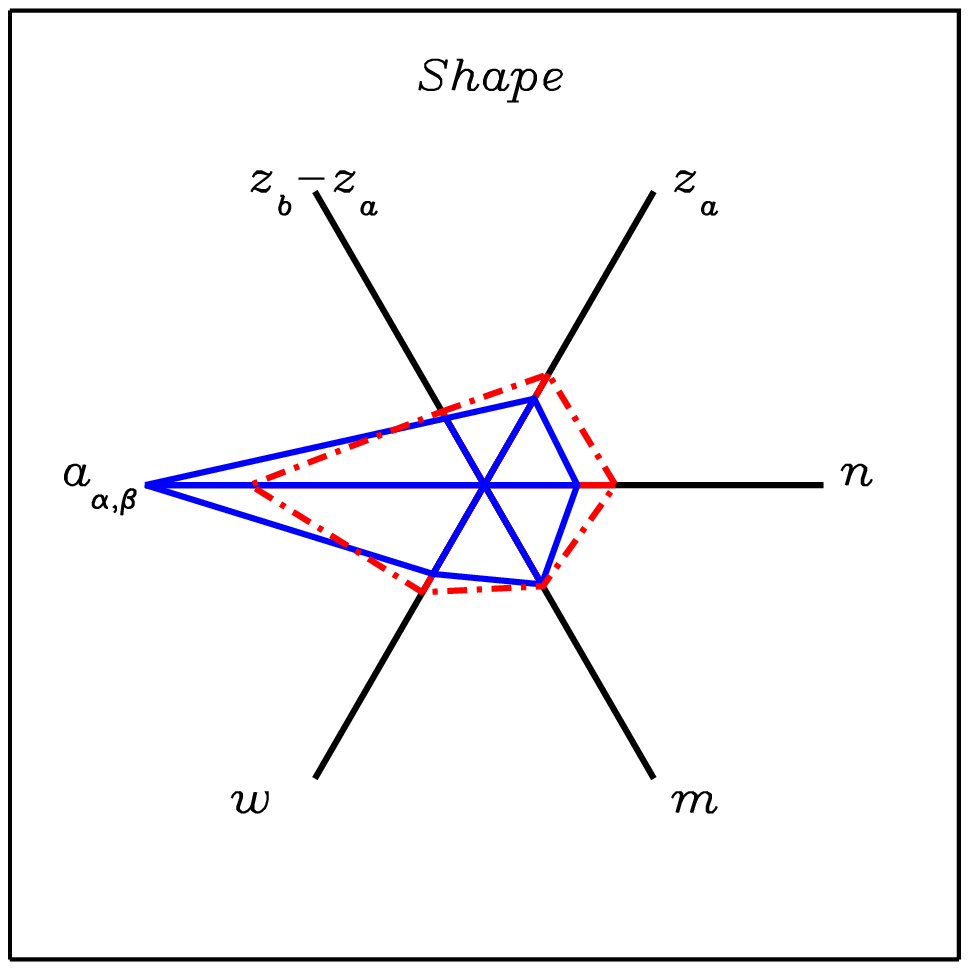}} }
\caption{The sensitivity of the model to various parameters. While the left
panel illustrates how much the variation of different parameters could
change the amplitude of the $850\,\mu$m source count curve, the right panel
shows their effect in changing the shape of the source count curve. To measure
the amplitude changes, we varied each parameter by $10\%$ around its best-fit
value and measured the difference in the total source counts. The shape
sensitivity is probed by measuring the relative change each varying parameter
can cause in the ratio between two source counts at faint and bright flux
thresholds (i.e. $\frac{N_{1 {\rm{mJy}}}}{N_{{\rm{1 Jy}}}}$). The lengths of
the coloured segments on each axis shows how large those changes are.
The blue (red) segments which are connected by solid (dash-dotted) lines
represent the result as we increased (decreased) the value of each parameter.
We show parameters $a_{\alpha}$ and $a_{\beta}$ on one axis since their effect
in changing the source count properties is similar.} 
\label{fig:shape-amp}
\end{figure*}
\begin{figure*}
\centerline{\hbox{\includegraphics[width=0.25\textwidth]
             {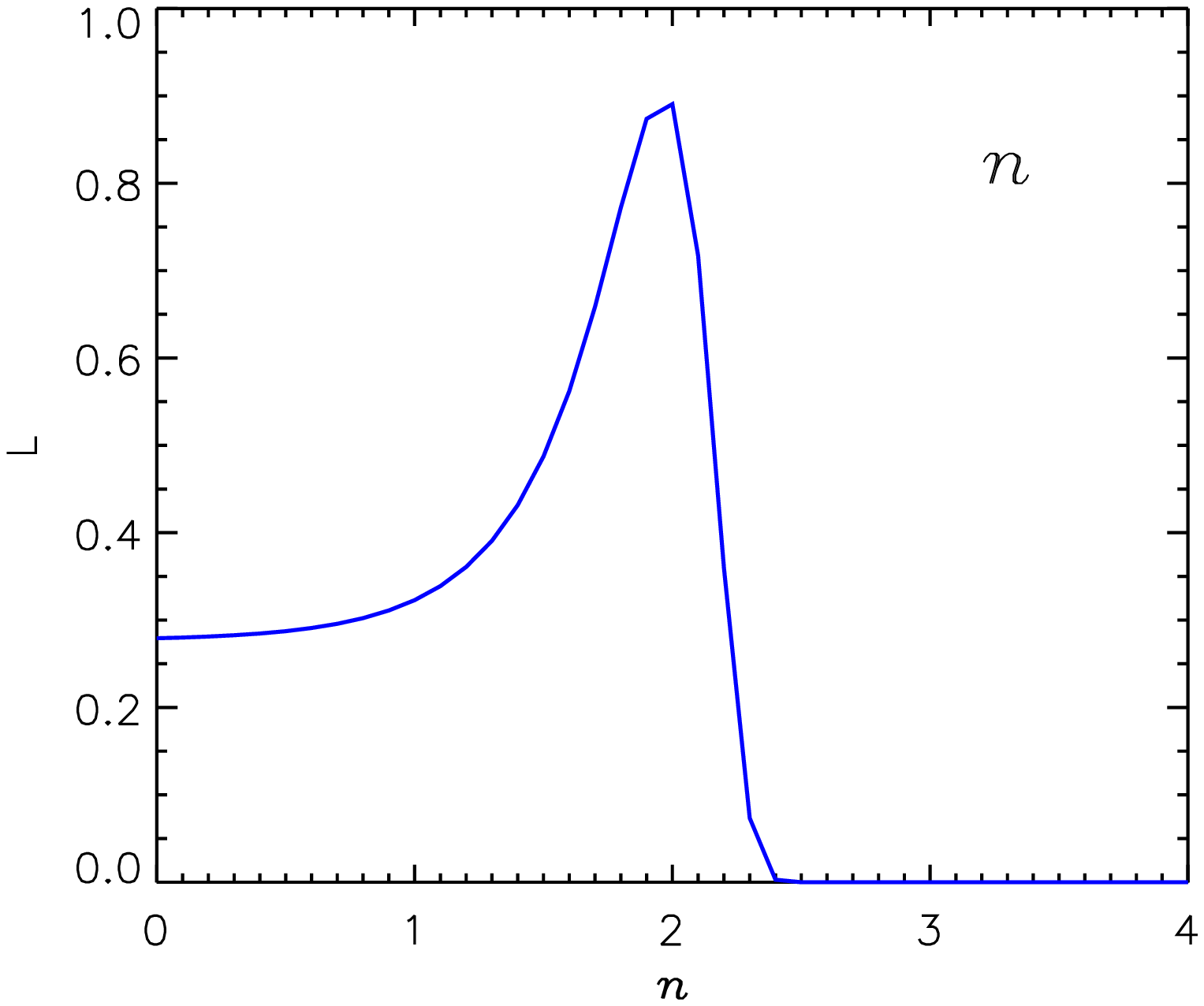}}
            \hbox{\includegraphics[width=0.25\textwidth]
             {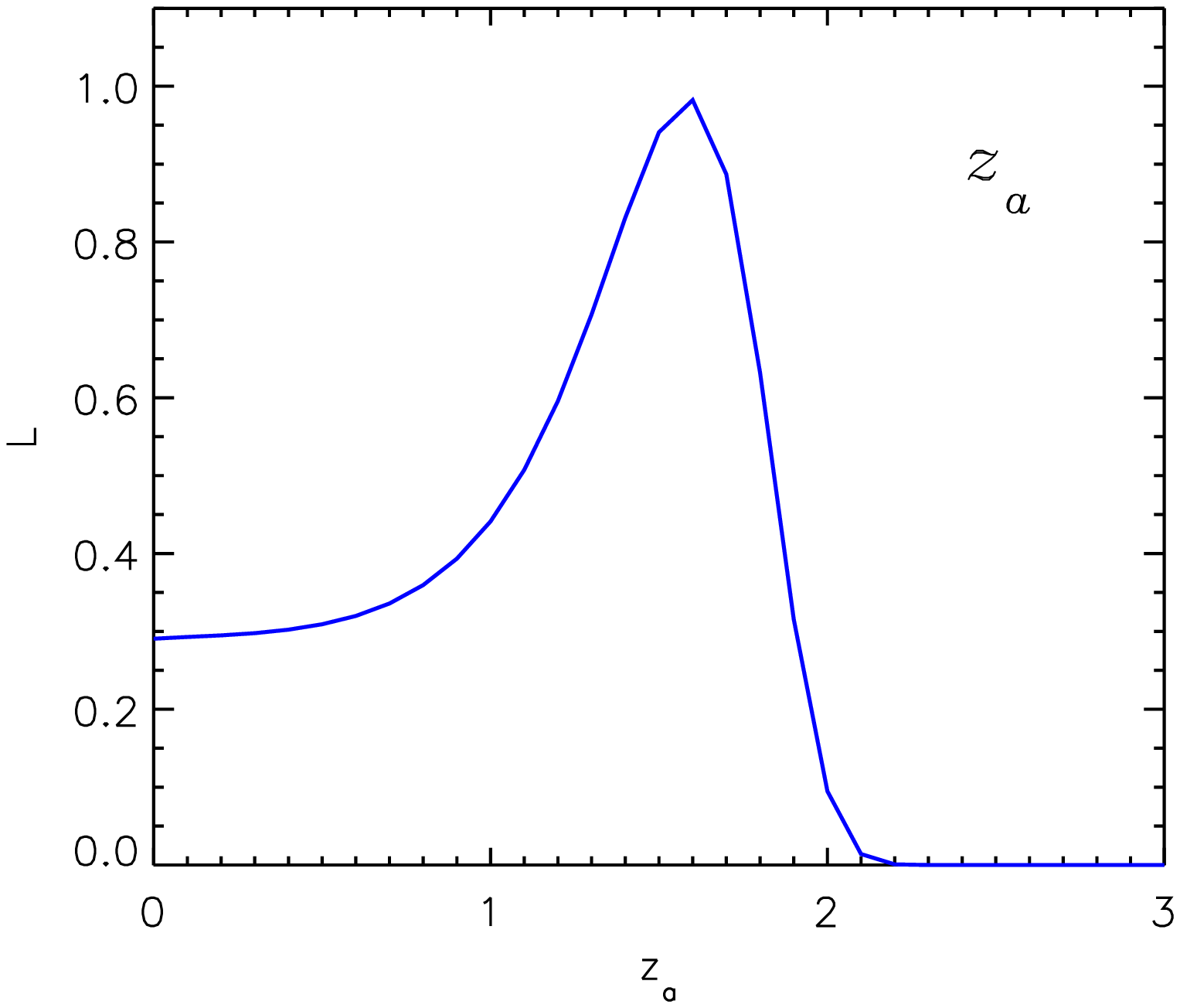}}
            \hbox{\includegraphics[width=0.25\textwidth]
             {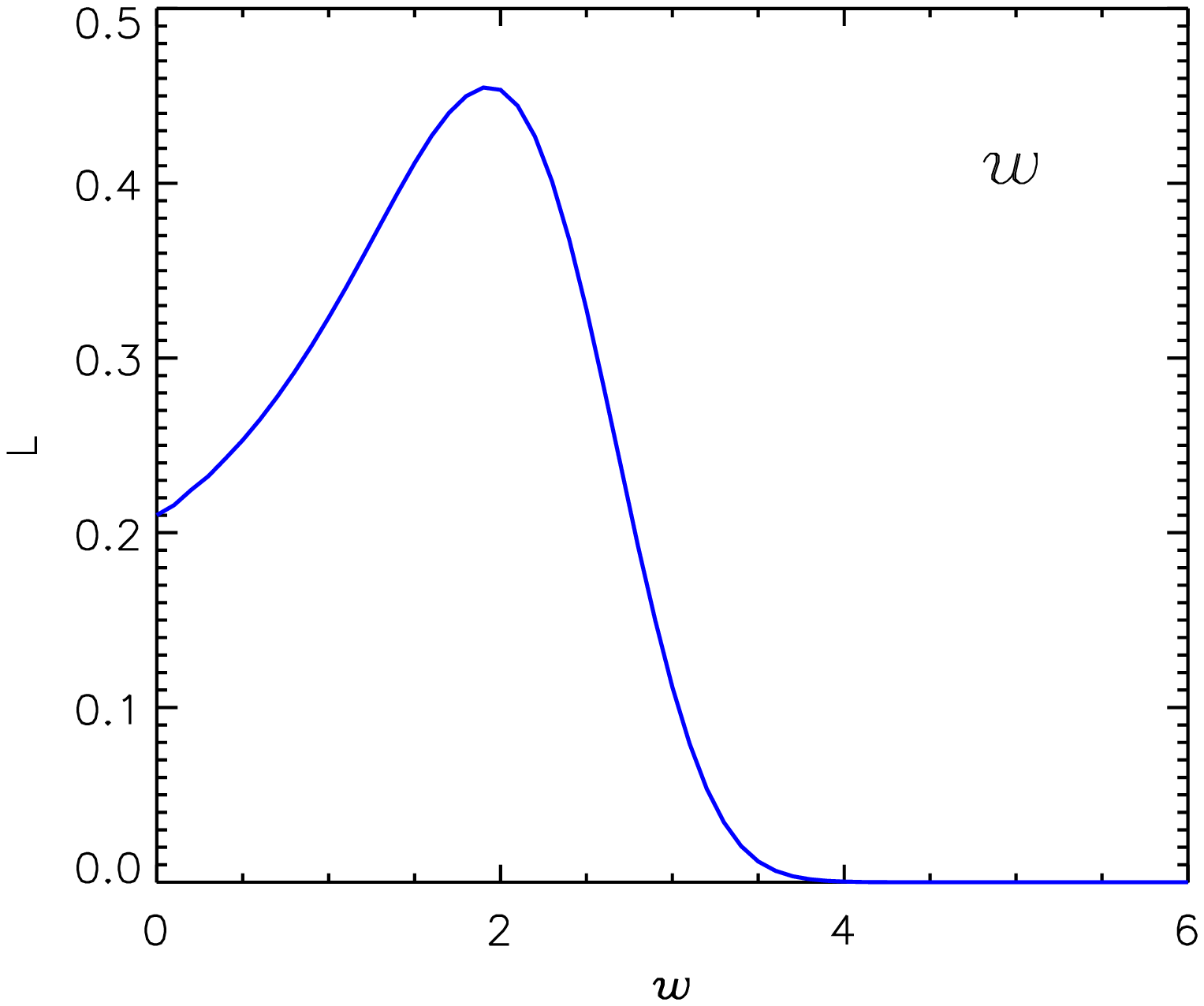}} }
\centerline{\hbox{\includegraphics[width=0.25\textwidth]
             {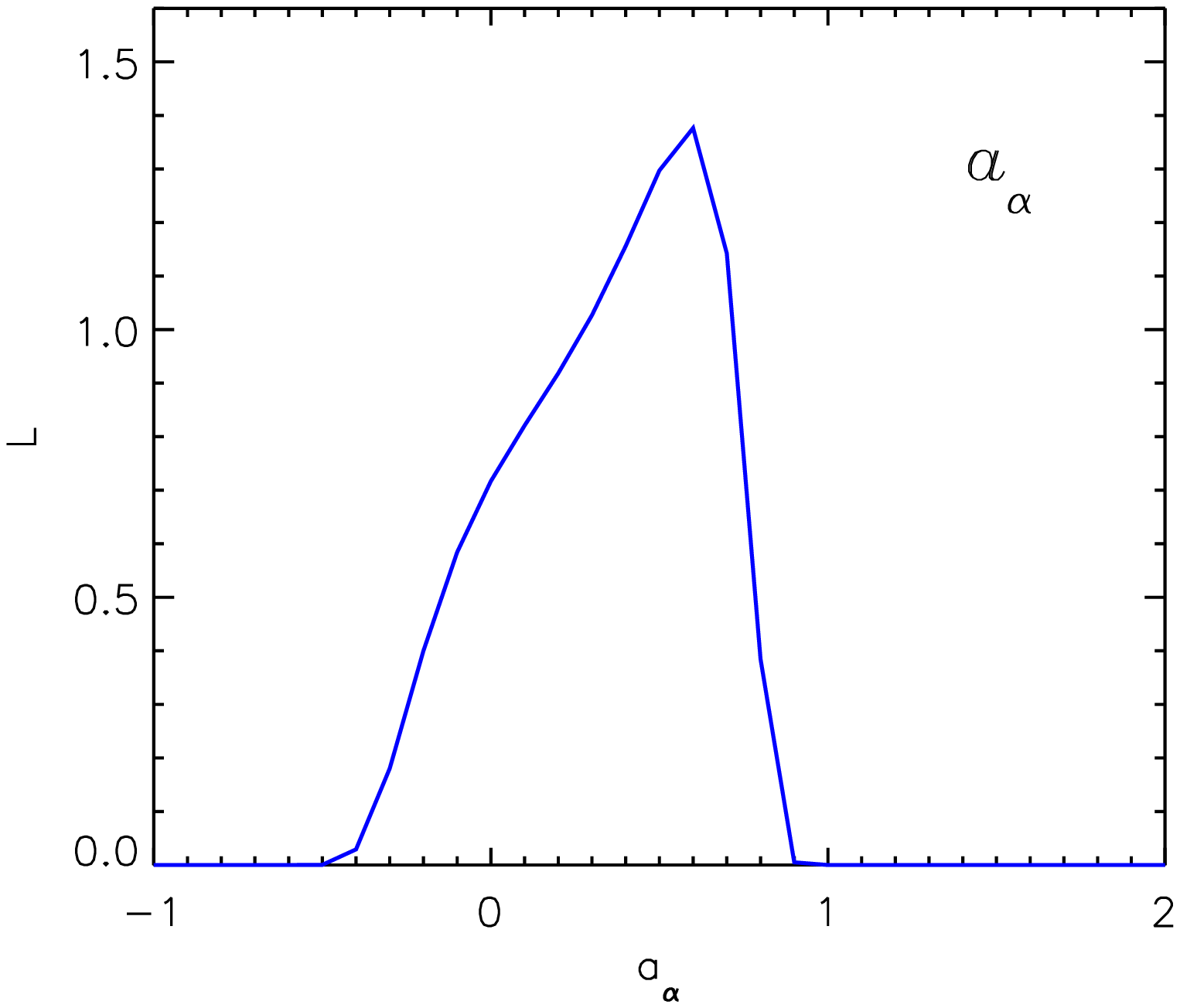}}
            \hbox{\includegraphics[width=0.25\textwidth]
             {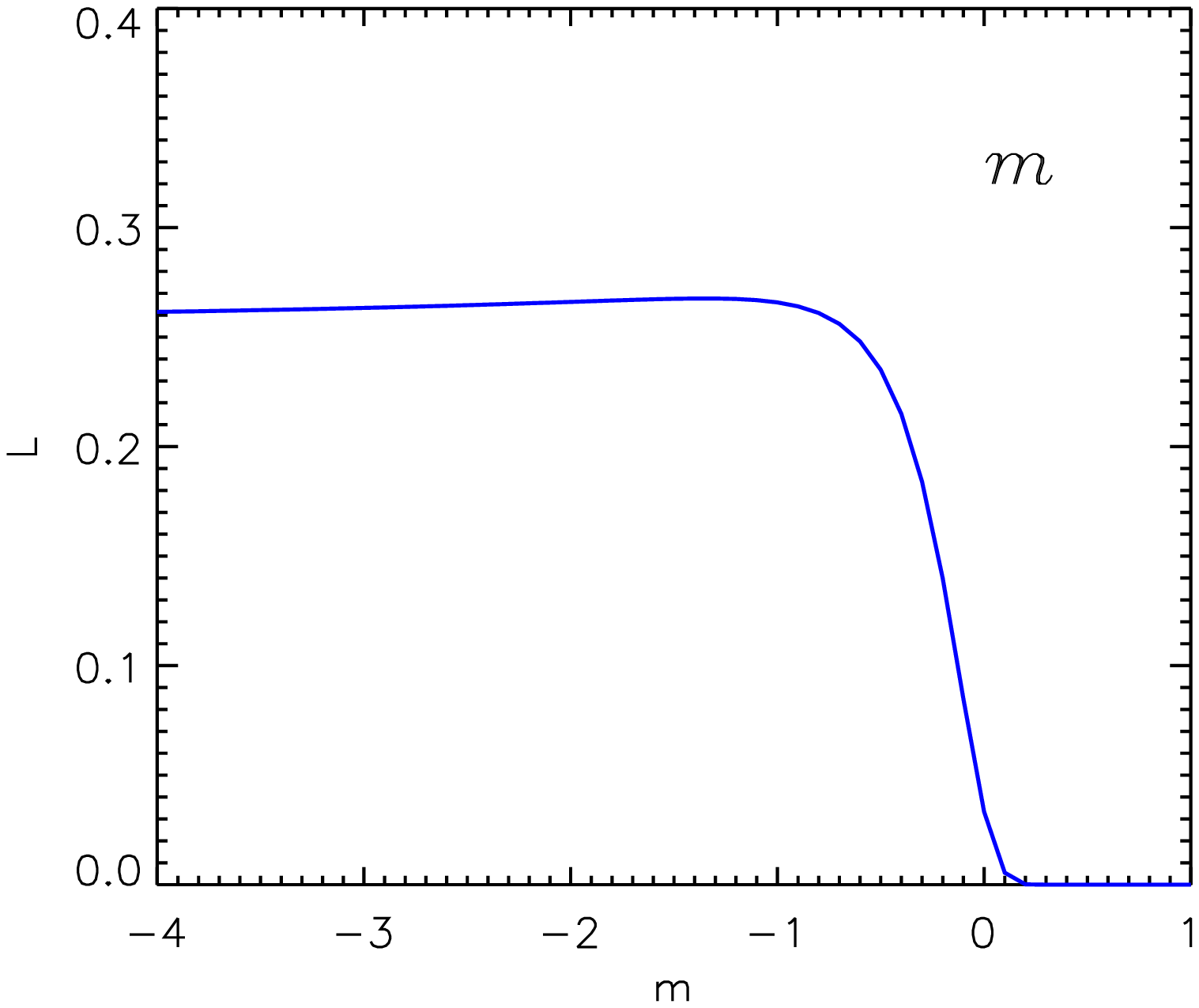}}
            \hbox{\includegraphics[width=0.25\textwidth]
             {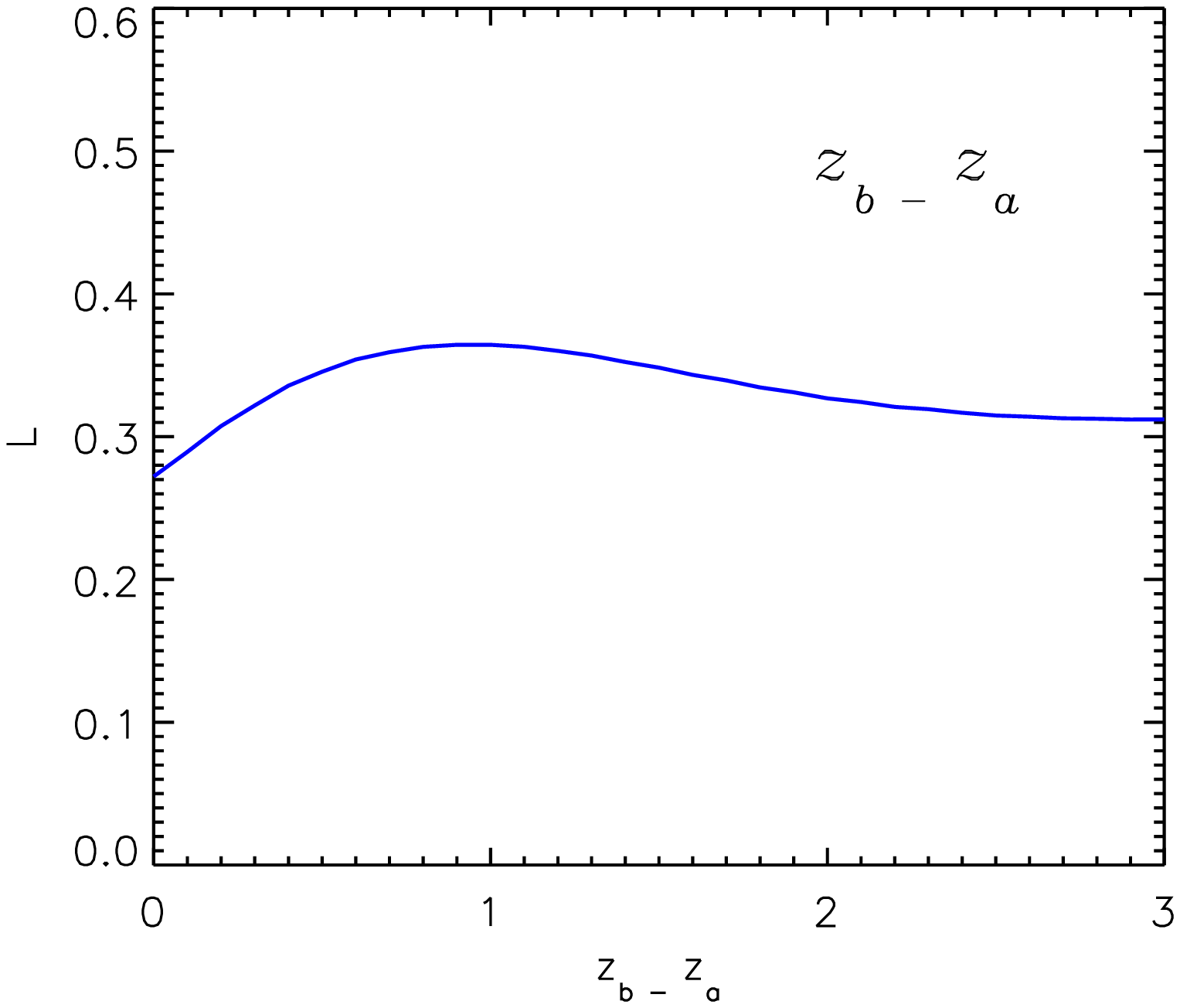}} }
\caption{The likelihood (i.e. $exp(-\frac{1}{2}\chi^2)$) distribution of
different parameters for the best-fit model. For each panel, we fixed all the
parameters at their best-fit values and changed only one parameter at a time.
By varying each parameter, the quality of the fit and hence the likelihood is
changing. We normalized the amplitude of curves such that the area under each
curve (i.e. the total probability) be unity.} \label{fig:all-param}
\end{figure*}
\subsection{The source count curve: Amplitude vs. Shape}
\label{sec:amp-shape}
To match the observed source counts, our model should be able to reproduce both
the typical number of sources (i.e., the amplitude of the source count curve as
a function of flux threshold) and the ratio between the source counts at faint
and bright flux thresholds (i.e., the shape of the source counts curve).
Following this line of argument, we can categorize our model parameters into two
different groups: those which play a stronger role in forming the amplitude of
the curve and those which mainly affect its shape.
As mentioned earlier, the luminosity evolution (see equation  \ref{eq:g_z})
has the dominant role in determining the amplitude of the source count curve
while its shape is controlled by other ingredients of our model which are
colour evolution and the evolution of LF slopes.
 
These trends are illustrated in Figure~\ref{fig:shape-amp}. In the left panel,
the relative contribution of each parameter in changing the amplitude of the
source count curve is visualized; the length of the coloured segments along each
axis (coloured segments along different axis are connected to each other)
represents the sensitivity of source count amplitude to that parameter. The
length of coloured segments is computed by changing all the model parameters
(one at a time) by the same fraction, for instance $10\%$, and measure the
change caused in the total source count. The segments which are connected by
dot-dashed (red) and solid (blue) lines correspond respectively to a decreasing
and an increasing parameter. In the right panel, the relative contribution of
different parameters in controlling the shape of the source counts curve is
shown.  In this figure, the length of coloured segments represents the change in
the ratio between faint and bright source counts. This time, we measured how
much the ratio between a very faint source count, like $1$ mJy, and a very bright
one, like $1$ Jy, is changing due to a fixed change in different parameters
(e.g., $10\%$). The dot-dashed (red) and solid (blue) lines connect the coloured
segments which correspond to an increase or a decrease in the values of the
parameters.

As is evident from these diagrams, the amplitude of the source count curve is
strongly sensitive to the parameters of the luminosity evolution, particularly
$z_a$ and $n$, while its shape is determined mainly by an evolution in the shape
of LF and/or the strength of the colour evolution.

\subsection{The luminosity evolution}
\label{sec:lum_evol}
The luminosity evolution is the backbone of our model and has the most
important role in producing the observed source counts and the redshift
distribution of $850\,\mu$m objects. However, it is not surprising that the
calculated source count is not strongly sensitive to the model properties at
high redshifts. This is mainly because objects at very high redshifts 
have decreasing fluxes (despite the advantageous $K$-correction).
Consequently, the chosen value of $m$ has a negligible effect on the
amplitude and shape of the source count curve (see Figure~\ref{fig:shape-amp}). 
Nevertheless, one should note that this argument would not necessarily work if
the population of bright IR galaxies continued to evolve with look-back time
for all redshifts. In other words, the exact value of $m$ is not strongly
constrained by the observed source counts and all the negative values can
produce similar results (see the middle panel in the bottom row of Figure~\ref{fig:all-param}).
\begin{figure}
\centerline{\hbox{\includegraphics[width=0.5\textwidth]
             {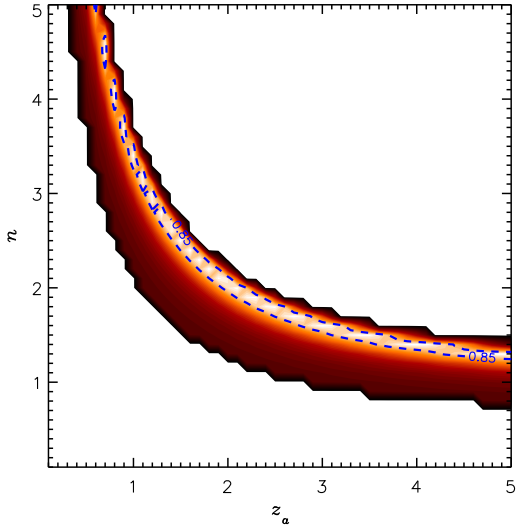}}}
\caption{Map of likelihoods for models with different $n$ and $z_a$
where $[w, a_{\alpha}, a_{\beta}] = [0, 0, 0]$. The region which contains
the best-fit models is in the middle of the coloured band and surrounded by
the dashed blue contours which indicate the likelihood of $0.85$.}
\label{fig:degeneracy}
\end{figure}

Contrary to $m$, the source count curve is very sensitive to the exact values
of $n$ and $z_a$ which respectively control the rate by which the
characteristic luminosity of IR galaxies increases with redshift and up to
which redshift this growth continues. However, there are two solutions for
reproducing the same source counts: one is to increase the characteristic
IR luminosity rapidly up to a relatively low redshift and the other one is to
increase the characteristic IR luminosity by a moderate rate but for a longer
period of time (i.e., up to higher redshifts). Therefore, as illustrated in
Figure~\ref{fig:degeneracy}, there is a degeneracy between $n$ and $z_a$ and
one cannot constrain them individually by looking at the observed source
counts.  However, additional information about the redshift distribution of
submm galaxies or the slope by which the characteristic IR luminosity is
growing at low redshifts could resolve this degeneracy. As we will discuss
later, the former constraint is used in finding our best-fit model.

Unlike $n$ and $z_a$, the length of the redshift interval during which the
luminosity evolution remains constant before starting to decline,
$z_b - z_a$, is not an essential part of our model. We only introduced this
feature to have a smoother transition between growing and declining
characteristic IR luminosity and also redshift distribution of submm galaxies.
As Figures \ref{fig:shape-amp} and \ref{fig:all-param} show, variations in the
adopted value for $z_b - z_a$ do not change the source count significantly.
Moreover, any change in its value could be compensated by a very small
change in $n$ and/or $z_a$.\\

\begin{figure*}
\centerline{\hbox{\includegraphics[width=0.55\textwidth]
             {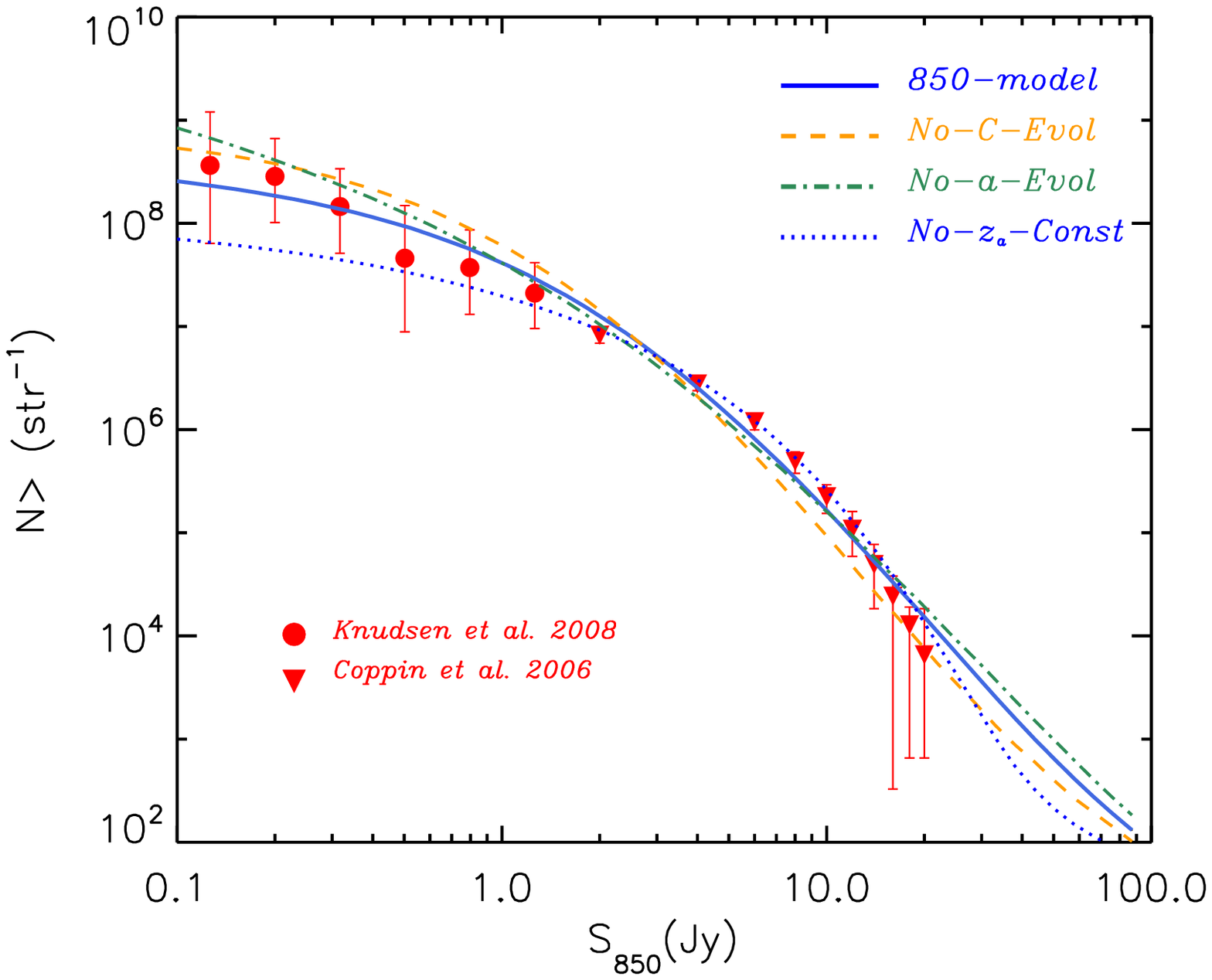}}
            \hbox{\includegraphics[width=0.55\textwidth]
             {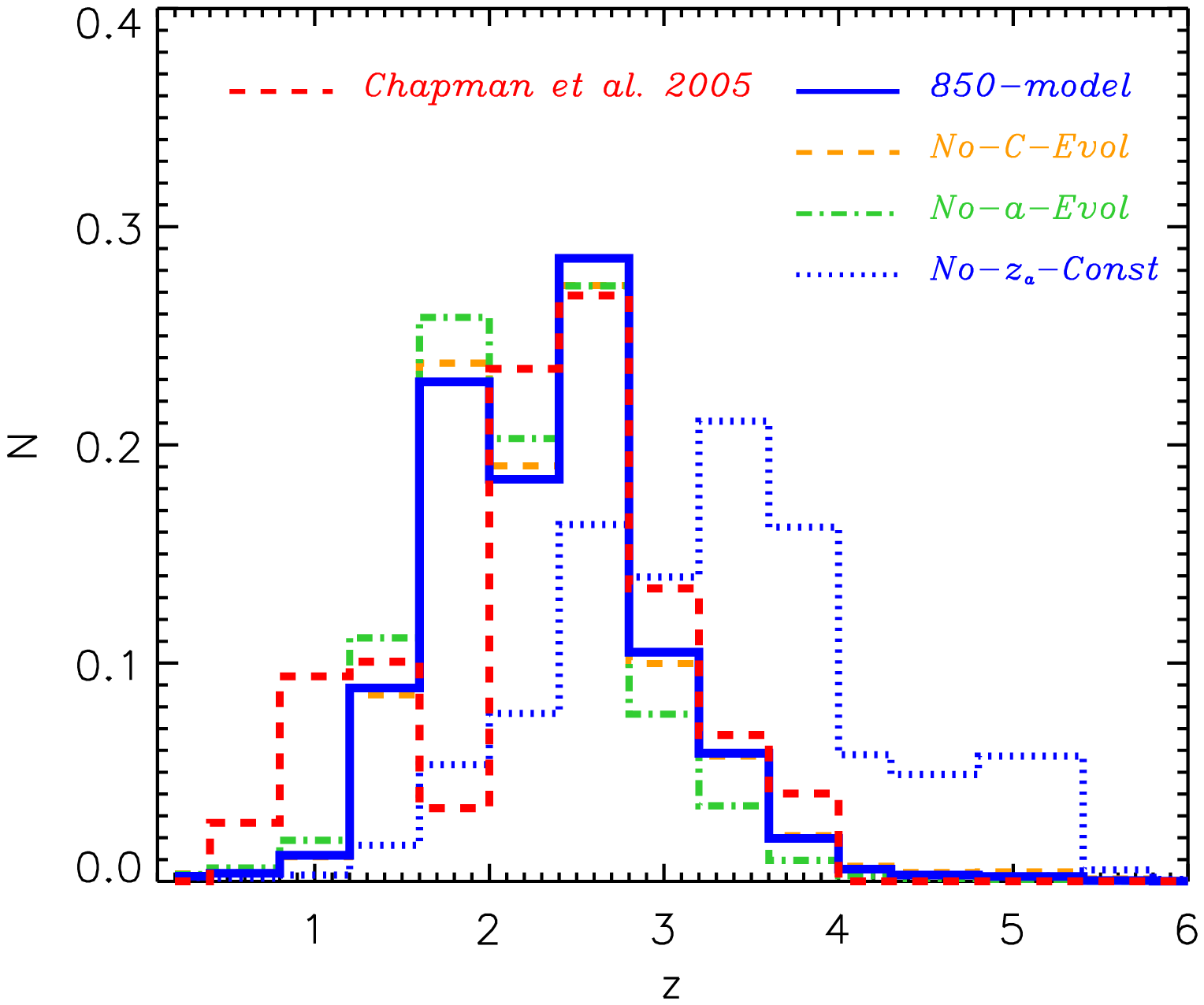}} }
\caption{Left panel: The best-fit model which is constrained by
$850\,\mu$m source count and the redshift distribution of submm galaxies
(the blue solid line) is illustrated next to the reference observed data
points (see Section~\ref{sec:observed}). For comparison, three other best-fit
models are also plotted: the best-fit model without colour evolution shown by
the dashed orange line (i.e., ``No-C-Evol''), a model without evolving LF
slopes shown by the dot-dashed green curve (i.e., ``No-a-Evol'') and finally
the best-fit model which is only constrained by the source count is
shown using the dotted blue curve. Right panel: the comparison between the
observed redshift distribution of $850\,\mu$m galaxies which are brighter than
$5$\,mJy \citep{Chapman05} and what our best-fit model implies. The histogram
shown in red with dashed line is the observed probability distribution and the
histogram with solid blue line shows the probability distribution of similar
objects in our model. For comparison, the best-fit model without constrained
redshift distribution, is shown with dotted blue histogram ("No-z$_{\rm{a}}$-Const"). The other two models, ``No-C-Evol'' and ``No-a-Evol'' are respectively shown by dashed orange and dot-dashed green lines.}
\label{fig:850-best}
\end{figure*}

Based on these considerations and as we already mentioned in Section~\ref{sec:CLFz}, we reduce the number of free parameters we use in our model by
choosing $m = -1$ and $z_b - z_a = 1$. Furthermore, the observed redshift
distribution of submm galaxies which peaks around $z\sim 2$ \citep{Chapman05,Wardlow11},
limits the acceptable values of $z_a$ to $\sim 1.6$ (see the right panel in
Figure~\ref{fig:850-best}).

\subsection{Other necessary model ingredients}
\label{sec:col-shape-evol}
Although the luminosity evolution is necessary for producing correct number of
observable sources, it is not sufficient to provide a correct shape for the
source count curve. Based on our experiments by varying different parameters of
the luminosity evolution, one can either get a good fit at the faint number
counts and under-produce the bright end or produce correctly the bright source
counts and over-estimate the faint objects. In other words, other model
ingredients like the colour evolution and/or the evolution of LF slopes are
required to adjust the shape of the source count curve appropriately in order to
fit the faint and bright source counts at the same time. For instance, the colour
evolution can be used to help the model with under-production of bright sources
and the evolution of LF slopes can compensate for the over-production of the
faint sources. 

As a starting point, we have first explored fits using either the colour
evolution or the evolution of the LF slopes, in order to keep the model as
simple as possible.  We did this by trying to fit the $850\,\mu$m source
counts once using a combination of the luminosity evolution together with the
colour evolution when the slopes of LF do not evolve (i.e., No-a-Evol) and once
using the luminosity evolution combined with the evolving LF slopes, without
evolving the colour distribution (i.e. No-C-Evol). The results are shown in
Figure~\ref{fig:850-best} and are compared with the best-fit result when the
colour evolution and the evolution of LF slopes are both active (i.e.,
850-model). The quality of the fit for ``850-model'' is better than both of the
other cases which may be attributed to the additional free parameter used
in the ``850-model''. However, both ``No-a-Evol'' and ``No-C-Evol''
models have unfavourable implications. The best-fit for the ``No-C-Evol''
  case (see Table \ref{table:850models}) requires a large increasing slope in the luminosity evolution function
  which gives rise to a violation of the counts of IR sources at short
  wavelengths (e.g. at $70\,\mu$m, see Section~\ref{sec:others}). The
  ``No-a-Evol'' fit on the other hand, requires a steep colour evolution which is too extreme to be acceptable, since it would imply that at redshifts $z>1$ all infrared sources, independent of their luminosities, have the same dust temperatures as low as  $T\sim 20$K. However, when both colour evolution and evolution of the bright and faint-end slopes of
the LF are allowed simultaneously, these problems disappear which makes this solution preferable.

\subsection{The $850\,\mu$m best-fit model}
\label{sec:bst-fit}

\begin{table*}
\begin{center}
\begin{tabular}{ccccccc}
\hline
Model  & $n$ & $z_a$ & $w$ & $a_{\alpha}$ & $a_{\beta}$\\
\hline
850-model &2.0 & 1.6 & 2.0 & 0.6 & 0.4\\

No-C-Evol & 2.8 & 1.6 & 0 & 0.6 & 0.6\\

No-a-Evol & 1.6 & 1.6 & 5.6 & 0 & 0\\

No-$z_a$-Const & 2.4 & 3.6 & 2.4 & 1.0 & 2.2\\
\hline
\end{tabular}
\caption{Parameters which define different best-fit models constrained to reproduce the observed source counts at $850\,\mu$m. All the models are using $z_b - z_a = 1$ and $m = -1$ and except the "No-$z_a$-Const" model, all of them are constrained to reproduce the redshift distribution of submm galaxies and therefore use $z_a = 1.6$. The predicted source count each of those models and their implied redshift distribution for submm sources is illustrated in Figure \ref{fig:850-best}}
\label{table:850models}
\end{center}
\end{table*}

As we showed, it is necessary to have an evolutionary model which
incorporates a luminosity evolution, colour evolution and also evolving LF
slopes, in order to fit the observed $850\,\mu$m source counts. We also argued
that we can fix some of the initial parameters of the model since they have
no significant effect on the results and chose proper values for them (i.e.,
$m = -1$ and $z_b - z_a = 1$). Moreover, we showed that we need to set the
redshift at which the luminosity evolution peaks to $z_a = 1.6$ to reproduce
the redshift distribution of submm galaxies correctly which also resolves the
degeneracy between $z_a$ and $n$. After taking into account all of those
considerations, we end up  with 4 free parameters in our model which are
needed to be  adjusted properly to reproduce our observed sample of $850\,\mu$m
source counts; those parameters are the slope of the colour evolution, $w$,
the rate by which the faint and bright end slope of LF is changing with
redshift, $a_{\alpha}$ and $a_{\beta}$, and finally the growth rate of the
characteristic IR luminosity, $n$. 

Since our algorithm for calculating the source count is fast and accurate,
contrary to much more cumbersome Monte-Carlo-based approaches, we can perform a
comprehensive search in the parameter space for the best-fit model instead of
choosing it ``by hand''. We split each dimension of relevant regions of parameter space into equally spaced grids and calculate the source count for modes associated with each node of our grid structure. Then we calculate the likelihood of each model for reproducing the observed source counts (i.e. 
$exp(-\frac{1}{2}\chi^2$)). Finally, we choose the model with maximum likelihood as our best-fit model. The parameters which define the best-fit model for $850\,\mu$m source counts, ``850-model'', is shown in Table~\ref{table:850models} together with those of "No-C-Evol" and "No-a-Evol" which we discussed in the previous section. Those models are also compared with observational data sets used for constraining them, in Figure~\ref{fig:850-best} where also the redshift distribution implied by each model is illustrated (in the right panel). 

It is also interesting to inspect the properties of a best-fit model in which
the peak of the luminosity evolution, $z_a$, is not fixed. The parameters which define this model are presented in the last row of Table~\ref{table:850models}  and its source count and redshift distribution are shown by the blue dotted lines in Figure~\ref{fig:850-best}. The quality of the fit for this model is even better
than ``850-model'' except at very low flux thresholds (unsurprisingly, since it has an additional free parameter,
namely $z_a$).  The rising slope of luminosity evolution, $n$, and the colour evolution, $w$, in this model are not drastically different from the ``850-model'' but its peak of luminosity evolution happens at higher redshifts and the model requires much steeper slope evolutions to compensate for too many observable sources. Not only is this model unable to produce the observed redshift distribution ( see the right panel in Figure~\ref{fig:850-best}), it predicts too few observable sources at shorter wavelengths which makes it unfavorable.

\section{Other wavelengths}
\label{sec:others}
In the previous section, we discussed our model parameters and their role in
producing the observed $850\,\mu$m source counts and the redshift distribution of
submm galaxies. We used those observed quantities as constraints to find our
best-fit model, ``850-model''.
If we assume the evolution scenario that our best-fit model suggests is correct,
and also the set of SEDs we used for reproducing the $850\,\mu$m source count are
good representatives for real galaxies, then we expect the same model, together
with the same set of SEDs, to reproduce the observed number counts of IR sources at
other wavelengths. Moreover, one can use the source count at other wavelengths to find a best-fit model for that specific wavelength and again expects to  reproduce the source count at other wavelengths correctly.

\begin{figure*}
\centerline{\hbox{\includegraphics[width=0.5\textwidth]
             {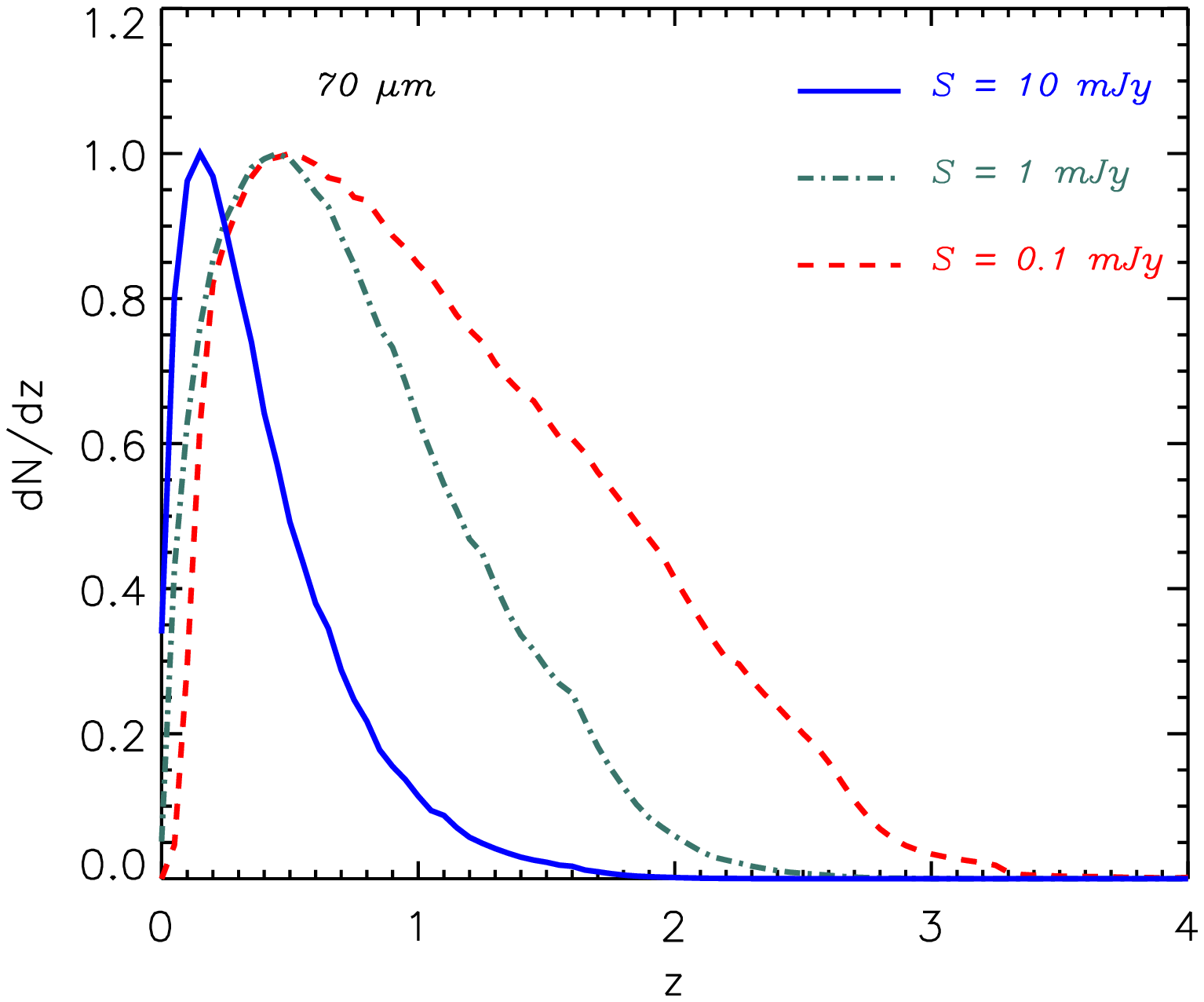}}
            \hbox{\includegraphics[width=0.5\textwidth]
             {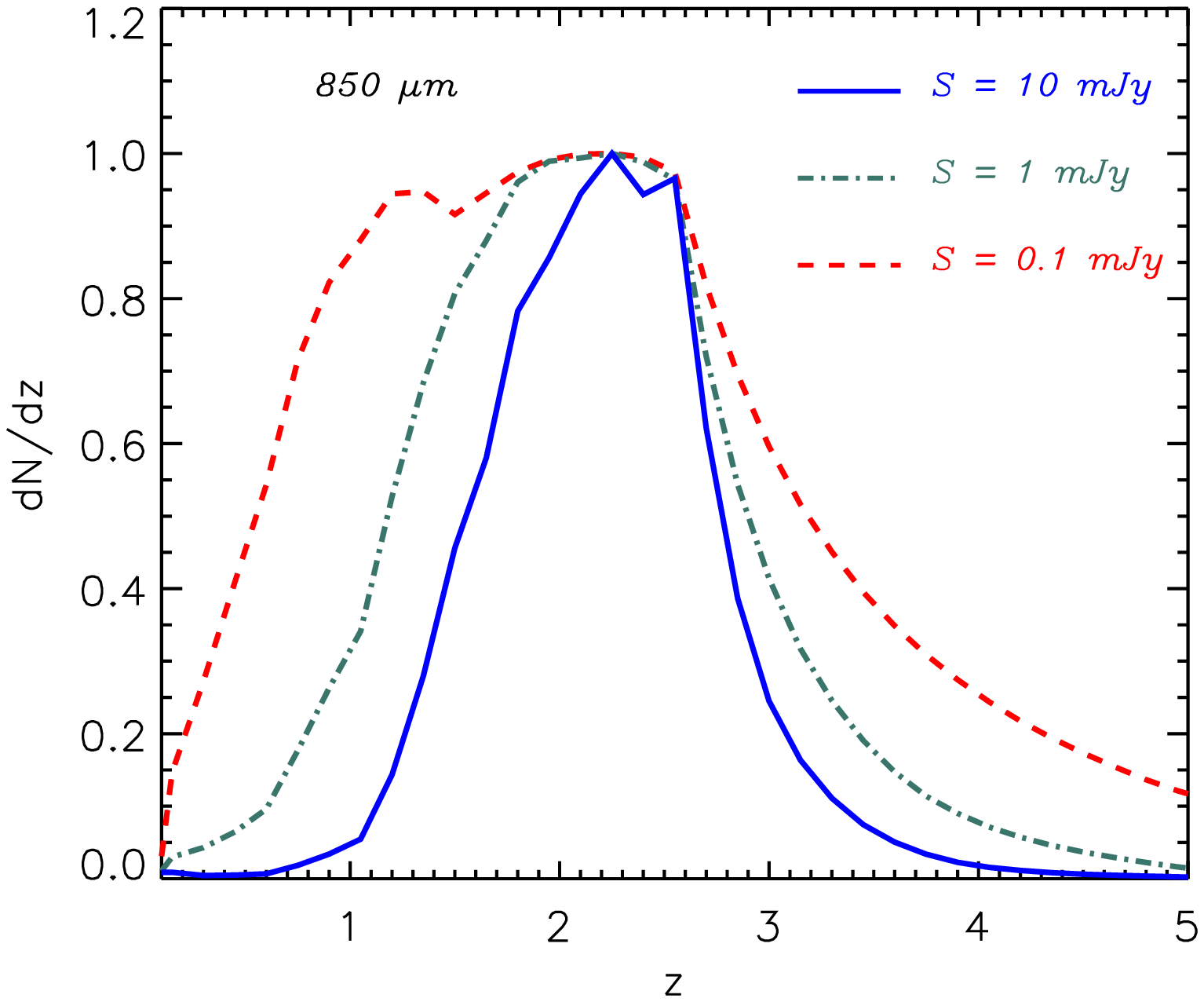}} }
\caption{The modeled redshift distribution of objects which are observed with
different flux thresholds at $70\,\mu$m (in the left panel) and $850\,\mu$m
(in the right panel). Three different flux thresholds which are $0.1$, $1$
and $10$mJy are shown respectively using solid(blue), dot-dashed (green)
and dashed (red) lines.}
\label{fig:short-long-fluxes}
\end{figure*}

Before examining those expectaions, it is important to note that there are some fundamental differences between the
source counts at shorter wavelengths and in the submm. Most importantly, as
illustrated in Figures~\ref{fig:short-long-fluxes} and \ref{fig:z_dist}, objects observed at short
wavelengths (e.g., $70\,\mu$m) are mainly at low redshifts while the observed
submm galaxies are distributed in a wider redshift interval and at higher
typical redshifts. Also, the brightest objects at $70\,\mu$m are the closest
ones which is not the case for brightest submm galaxies\footnote{In fact the wide
distribution of observed submm galaxies in redshift space, combined with the availability of a measured redshift distribution are the primary reasons we chose their observational properties to constrain our model. } \citep{Berta11}.
Moreover, there are additional physical processes, like AGNs, which can change the energy output
of galaxies at shorter wavelengths. Since we do not take into account AGNs as a
separate population, our model is not expected to reproduce necessarily good results for
wavelengths shorter than $60-70 \,\mu$m (see also the discussion in Section~\ref{sec:SED}).  Going from longer to shorter
wavelengths, the observed sources are typically at lower and lower redshifts.
Consequently, the source counts at short wavelengths are only sensitive to the
very low redshift properties of our model. In fact, the $70\,\mu$m source count is
mainly sensitive to the growth rate of the luminosity evolution, $n$, and the
evolution in LF slopes; but by going to longer wavelengths, the colour evolution
and the redshift at which the luminosity evolution is peaking, $z_a$, become
more important. In other words, if a model reproduces the observed source
counts at short wavelengths, this forms a confirmation that the evolution at low
redshifts is represented correctly but for a best-fit model constrained by observations at shorter wavelengths, the properties of model at intermediate and high redshifts are not strongly constrained.

\begin{table*}
\begin{center}
\begin{tabular}{ccccccc}
\hline
$\lambda$  & $n$ & $w$ & $a_{\alpha}$ & $a_{\beta}$\\
\hline
$850\,\mu$m & 2.0 & 2.0 & 0.6 & 0.4\\

$500\,\mu$m & 2.6 & 3.8 & 0.0 & 0.6\\

$350\,\mu$m & 3.0 & 4.4 & -0.2 & 1.4\\

$250\,\mu$m & 3.0 & 4.4 & 0.0 & 1.4\\

$160\,\mu$m & 2.2 & 4.2 & 0.8 & 0.2\\

$70\,\mu$m & 2.2 & 2.0 & 0.6 & 0.6\\
\hline
\end{tabular}
\caption{Parameters which define different best-fit models constrained to reproduce the observed source counts at different wavelengths. The first column, $\lambda$, indicates the wavelength for which the model is constrained to produce the best fit to the observed source counts. All the models are forced to reproduce the redshift distribution of submm galaxies and therefore use $z_a = 1.6$, $z_b - z_a = 1$ and $m = -1$. The predicted source count each of those models is implying for different wavelengths is illustrated in Figure \ref{fig:all-wavelengths}}
\label{table:models}
\end{center}
\end{table*}
\begin{figure*}
\centerline{\hbox{\includegraphics[width=0.45\textwidth]
             {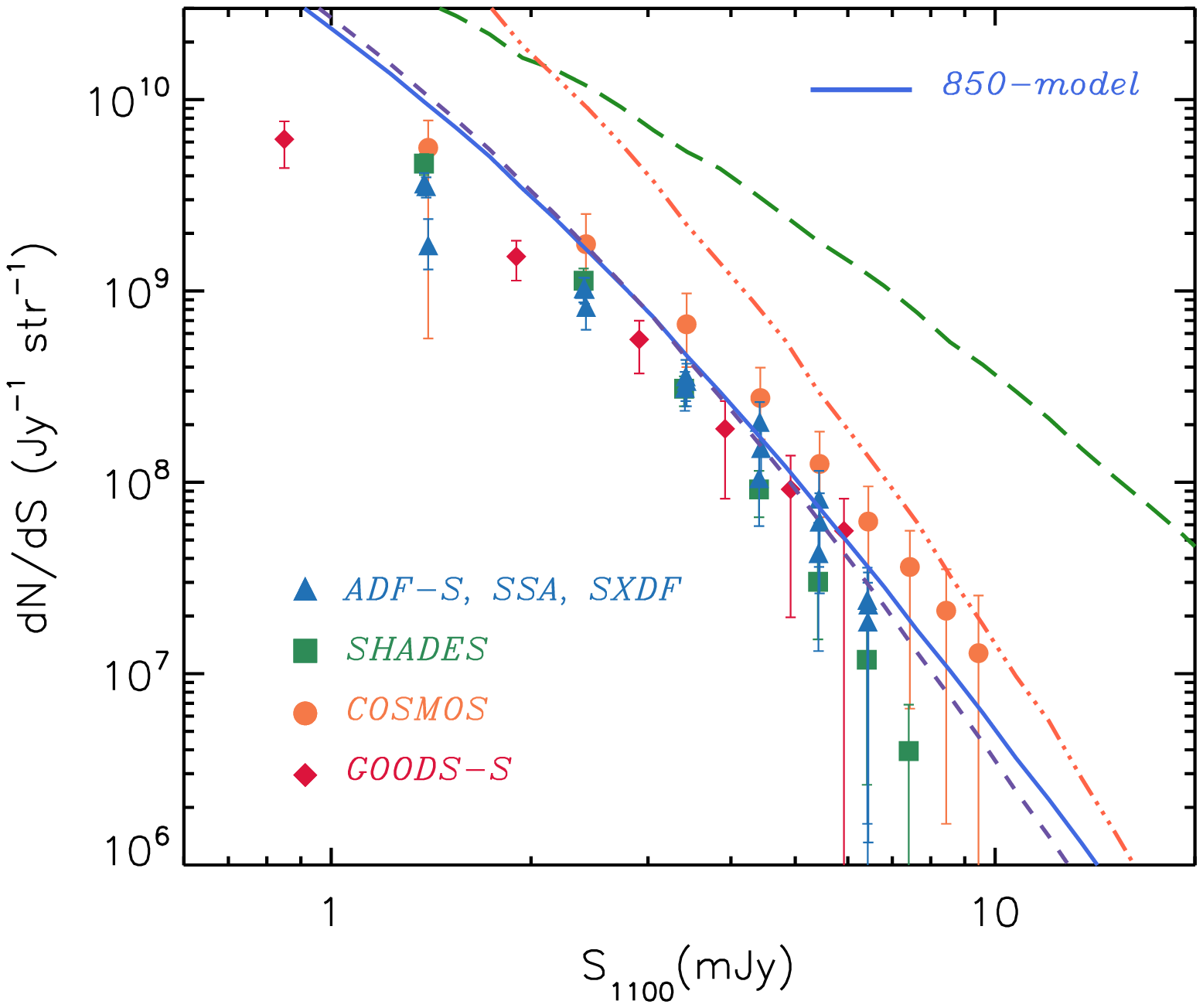}}
            \hbox{\includegraphics[width=0.45\textwidth]
             {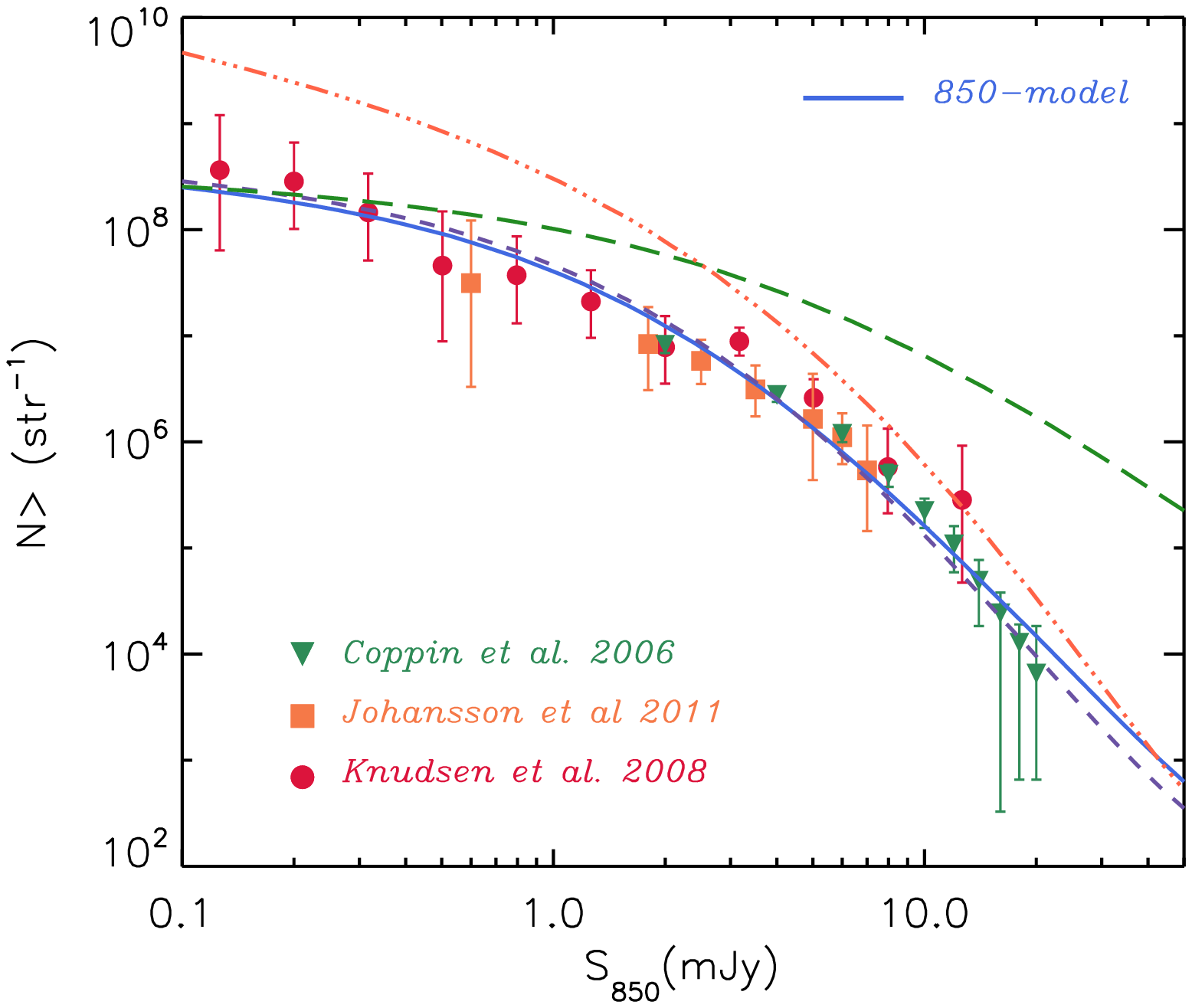}} }
\centerline{\hbox{\includegraphics[width=0.45\textwidth]
             {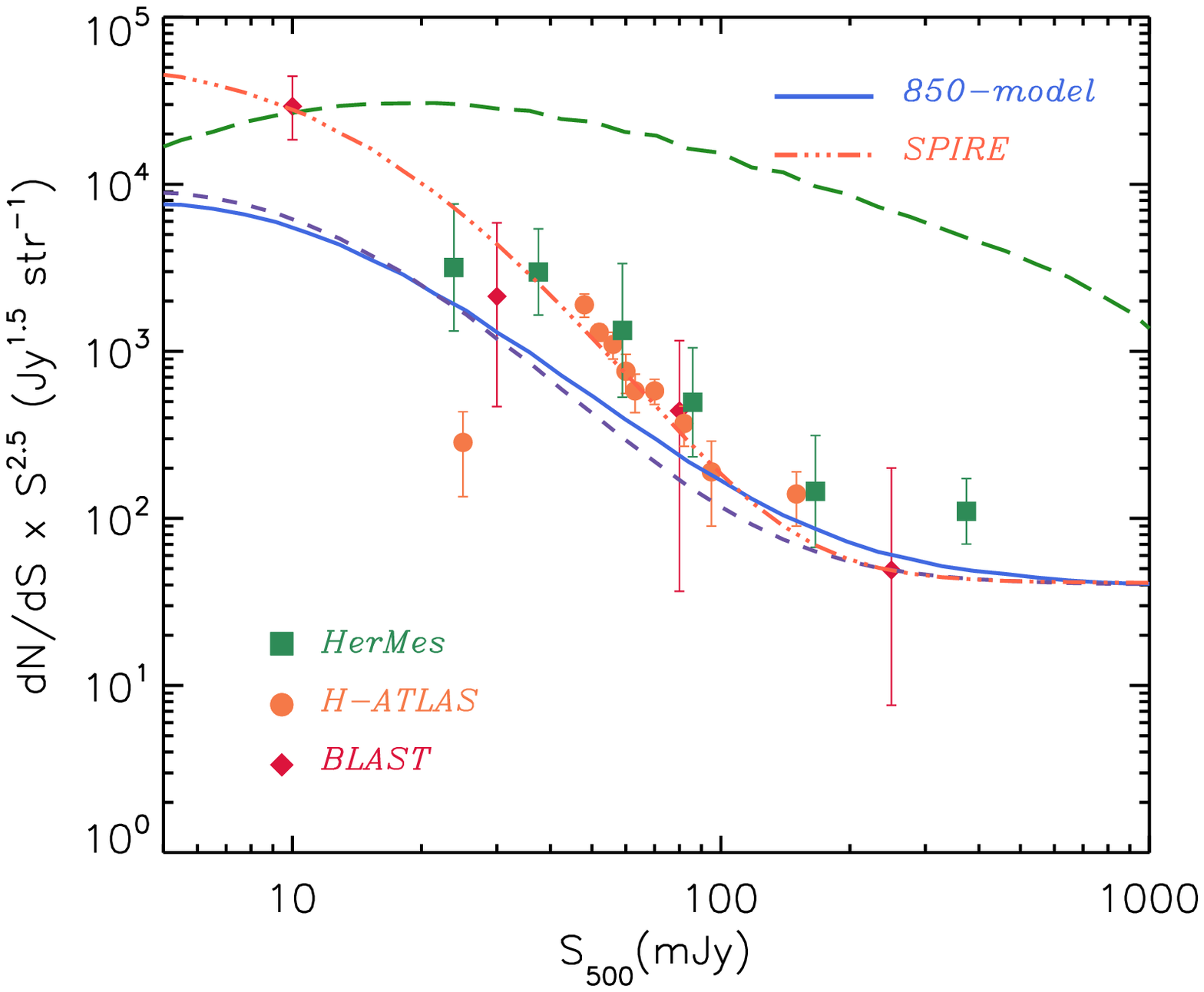}}
            \hbox{\includegraphics[width=0.45\textwidth]
             {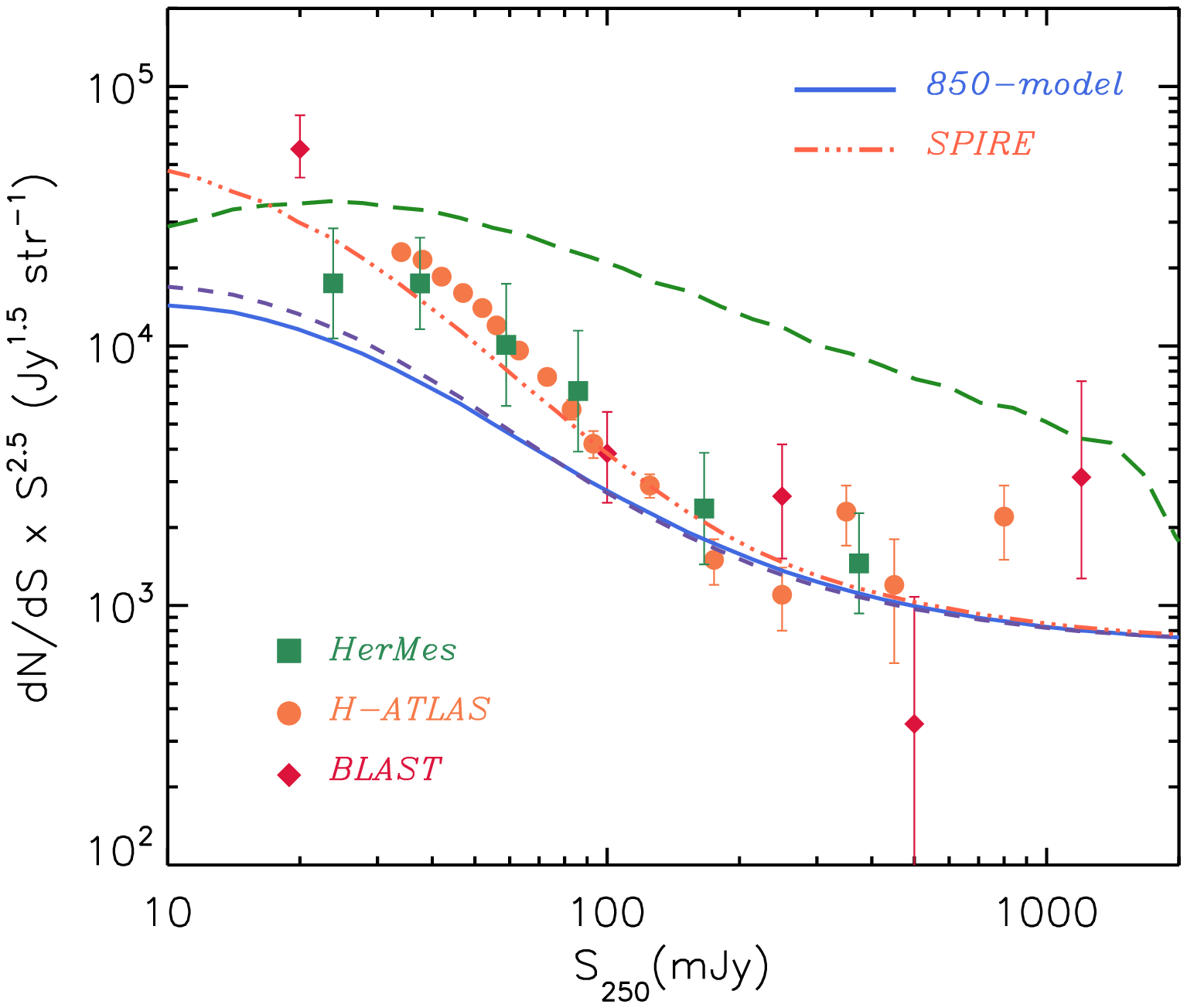}} }
\centerline{\hbox{\includegraphics[width=0.45\textwidth]
             {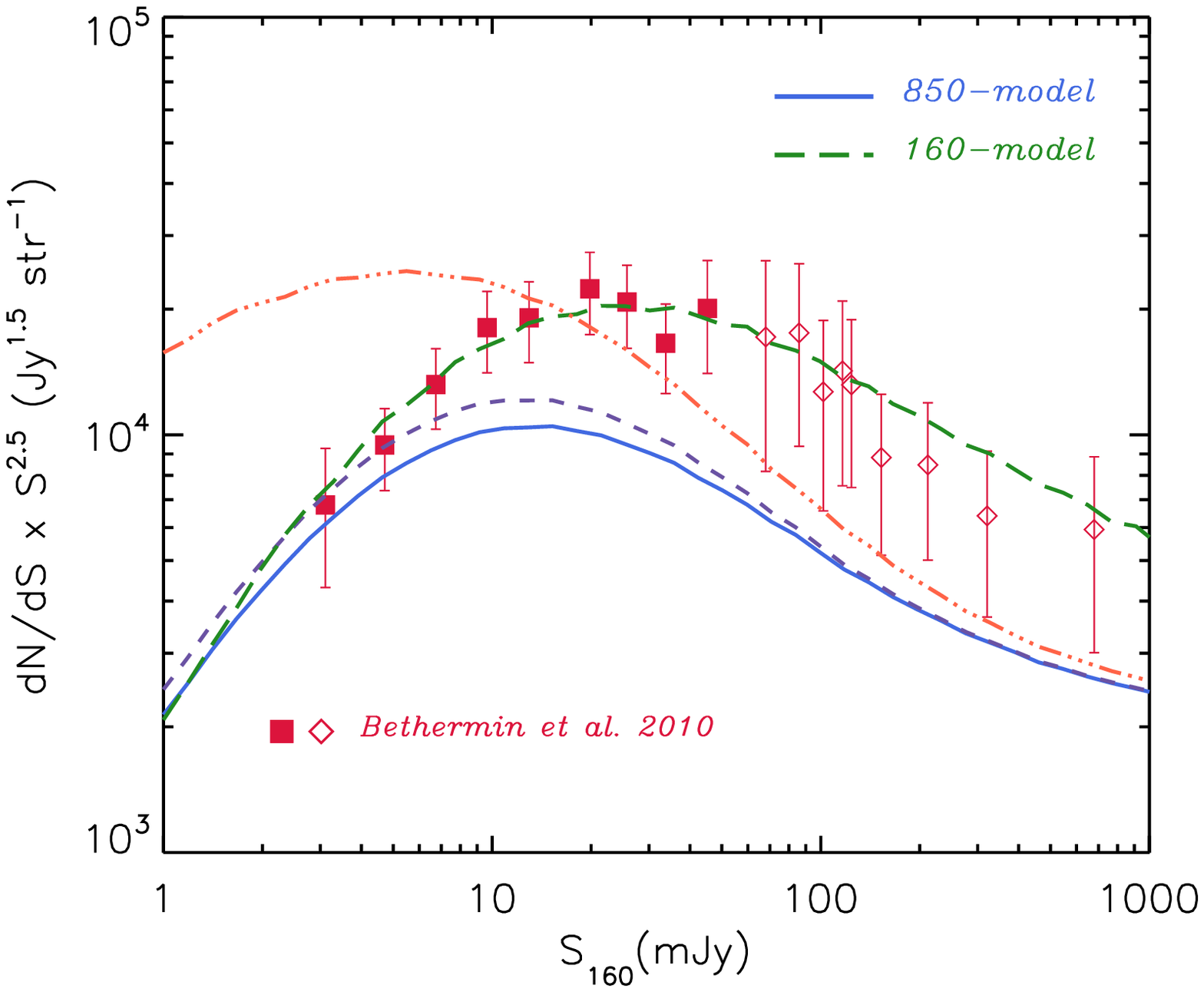}}
            \hbox{\includegraphics[width=0.45\textwidth]
             {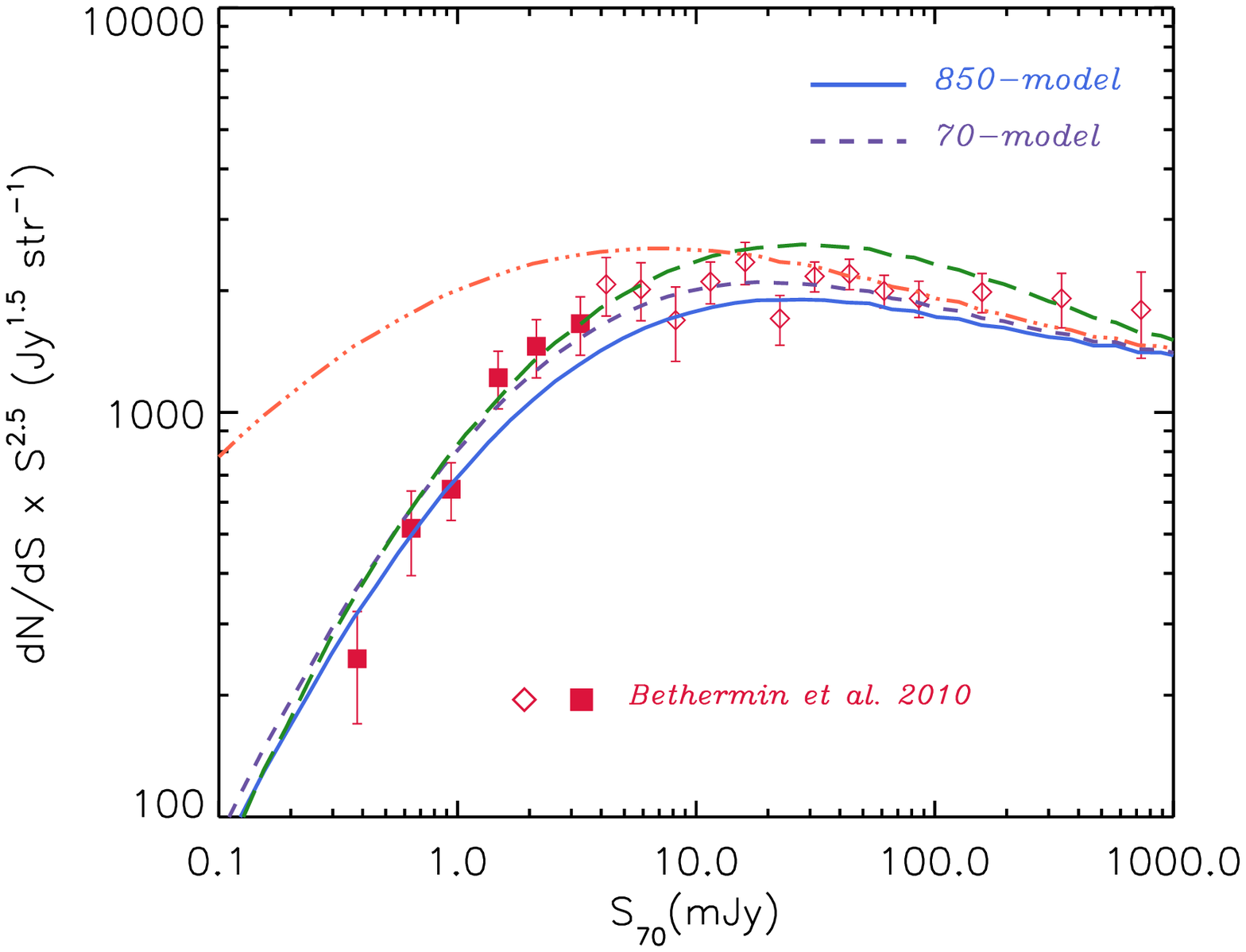}} }
\caption{In different panels, the best-fit model predictions for cumulative
source counts at $850\,\mu$m and differential source counts at $1100$, $500$,
$250$, $160$ and $70\,\mu$m are plotted next to the observed data. The $1100\,\mu$m observational data sets are from \citet{Scott10} (red diamonds) \citet{Austermann09,Austermann10} (orange circles and green squares respecitvely) and \citet{Hatsukade11} (blue triangles); data at
$850\,\mu$m is from \citet{Coppin06}(green triangles),
\citet{Knudsen08} (red circles) and \citet{Johansson11} (orange squares).
The data for $500$ and $250\,\mu$m source counts are from the BLAST
experiment taken from \citet{Patanchon09} (red diamonds), Herschel data taken
from \citet{Oliver10} (green squares) and \citet{Clements10} (orange circles).
We also used one data point at $350\,\mu$m from the SHARC2 survay (green
triangle)\citep{Khan07}. The data at $160$ and $70\,\mu$m are based on Spitzer
observations taken from \citet{Bethermin10} where the data points shown by
filled squares represent stacking results. Different lines correspond to the source counts produced by various models which are constrained to fit the observed source counts at different wavelengths: blue solid line shows the "850-mode", purple dashed line shows the "70-model" and green long-dashed and orange dot-dot-dashed lines are for "160-model" and "500-model" respectively (see Table~\ref{table:models})} \label{fig:all-wavelengths}
\end{figure*}
To investigate the performance of different best-fit models constrained by source count observations of wavelengths other than $850\,\mu$m, we use the same procedure we incorporated in finding the ``850-model'' and only vary the effective parameters of the model, namely $n$, $w$, $a_{\alpha}$ and $a_{\beta}$, to fit the observed data. The parameters which define best-fit models at different wavelengths are shown in Table~\ref{table:models} and the source counts they produce at different wavelengths are illustrated next to the observational data points in Figure~\ref{fig:all-wavelengths}. Different panels are for source count at different wavelengths ranging from $1100\,\mu$m on the top left  to $70\,\mu$m on the bottom right. We do not show the results for $350\,\mu$m to maintain the symmetry of figures since for this wavelength models and their comparison with observed data are in many respects identical to the case of $250\,\mu$m. In Figure~\ref{fig:all-wavelengths}, the source counts produced by different models are also shown using lines with different styles and colours: purple short-dashed for $70\,\mu$m, green long-dashed for $160\,\mu$m, orange dot-dot-dashed for $500\,\mu$m and finally solid blue lines for $850\,\mu$m (i.e the "850-model"). In the following we discuss those results by categorizing them in different wavelength ranges, namely $850\,\mu$m and $1100\,\mu$m as long submm wavelengths, $500\,\mu$m, $350\,\mu$m and $250\,\mu$m as SPIRE or intermediate wavelengths and finally $70\,\mu$m and $160\,\mu$m as short wavelengths.

\subsection{Long submm wavelengths: $850\,\mu$m and $1100\,\mu$m}
\label{sec:submm}
As we mentioned earlier, the SED of star forming galaxies at long submm ranges is essentially controlled by the Rayleigh-Jeans tail of the dust emission which simply falls off smoothly (see Figure~\ref{fig:SED}). This means that if a model reproduces the observed counts at $850\,\mu$m, a good fit to the observed counts at similar and longer wavelengths is guaranteed that given the distribution of sources which produce those counts have a similar redshift distributions, which in turn makes those counts equally sensitive to different parameters in our model. As illustrated in the top panels of Figure~\ref{fig:all-wavelengths}, both models which are constrained by observed $70\,\mu$m and $850\,\mu$m counts and produce a good fit to $850\,\mu$m data, also produce a good fit to $1100\,\mu$m source counts. Our experiments with other models also confirm that the quality of the fit they produce for observed counts at $850\,\mu$m and $1100\,\mu$m is highly correlated. Both of the models constrained by $160\,\mu$m and SPIRE (e.g. $500\,\mu$m) source counts over produce the long submm wavelength source counts: the "160-model" over-produces submm source counts mainly at bright flux thresholds due to its extreme colour evolution (see Table~\ref{table:models}) but fits the observations at faint fluxes, thanks to its luminosity evolution which is similar to those of "850-model" and "70-model". However, SPIRE constrained models (e.g. "500-model") have stronger evolutions both in terms of luminosity and colour and over-produce the data in all observed fluxes.  

One should note that the $1100\,\mu$m data points are highly incomplete below $3$mJy \citep{Hatsukade11} and are prone to large field to field variations and relatively large errors in the whole range of observed fluxes. Moreover, the observed counts are at bright flux thresholds which makes them sensitive to a narrower redshift interval in comparison to the fainter flux thresholds (see the right panel of Figure~\ref{fig:short-long-fluxes}) and because of those reasons they cannot constrain models better than what $850\,\mu$m source counts are capable of. Therefore, we only consider the "850-model" as a model constrained by long submm source counts.

\subsection{SPIRE intermediate wavelengths: $500\,\mu$m, $350\,\mu$m and $250\,\mu$m}
\label{sec:spire}
The three $500\,\mu$m, $350\,\mu$m and $250\,\mu$m wavelengths are close together and besides being in the middle of wavelength ranges we study, they have intermediate properties with respect to the redshift range each wavelength is mostly sensitive to: as it is shown in Figure~\ref{fig:z_dist}, while at the bright flux thresholds the source counts mainly consist of low redshift sources, at fainter fluxes they are sensitive to the intermediate (i.e. $1<z<2$) and high redshifts (i.e. $2<z$). Naturally the general behavior of models for $500\,\mu$m is closer to $850\,\mu$m while $250\,\mu$m is close to shorter wavelengths. However, those intermediate wavelengths are closely similar to each other more than being similar to other wavelengths. Consequently, the best-fit models constrained by SPIRE wavelengths have similar parameters ( 250-model and 350-model have almost identical parameters) and any of those models agrees with the observed source counts of the other two. However, SPIRE-constrained models all require steep luminosity evolution and strong colour evolution in addition to a steepening bright end slope of the LF with redshift (i.e. positive $a_{\alpha}$ together with almost zero $a_{\beta}$, see Table~\ref{table:models}). The first two mechanisms produce too many objects in comparison to what is needed for the source counts at other wavelengths while the steeper bright end of the LF partially compensate the over-production of bright objects: as is shown in Figure~\ref{fig:all-wavelengths}, SPIRE-constrained models over-produce the faint source counts both at longer and shorter wavelengths but the steep bright end of LF produces results which are close to the observed faint source counts at very long wavelengths (i.e. $850\,\mu$m and $1100\,\mu$m) which in turn is too strong to leave enough sources required for the bright $160\,\mu$m sources. As we will discuss later, it is not surprising that SPIRE-models agree with bright $70\,\mu$m counts since they are essentially $z\sim0$ objects for which evolutionary mechanisms in the model barely have any effect.

While the SPIRE models fail to agree with observations at shorter and longer wavelengths, the two models which are successful at long submm ranges (i.e. the "850-model" and "70-model") also have a reasonable agreement with the SPIRE data. However, they under-produce the counts at $S_{th} < 100$mJy by a factor of $\sim2$. As we will show later, assuming the "850-model" to be correct, this under-production is a hint for a population of cold luminous IR galaxies residing only in low and intermediate redshifts (i.e. $z \sim 1$). This is also consistent with the steep coluor evolution which is implied by best-fit models at those wavelengths which produces enough cold galaxies at lower redshifts. 

\subsection{Short wavelengths: $160\,\mu$m and $70\,\mu$m}
\label{sec:short}
At short wavelengths like $70\,\mu$m, the K-correction is not strong enough to counteract the effect of cosmological dimming which makes the source counts at these wavelengths almost insensitive to the properties of IR galaxies at high redshift. Since the starting point of our model is the observed distribution of IRAS galaxies which are selected to be local galaxies with $S_{60\,\mu {\rm{m}}}>1{\rm{Jy}}$, and the SED templates we use are extensively tested to match these galaxies, any variation of our model by construction cannot disagree with bright $70\,\mu$m counts \footnote{Unless the luminosity of IR galaxies increase with redshift extremely steep to produce high-z objects which are observed in $70\,\mu$m band as bright as the brightest local IR galaxies despite the cosmological dimming and negative K-correction.}. However, at fainter flux thresholds (e.g. $S_{th} < 100$mJy) the sensitivity of $70\,\mu$m counts to low and intermediate redshifts, $0.5 < z < 2$, increases (see Figure~\ref{fig:z_dist}) which makes them more sensitive to the slope of the luminosity evolution. As a result, all the models agree with $70\,\mu$m observations at bright flux threshold while the SPIRE-model which has steeper luminosity evolution (i.e. greater $n$, see Table~\ref{table:models}) diverges from observed counts and over-produces them. It should be noted that the "70-model" which is tuned to agree with mainly low-z IR sources performs as good as "850-model" in longer wavelengths.

At $160\,\mu$m, the redshift distribution of sources is almost identical to the distribution of galaxies which are responsible for $70\,\mu$m: the bright end of the counts is governed by local IR galaxies while for flux thresholds $S_{th} < 100$mJy the intermediate redshifts (i.e. $0.5 < z < 2$) become important. However, the flux threshold at which the turn over between domination of local and farther sources happens is one order of magnitude higher for $160\,\mu$m sources (see bottom row in Figure~\ref{fig:z_dist}). Moreover, at $160\,\mu$m the domination of intermediate redshift sources in shaping the faint source counts is slightly stronger than at $70\,\mu$m. In other words, the $160\,\mu$m source counts are slightly more sensitive to the distribution and evolution of distant IR galaxies in comparison to $70\,\mu$m counts. 

The best-fit model which is constrained by $160\,\mu$m source counts has similar slope of luminosity evolution, $n$, compared with the "850-model" and "70-model" but requires much stronger colour evolution and larger fraction of bright objects (i.e. steeper faint end slope and flatter bright end slope of LF at higher redshifts). 

The SPIRE-models on the other hand, violate the observed $160\,\mu$m counts by under-producing them at bright end and over-producing them at faint fluxes. The "850-model" and "70-model" on the other hand, match the faintest $160\,\mu$m count but under produce the brighter objects by a factor of $\sim2$ which is the same factor showing up in the difference between what those two models produce and observed faint SPIRE counts. A similar discrepancy between models and $160\,\mu$m counts has been pointed out in some recent works \citep{LeBorgne09,Valiante09}; while the \citet{LeBorgne09} best-fit model which produces correct source counts at $70\,\mu$m and $850\,\mu$m deviates from observations between $10{\rm{mJy}}<S_{160} < 100$mJy by a factor of $\sim2$, the \citet{Valiante09} model underestimates number counts at $160\,\mu$m by a factor of $\sim5$. We discuss this issue further in \ref{sec:all-the-best}.
\begin{figure*}
\centerline{\hbox{\includegraphics[width=0.45\textwidth]
             {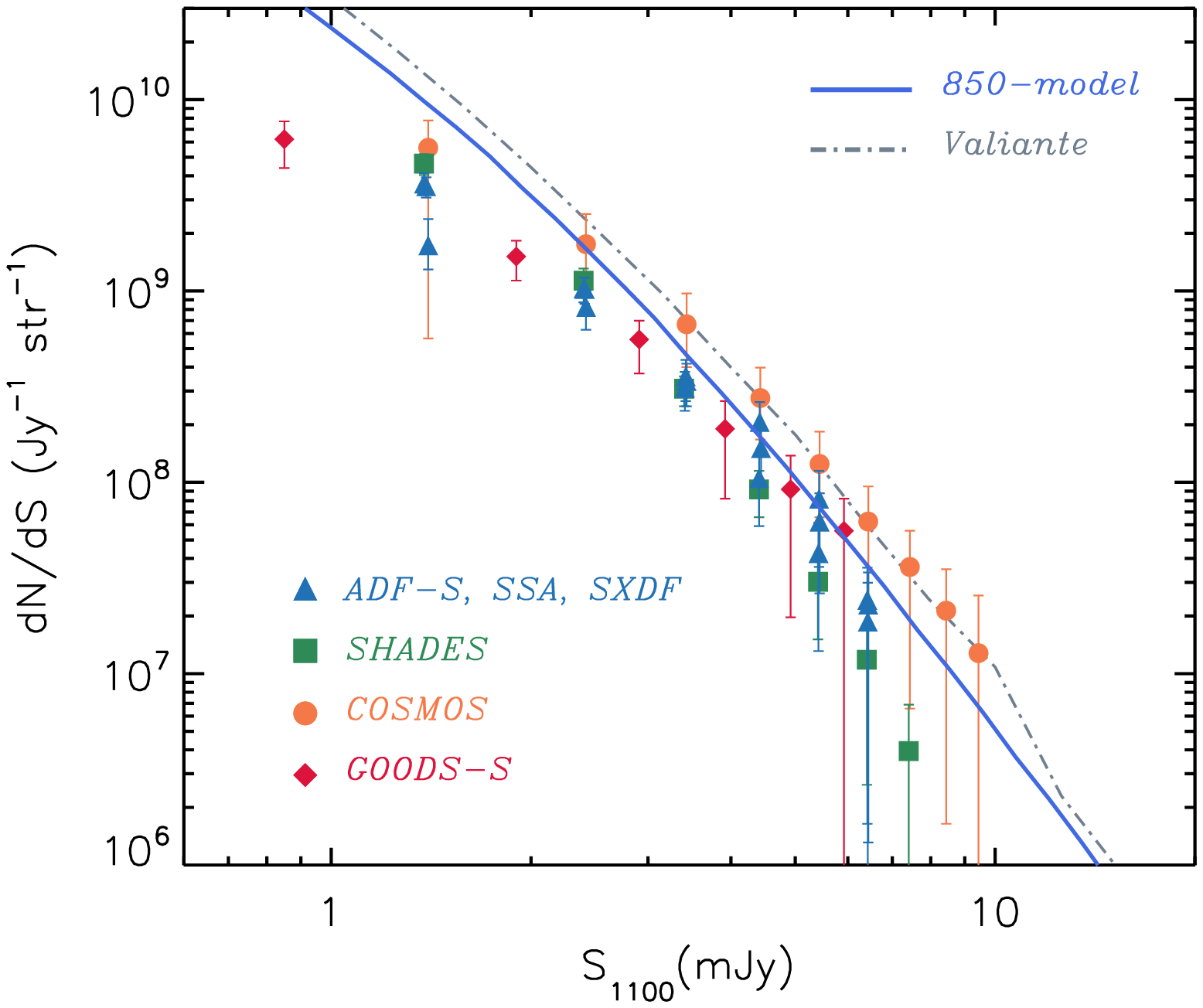}}
            \hbox{\includegraphics[width=0.45\textwidth]
             {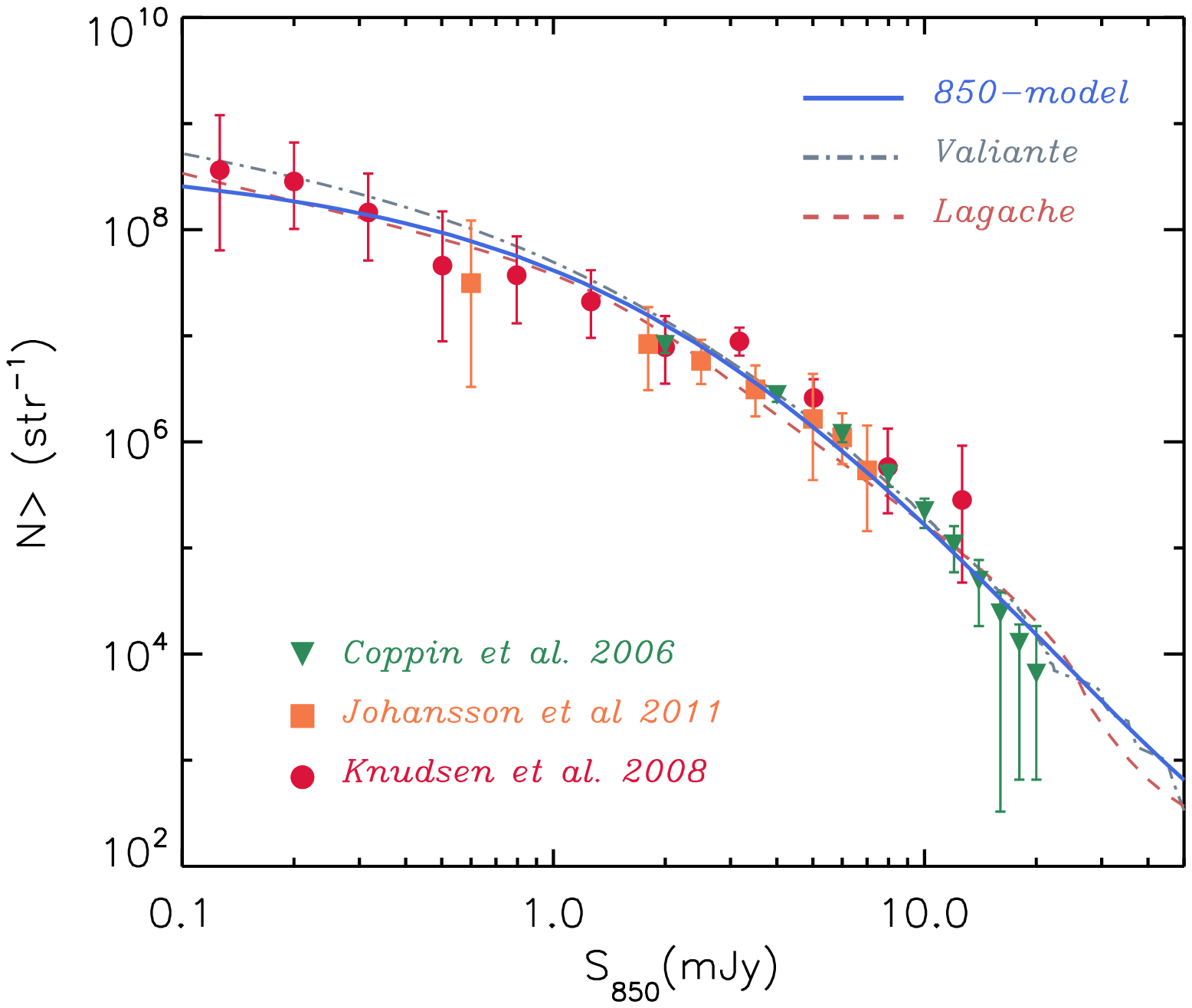}} }
\centerline{\hbox{\includegraphics[width=0.45\textwidth]
             {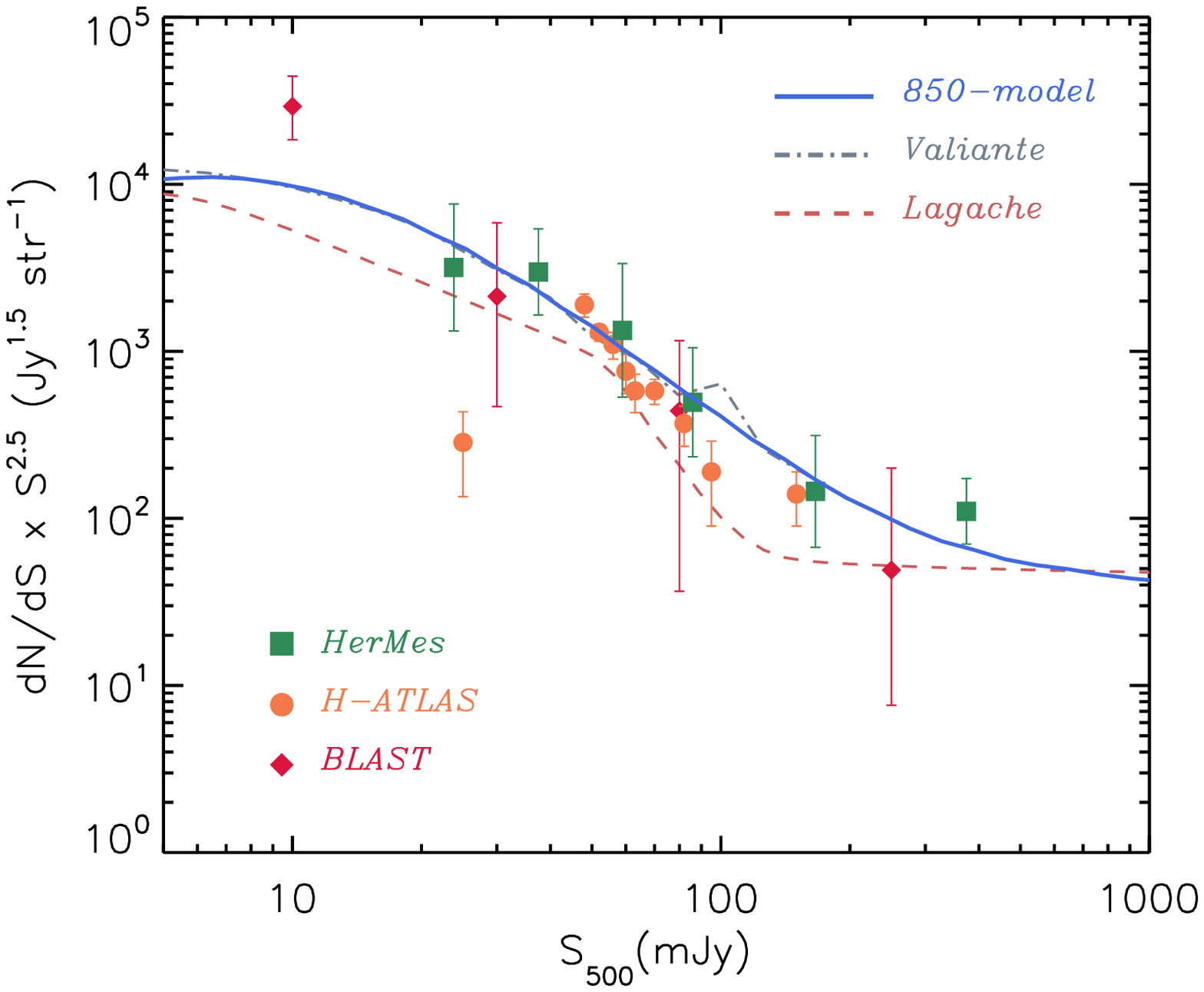}}
            \hbox{\includegraphics[width=0.45\textwidth]
             {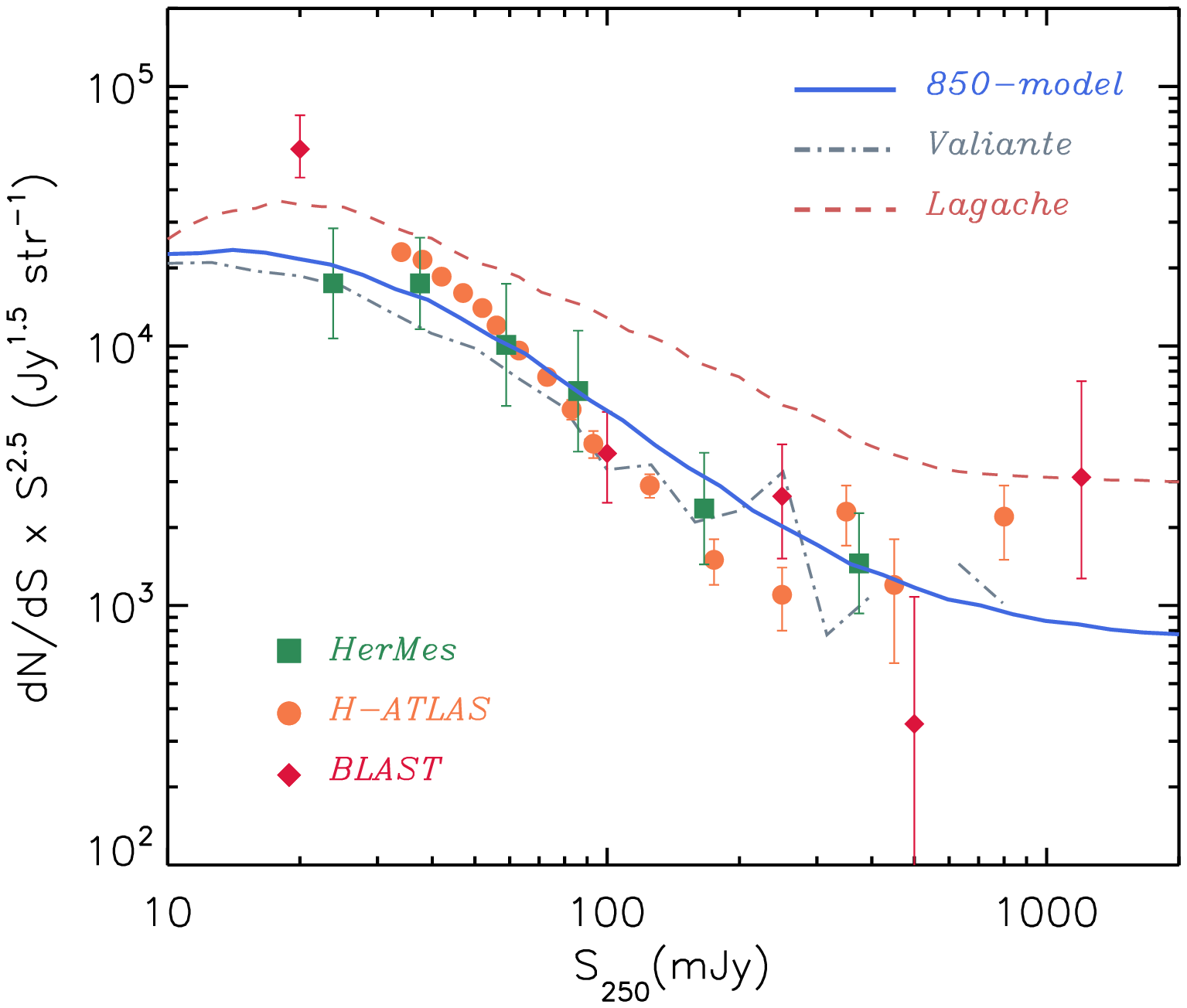}} }
\centerline{\hbox{\includegraphics[width=0.45\textwidth]
             {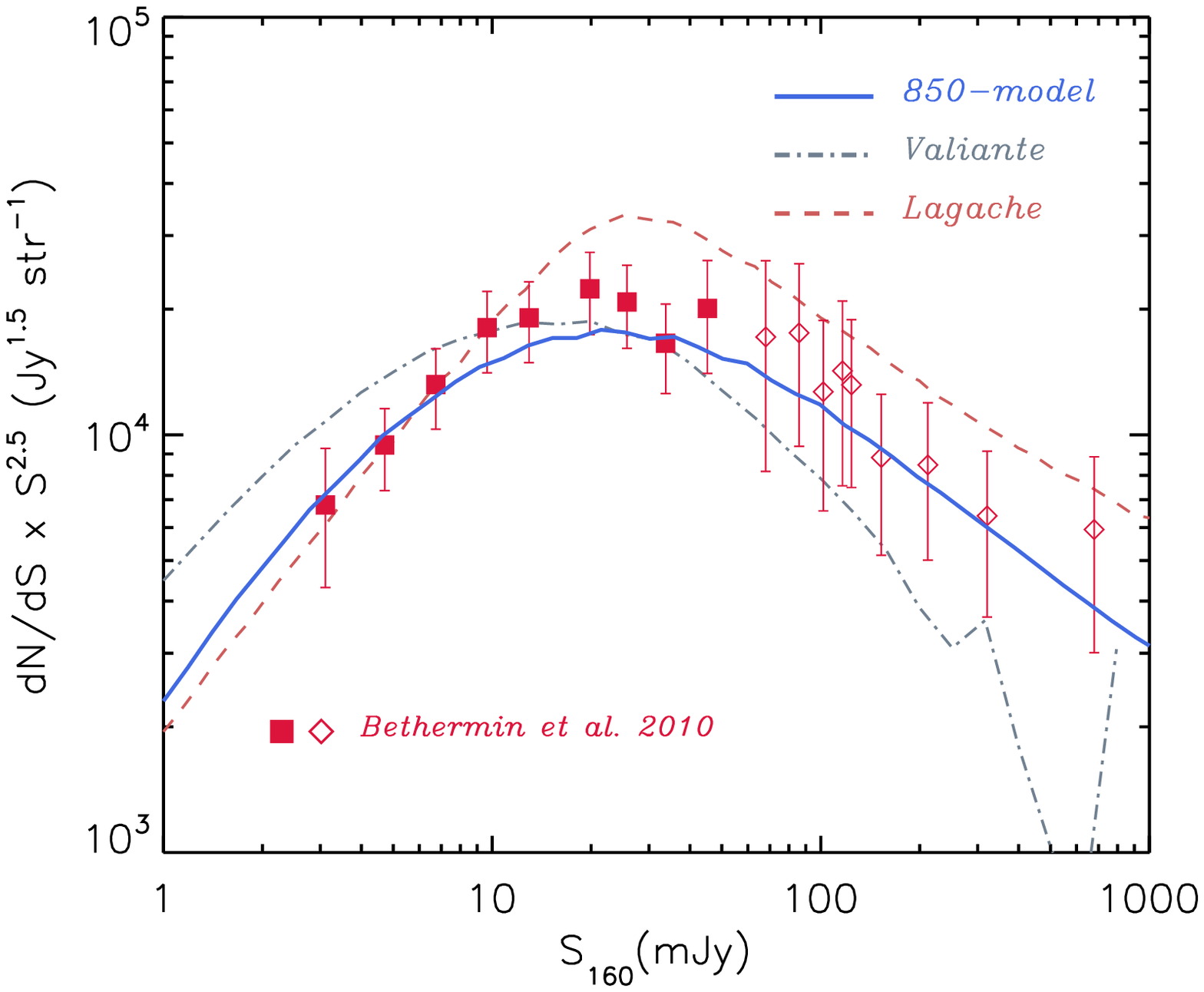}}
            \hbox{\includegraphics[width=0.45\textwidth]
             {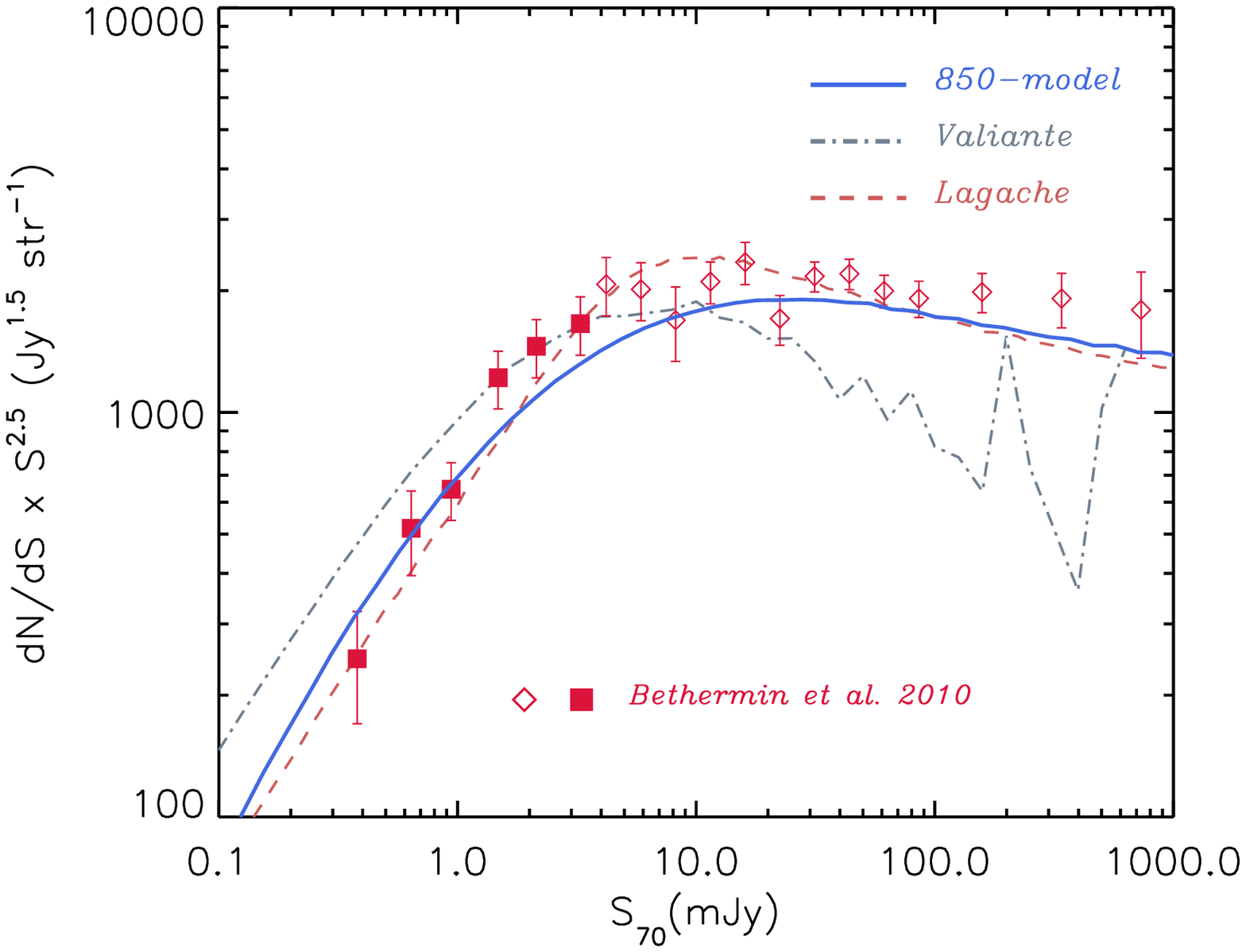}} }
\caption{The same set of observed source counts which is shown in Figure \ref{fig:all-wavelengths} is also illustrated here. The blue solid line shows the result of "850-model" using a modified set of SED (see Section~\ref{sec:all-the-best}) to correct under-production of bright  $160\,\mu$m sources and faint SPIRE source. Source count produced by two different models, \citet{Lagache04} and \citet{Valiante09}, are also respectively shown by orange dashed line and gray dot-dashed lines. While \citet{Valiante09} and "850-model" counts at $160\,\mu$m are almost identical, we did not find \citet{Lagache04} model predictions at $1100\,\mu$m to include them in the top-left panel.} \label{fig:all-wavelengths-comp}
\end{figure*}
\begin{figure*}
\centerline{\hbox{\includegraphics[width=0.48\textwidth]
             {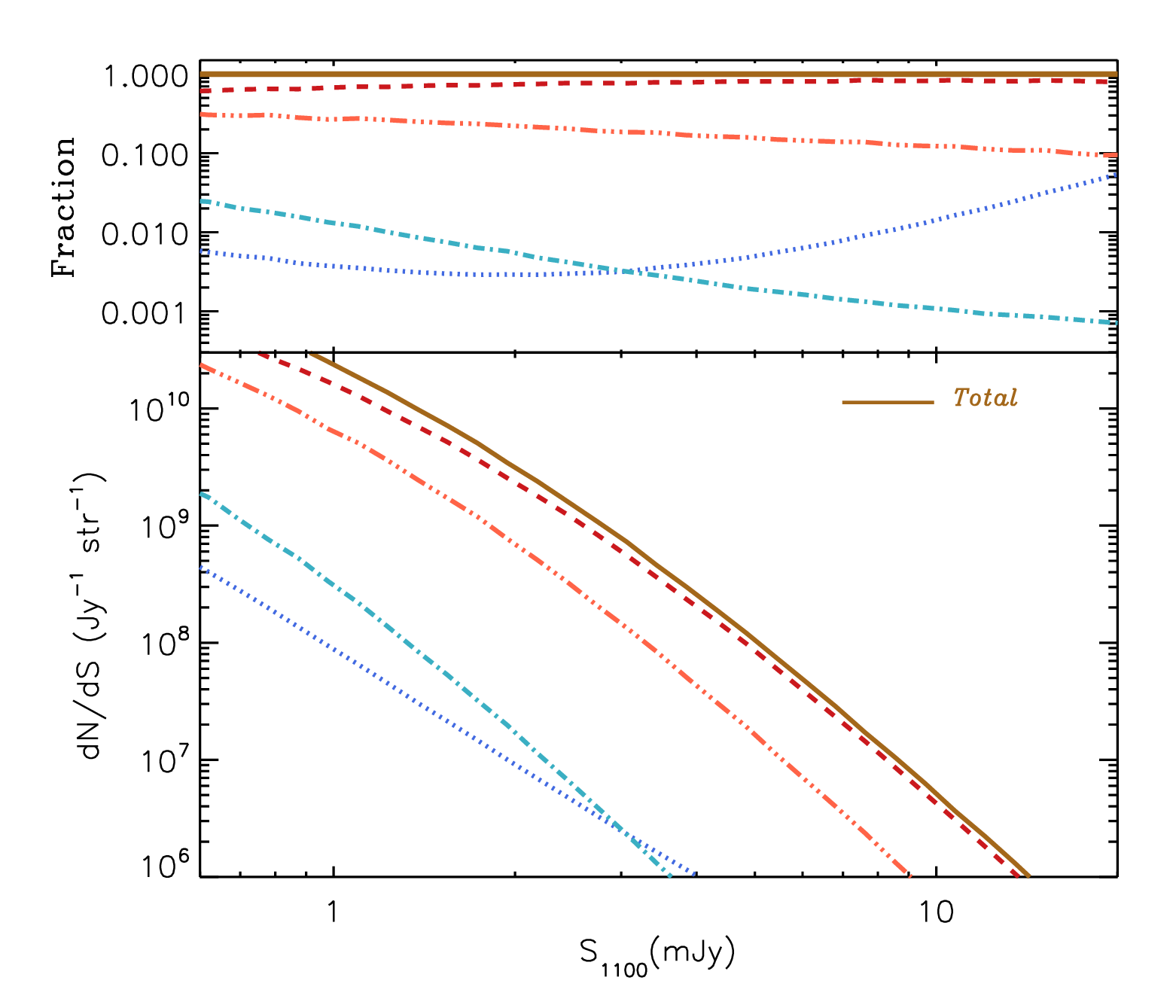}}
            \hbox{\includegraphics[width=0.48\textwidth]
             {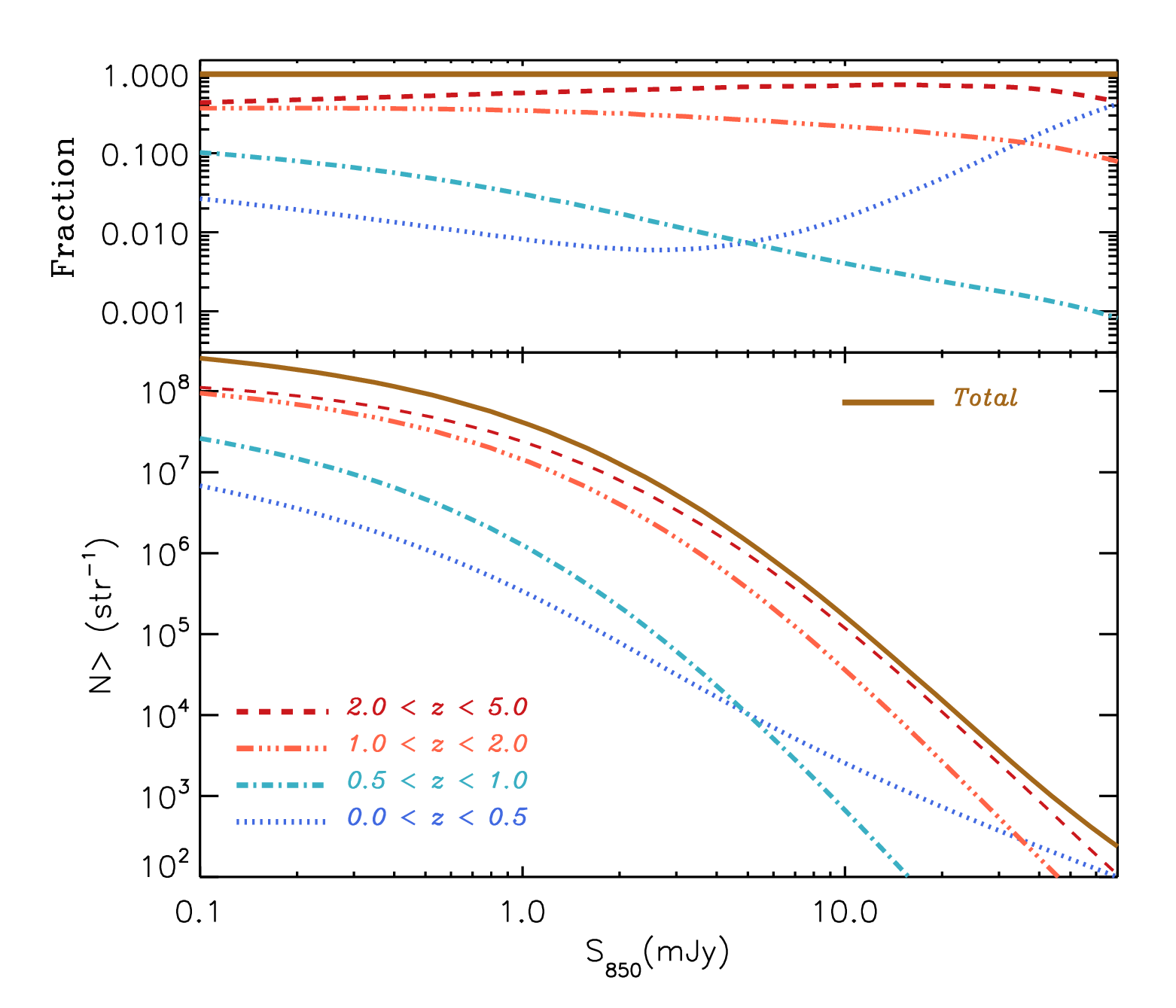}} }
\centerline{\hbox{\includegraphics[width=0.48\textwidth]
             {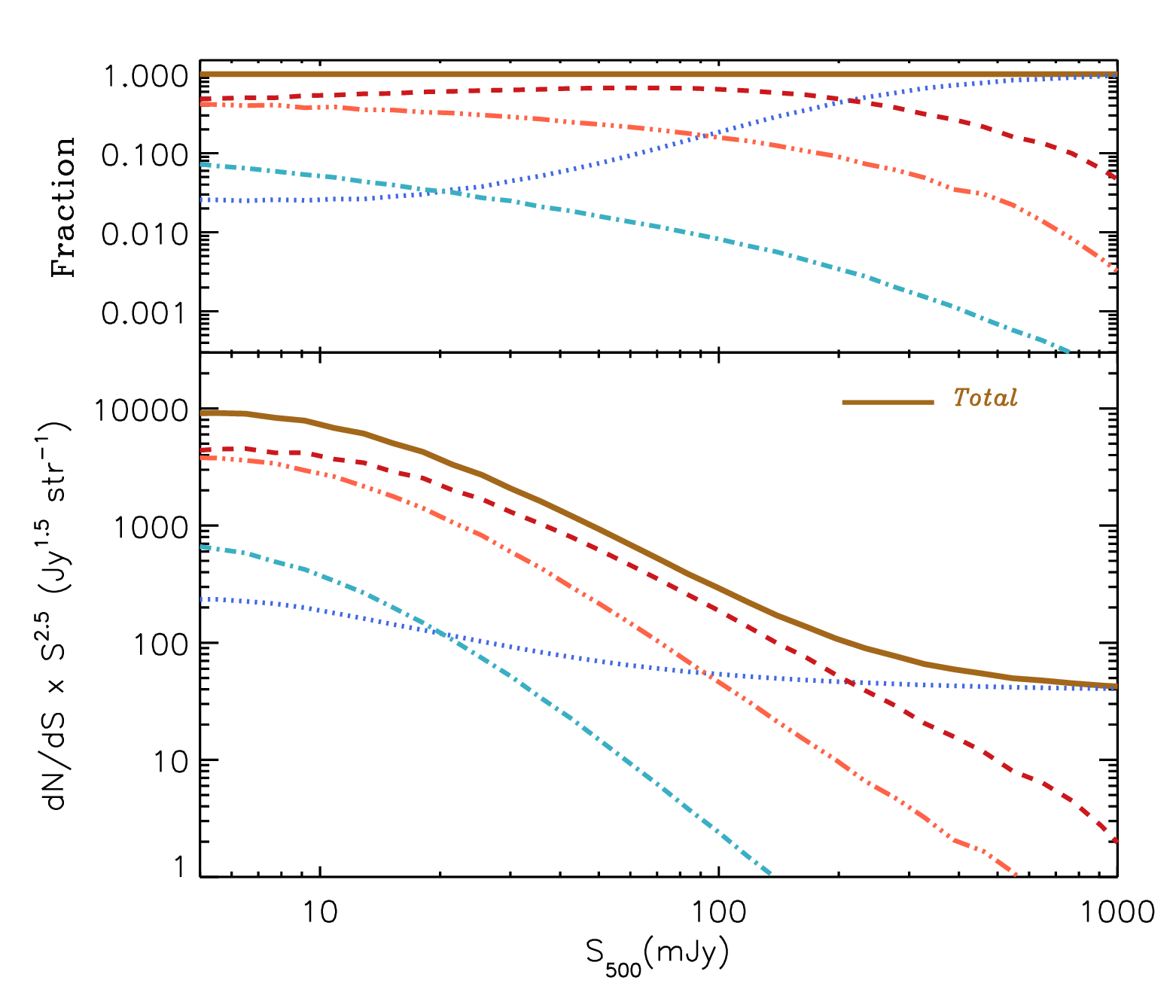}}
            \hbox{\includegraphics[width=0.48\textwidth]
             {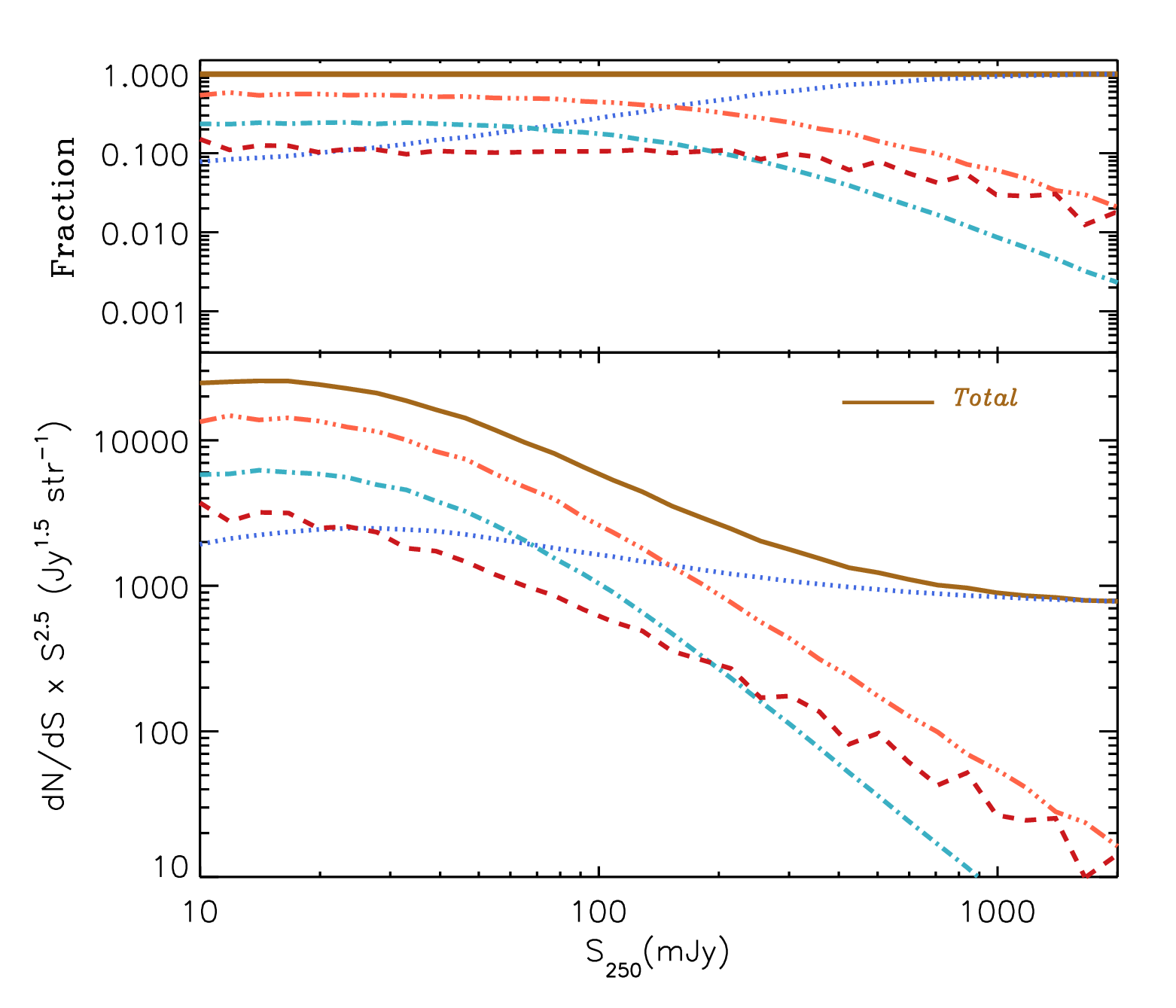}} }
\centerline{\hbox{\includegraphics[width=0.48\textwidth]
             {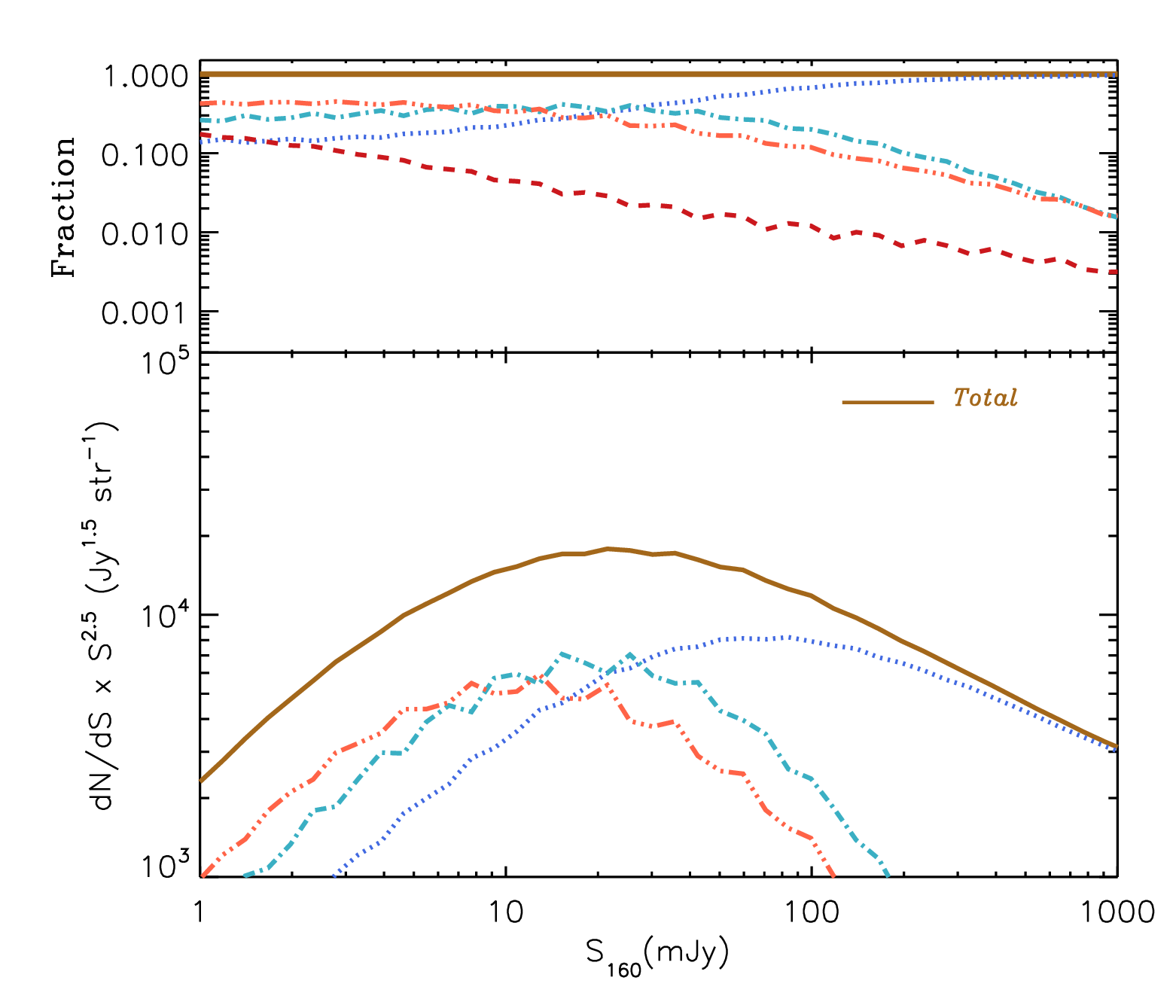}}
            \hbox{\includegraphics[width=0.48\textwidth]
             {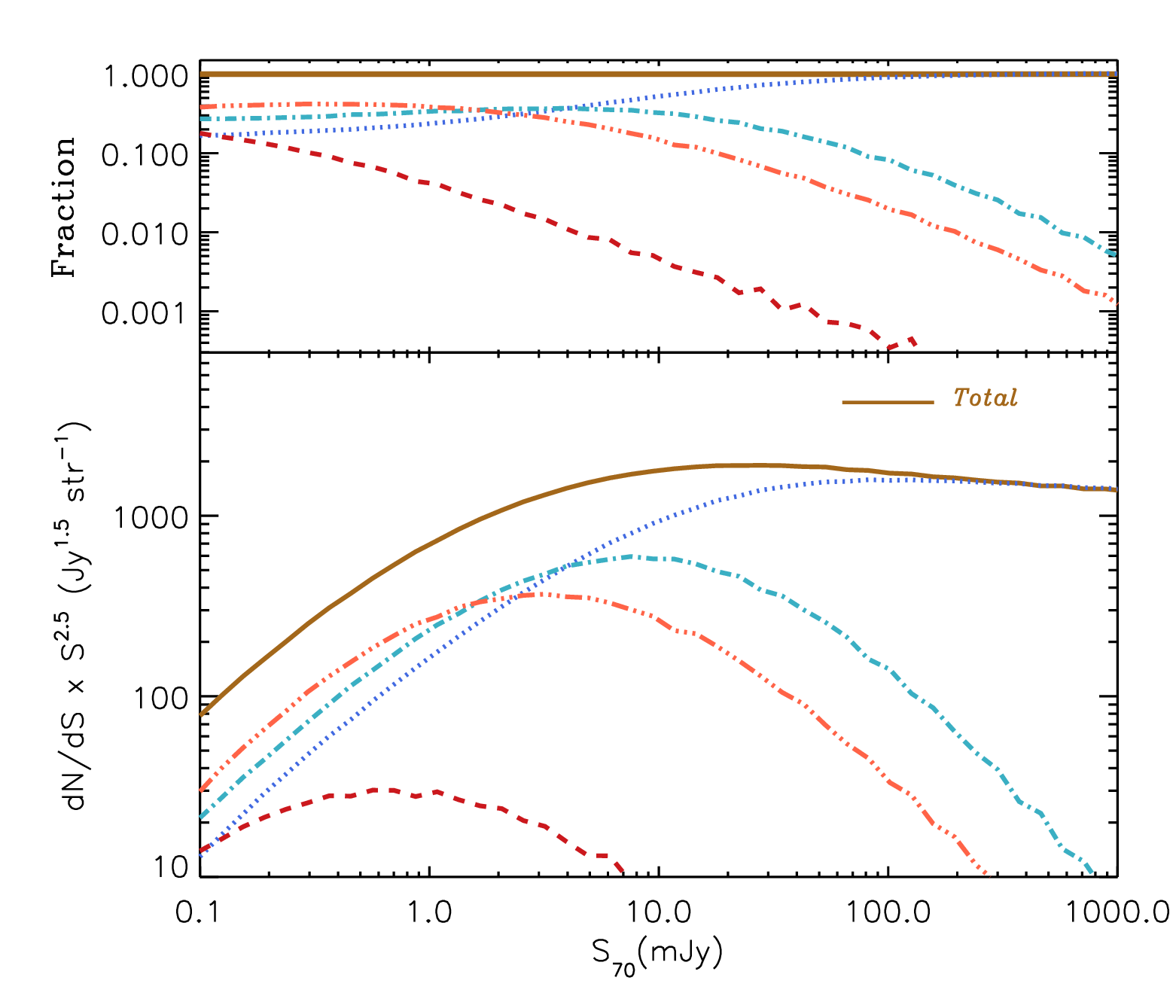}} }
\caption{The fractional contribution of sources in different redshift bins in producing the observed source counts at different wavelengths are illustrated. The blue Dotted, cyan dot-dashed, orange dot-dot-dashed and red dashed lines respectively show the source count produced by objects in $0 < z<0.5$, $0.5 < z<1$, $1 < z<2$ and $2 < z<5$ redshift intervals while the total source count is shown using the brown solid line. In the top section of each panel, the fraction of sources for each redshift bin in producing source counts at different flux thresholds is plotted.} \label{fig:z_dist}
\end{figure*}
\subsection{A best-fit model for all wavelengths}
\label{sec:all-the-best}
As we showed in previous subsections, the required evolution scenarios for different models which are constrained by various wavelengths is too diverse to be reconciled in a single model; however, the $70\,\mu$m and $850\,\mu$m source counts can be explained by the same model, either the "850-model" or "70-model" which are almost identical in terms of implied evolution and distribution of IR galaxies and predicted source counts at different wavelengths. While $70\,\mu$m source counts are produced mainly by local and low redshift objects, $850\,\mu$m sources are distributed in higher redshifts and in a wide redshift interval. This means the model which can explain the source count at both $70\,\mu$m and $850\,\mu$m gives a consistent distribution of IR galaxies from local Universe to very high redshifts. In other words, the same evolution scenario for star-forming galaxies which we observe locally, can also produce the observed $850\,\mu$m counts. This scenario, contrary to what is implied by the best-fit models constrained at other intermediate wavelengths, does not strongly violate observed source counts at other bands, though it under-produces them at some flux thresholds. On the other hand, models which agree with observations at intermediate  wavelengths require extreme color evolutions to produce larger number of cold objects in intermediate redshift ranges which is relevant for the source count at those wavelengths (i.e. $1<z<2$, see the left panel in the middle of Figure~\ref{fig:z_dist}). However, this steep colour evolution at higher redshifts (i.e. $z>2$) produces too many observable sources at $850\,\mu$m. 

Moreover, as we mentioned earlier different authors with fundamentally different models have reported inconsistencies in their best-fit models (which fit the long and short wavelengths) and intermediate source counts \citep{LeBorgne09,Valiante09} \footnote{Those works point out this issue only for the source counts at $160\,\mu$m since the observations at $250\,\mu$m, $350\,\mu$m and $500\,\mu$m have been available only recently.} and the underestimation of $160\,\mu$m (and SPIRE wavelengths) seems to be model independent.

There are two main possibilities which can explain the origin of discrepancy between the "850-model" and source counts at SPIRE band and $160\,\mu$m, besides doubting the validity of observed counts: {\bf{(i)}} the IR SEDs in our model are not representative and should be modified in a certain way to produce more flux at observed wavelengths $\lambda_{obs} \sim 160$ - $500\,\mu$m or {\bf{(ii)}} existence of a population of cold galaxies at relatively low redshifts. Indeed, there is a body of evidences supporting a population of cold galaxies which are underrepresented in IRAS galaxies and hence $70\,\mu$m source counts \citep{Stickel98,Stickel00,Chapman02,Patris03,Dennefeld05,Sajina06,Amblard10}. However, based on available data there is no possible way to disentangle between the two mentioned possibilities \citep{LeBorgne09}. Therefore, due to its simplicity, we try to find a modification in SED amplitudes which could improve the agreement between the "850-model" and observed source counts at SPIRE wavelengths and $160\,\mu$m, without affecting other wavelengths. In the following we first introduce the desired SED modification and then based on the redshift distribution of sources responsible for different source counts in our modified model, we try to constrain different properties a population of cold IR galaxies should possess to be equivalent to that SED modification.

Since the "850-model" is capable of fitting the observed $70\,\mu$m source counts, any modification of SED templates should be at wavelengths longer than $70\,\mu$m to leave this agreement intact. The $850\,\mu$m source counts that model produces should also remains the same which put an upper limit for the allowed wavelength range of any modification: since a significant fraction of observed counts at $850\,\mu$m is produced by sources at high redshifts, typically $z > 2$ (see Figures~\ref{fig:short-long-fluxes} and \ref{fig:all-wavelengths-comp}), SED templates should not change at rest-frame wavelengths longer than $\sim 200\,\mu$m. Based on this ansatz, we searched for the amplitude and the wavelength range of SED changes which can bring the "850-model" intermediate wavelength counts close to the observed values. We found that if the SED templates we use in our model be amplified by a factor of $1.6$ in the rest-frame range of $70 < \lambda <150\,\mu$m, the "850-model" can fit the SPIRE and  $160\,\mu$m source counts, without any change in already decent results at $70\,\mu$m, $850\,\mu$m  and $1100\,\mu$m. Figure~\ref{fig:all-wavelengths-comp} illustrates the results produced by this modified "850-model" at different wavelengths (blue solid line) together with the results from \citet{Lagache04} (orange dashed line) and \citet{Valiante09} (gray dash-doted line) models. The redshift distribution of sources responsible for those results are also shown in Figure~\ref{fig:z_dist}.

The SED boost we implemented, recovers the observed $160\,\mu$m counts by doubling the number of observable sources with $S_{160} > 10{\rm{mJy}}$; as illustrated in Figure~\ref{fig:all-wavelengths-comp}, a significant fraction of those sources is distributed at redshifts $z < 1$ which is also in agreement with recent observations \citep{Jacobs11,Berta11}. However, for the SPIRE source counts, the SED modification increases the number of observed sources with $S_{\rm{SPIRE}} < 100{\rm{mJy}}$ which are mainly at redshifts $0.5 < z <2$. Those redshift distributions together with the fact that the emission from cold dust with temperatures $T \lesssim 30$K peaks at $\sim 100\,\mu$m, make our SED modification equivalent to adding a population of preferentially cold dusty galaxies which do not exist at high redshifts. The width of redshift range in which those objects can exist depends on their temperature range: a warmer population can have a wider redshift distribution but should have a typically higher redshifts to be invisible at $70\,\mu$m while a colder population can only exist in low redshift ranges in order to not interfere with $850\,\mu$m counts. Interestingly, this is in agreement with the required steep colour evolution implied by models constrained by $160\,\mu$m and SPIRE band counts (see Table~\ref{table:models}).

As one can see in Figure~\ref{fig:all-wavelengths-comp}, some models like \citet{Lagache04} and \citet{Valiante09} not only produce enough observable sources in SPIRE and  $160\,\mu$m bands, but at some flux thresholds produce too many sources. However, we notice that both of those models need an additional mechanism to compensate for under-production of visible sources at intermediate wavelengths which mimic a population of cold sources only in low redshifts. For instance, \citet{Lagache04} use a class of "cold galaxies" in their evolutionary model which are present only at low redshift, $z < 0.5$; \citet{Lagache03} show that this cold population is producing up to $\sim50\%$ of the observed $170\,\mu$m flux which means without them the source count is reduced by the same factor, consistent with our finding and also other models \citep{LeBorgne09,Valiante09}. Similarly, \citet{Valiante09} need to strongly modify the colour distribution of low redshift galaxies to correct for under-producing $160\,\mu$m sources by a factor of $\sim5$; they modified the observed colour distribution of IRAS galaxies which they use as starting point (similar to our model, see Section~\ref{sec:CLF0}) only for low redshift (i.e. $z<1$) objects, by assuming an asymmetric Gaussian distribution which is 7 times broader on the "cold" side of distribution in comparison to the "warm" side (see Section~4.4 in \citet{Valiante09}); even with this extreme modification their model does not completely match the observed $160\,\mu$m counts in addition to under-producing luminous $70\,\mu$m sources. 

Finally, it is worth mentioning that our SED modification is not expected to change the best-fit models which are based on  $70\,\mu$m and $850\,\mu$m since we used those source counts as constraints for our SED change search. However we double checked this issue by using the modified SED set and repeating the search in parameter space for a best-fit model which is constrained by source counts at $70\,\mu$m, $160\,\mu$m, SPIRE bands and $850\,\mu$m and recovering the parameters which define the "850-model", for the best-fit model.

\section{Discussion}
\label{sec:discussion}
 In this section, we discuss different properties of our best-fit model. First we discuss different implications of our model for the evolving properties of the IR galaxies like their colour and luminosity distributions. Then, we compare our model with other existing models for IR and submm source counts.

\subsection{The implied evolution scenario for dusty galaxies}
\label{sec:evol}
\begin{figure*}
\centerline{\hbox{\includegraphics[width=0.48\textwidth]
             {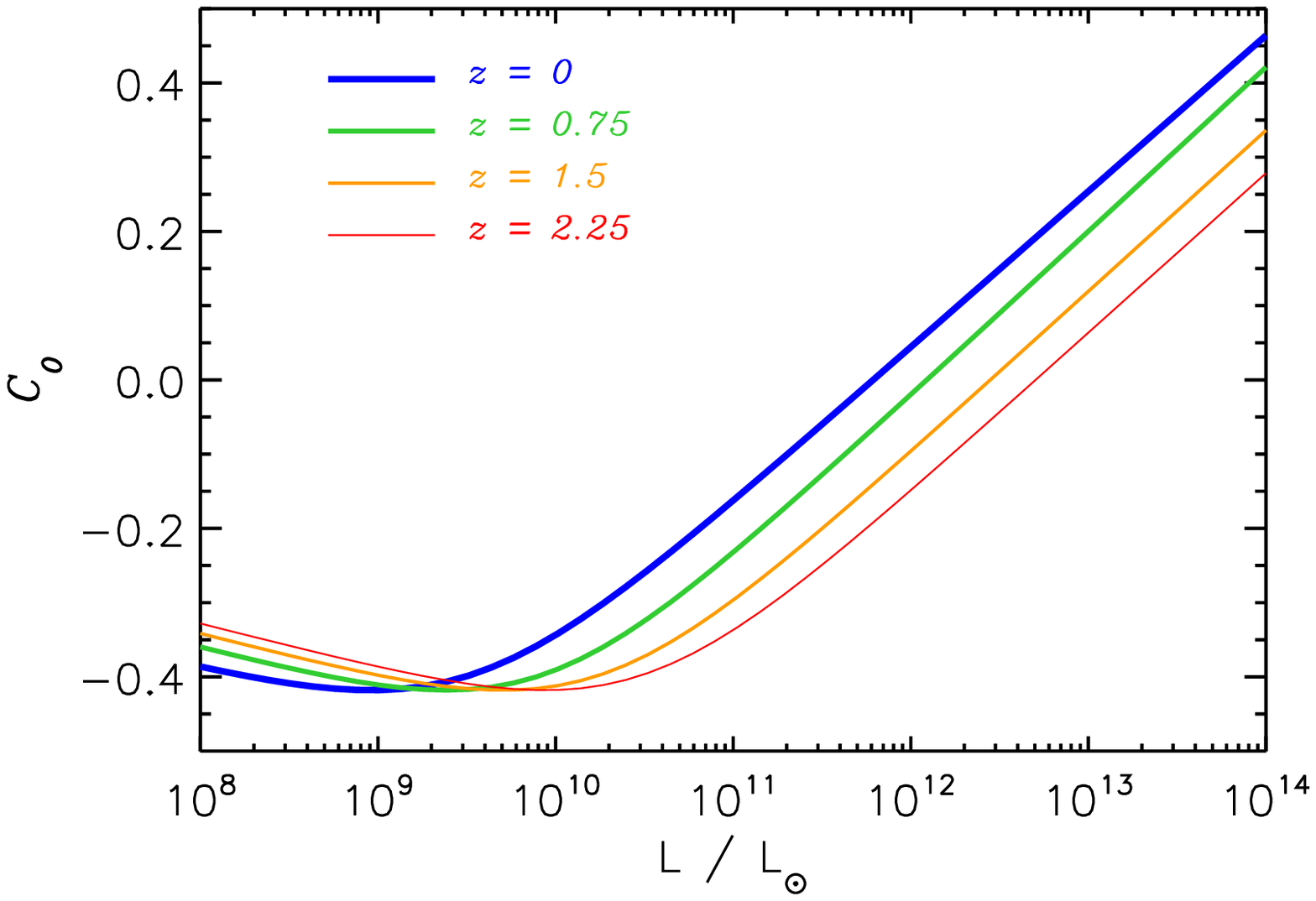}}
            \hbox{\includegraphics[width=0.48\textwidth]
             {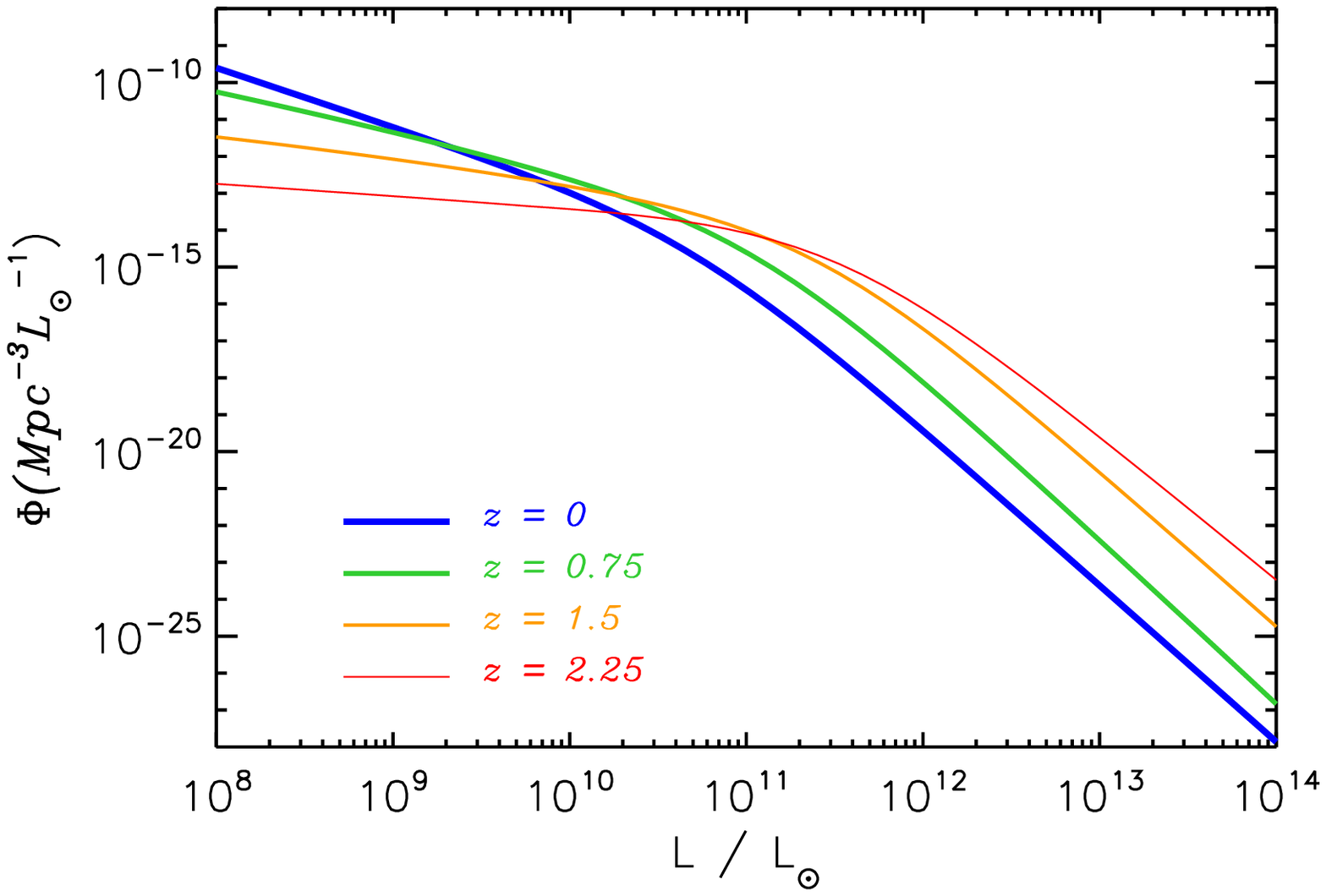}} }
\centerline{\hbox{\includegraphics[width=0.48\textwidth]
             {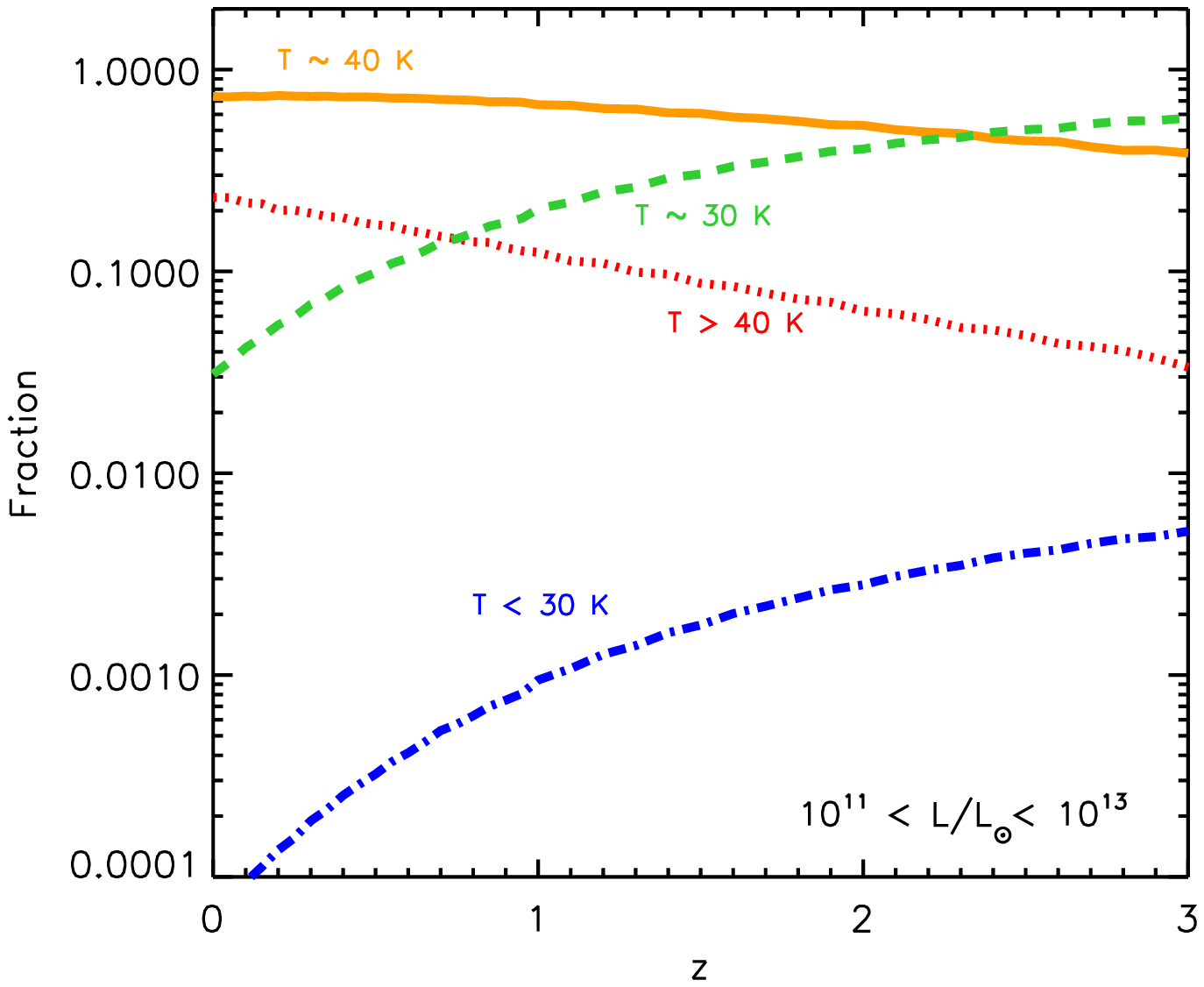}}
            \hbox{\includegraphics[width=0.48\textwidth]
             {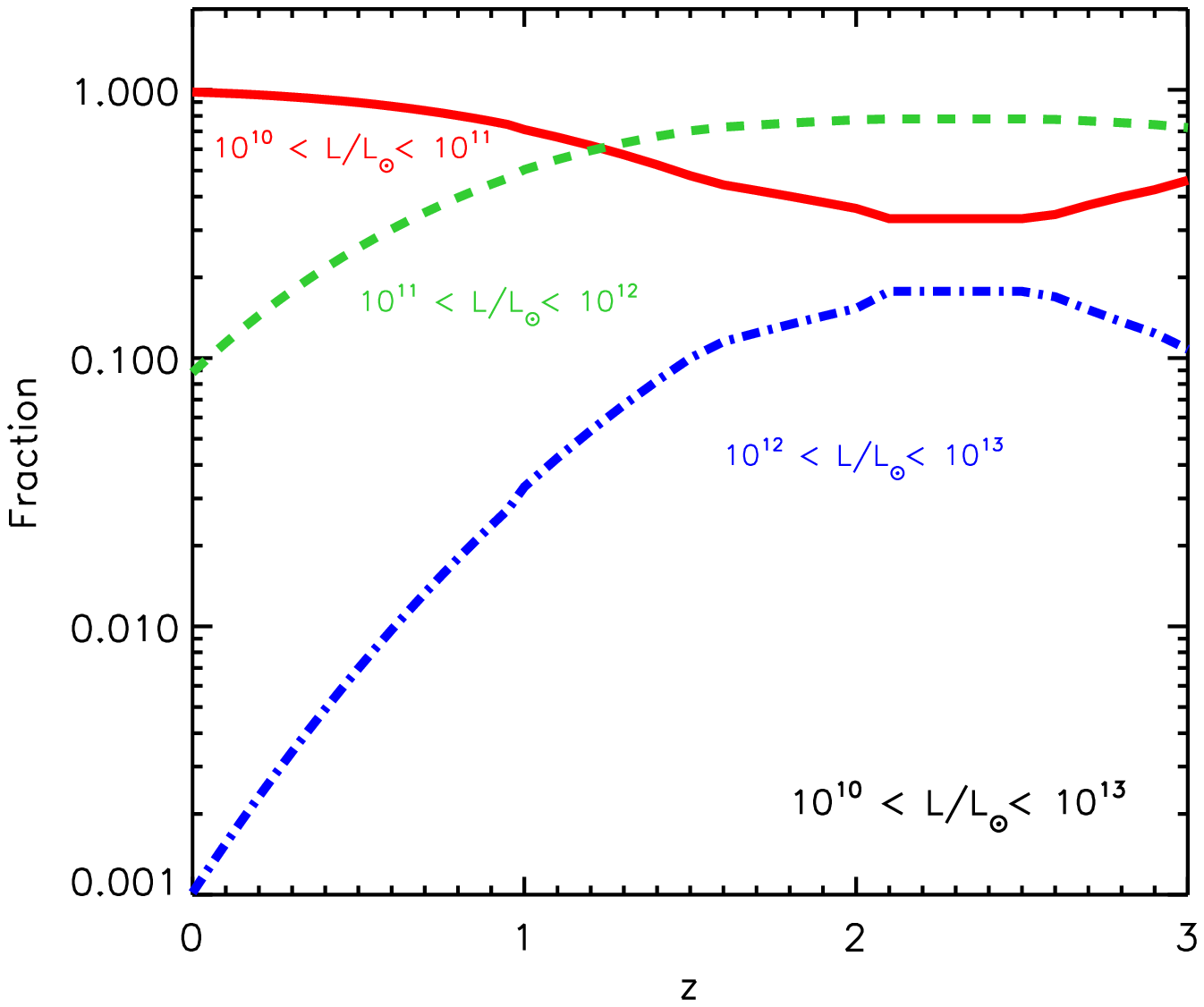}} }
\caption{Top: The evolution of CF (left) and LF (right) with redshift.
The lines from thick to thin are respectively for redshifts $z = $ 0 (blue),
0.75 (green), 1.5 (orange) and 2.25 (red). Bottom: The change of the relative
contribution of different populations of IR galaxies to their total density
with redshift, as required to reproduce the observed source counts
correctly. {\it{left:}} The fraction of objects with luminosities
$10^{11} \le L/L_{\odot} \le 10^{13}$ which have typical temperatures of
$T<30$ K, $\sim 30$ K, $\sim 40$ K and $T>40$ K shown respectively with
blue, green, orange and red lines. {\it{right:}} The fraction of total IR
luminosity generated by
objects with luminosities $10^{10} \le L/L_{\odot} \le 10^{13}$, which have
luminosities of $10^{10} \le L/L_{\odot}<10^{11}$,
$10^{11} \le L/L_{\odot}<10^{12}$ and $10^{12} \le L/L_{\odot} \le 10^{13}$
shown respectively with red, green and blue lines.} \label{fig:frac_evol}
\end{figure*}

Our best-fit model mimics the number density evolution of IR galaxies by employing a
luminosity evolution together with changing slopes of LF with redshift. While
the former changes the amplitude of LF, the latter acts to change the shape of
LF properly to reproduce a number density history which matches the observed FIR
and submm source counts.  Although the good agreement between source
counts which our best-fit model produces at different wavelengths and the
observed numbers firmly supports our model, the implied results should be treated
carefully, mainly due to the simplicity of the model: for instance, in our model
we adopted the local LF of IR galaxies which is a simple dual power law function, and
evolved its characteristic luminosity and slopes with redshift to distribute
objects with different luminosities in redshift space correctly. 

Although the functional forms we chose for evolution of those parameters are rather
arbitrary, the actual distribution they produce at different redshifts is
necessary to explain the properties of the observed sources. To emphasize this
point, in Figure~\ref{fig:frac_evol} we illustrate how the fractions of galaxies
in different luminosity bins are evolving with redshift, together with the exact
shape of the LF our model requires at different redshifts. As can be seen in the
bottom right panel of Figure~\ref{fig:frac_evol}, at low redshifts fainter
objects dominate the population of IR galaxies but at higher redshifts (i.e., $z
> 2$), objects which are in brighter luminosity ranges take over and become more
dominant. Specifically, this diagram shows that beyond $z\sim1$, objects in
  the Luminous Infrared Galaxies (LIRGs) class (with IR luminosities between
$10^{11}$ and $10^{12}\,L_\odot$) dominate the obscured cosmic energy
production (although in terms of numbers, fainter galaxies still dominate). The
diagram also shows that UltraLuminous Infrared Galaxies (ULIRGs, with IR
luminosities exceeding $10^{12}\,L_\odot$), although increasing in importance
towards high $z$, never dominate the cosmic energy production.
These conclusions are in agreement with analyses of Spitzer data by \citet{LeFloch05} for $z<1$ and by
\citet{Magnelli11} for $z<2.3$.\\

Our best-fit model implies a change in the shape of the LF with redshift and we
are the first to consider this possibility in a model of this type.  This slope
evolution implies that the faint end slope of the LF is flattening with
increasing redshift (see the top right panel of Figure~\ref{fig:frac_evol}). At
first sight it may seems surprising that our results are sensitive to the faint end
slope at high redshifts, but the strong luminosity evolution combined with the
strong negative $K$-correction brings sub-$L_*$ galaxies well within the region
of detectability at $850\,\mu$m. Nevertheless, this result depends strongly on
the sub-mJy counts at $850\,\mu$m which currently are derived from studies of
gravitationally lensing clusters, and therefore sample limited cosmic volume.
As such, these counts may be affected by cosmic variance and establishing their
levels more firmly (e.g., with ALMA), is necessary for confirming our
conclusion.  However, we note the non-parametric estimates of the IR luminosity
function at high redshifts \citep[e.g.,][]{Chapman05,Wardlow11} do indicate
flatter faint-end slopes than the local LF\null. These samples were selected at $850\,\mu$m and may not be complete in IR luminosity, in particular they may be deficient in objects with high dust temperatures
\citep[e.g.,][]{Magdis10}. In order to settle this point, high-$z$ luminosity
functions of IR luminosity-limited samples are required. Redshift surveys of
Herschel-selected samples will be needed to construct such luminosity functions,
and this will become feasible with facilities such as ALMA and CCAT\null.  If
confirmed, the flattening of the faint-end slope of the LF towards higher
redshifts requires a physical explanation, which perhaps could be found using
physically motivated models and simulations. One possibility is that a higher
metagalactic ultraviolet flux at higher redshifts would suppress the development
of a star-forming interstellar medium in low-mass galaxies. Another possibility
is a stellar-mass-dependent evolution in $M_{\rm dust}/M_{\rm stars}$ towards
higher redshifts. Such an evolution could result from the decreasing overall
metallicity towards high $z$ combined with the net effect over time of the
buildup of dust as a result of stellar evolution and its consumption by star
formation. The latter model can be tested observationally using a combination of
Herschel data and multi-band
optical imaging (Bourne et al., in prep.).

We also showed that our model requires a colour evolution to reproduce the
observed source counts. This colour evolution implies that objects with the same
luminosities have lower typical dust temperatures at higher redshifts. In the
bottom left panel of Figure~\ref{fig:frac_evol} we showed how the fraction of
objects with different temperatures is evolving with redshift for all the
objects which have intrinsic luminosities between $10^{11} < \frac{L}{L_{\odot}}
< 10^{13}$. While at low redshifts a population of objects which have typical
temperatures\footnote{We assign a single temperature to each SED based on its
  colour and finding a modified black body radiation with a fixed emissivity,
  $\beta = 1.5$, which can produce that colour.} of $40$K, dominates the colour
distribution, at higher redshifts the cooler typical temperature of $30$K is
dominating. Moreover, while at low redshifts objects with warm dust temperatures
are more common than the very cold objects, the situation is the opposite at
earlier times. Although it is difficult to observe the evolution of dust towards
cooler temperatures because of different selection effects, there is
observational evidence for the implied colour (i.e., temperature)
evolution \citep{Chapman05,Pope06,Chapin09,Symeonidis09, Seymour10,Hwang10,Amblard10}. 
  
The luminosity and colour evolutions discussed above, show up in redshift distributions of source counts at different wavelengths. Figure~\ref{fig:z_dist} illustrates the buildup of source counts at different wavelengths by contribution from different redshift bins. It is evident that the longer wavelengths are reflecting the distribution of higher redshift dusty galaxies while shorter wavelengths depend more on IR galaxies at lower redshifts. For instance $850\,\mu$m number counts mainly consist of galaxies with redshifts higher than $z\sim2$, especially at fluxes $\sim 10-20{\rm{mJy}}$; in addition, as shown in the bottom right panel of Figure ~\ref{fig:frac_evol}, at redshifts $z\sim2$ the luminosity distribution of IR galaxies is not evolving considerably and even at higher redshifts the fraction of fainter sources, which have lower dust temperatures, goes up again in comparison to the brightest objects. Moreover, for two objects with the same redshifts and intrinsic IR luminosities, the colder one will be brighter at observed $850\,\mu$m. This means the visible $850\,\mu$m sources at higher flux thresholds are preferentially colder than sources with fainter observed fluxes which are distributed typically at lower redshifts (see the increasing fraction of objects in $1 < z < 2$ redshift bin going towards lower fluxes in the top right panel of Figure~\ref{fig:z_dist}). This is also in agreement with our experiments which show that at the bright end, $850\,\mu$m number counts are very sensitive to the slope of the colour evolution in our model, which is in fact the dominant mechanism to produce those counts. Moreover, this implied broader temperature distribution of observed sources at fainter flux thresholds and shorter wavelengths is in agreement with observations which show bright $850\,\mu$m population (i.e. $> 4{\rm{mJy}}$) is highly biased towards cold dust temperatures while fainter sources (i.e. $1-4{\rm{mJy}}$) also contain an increasing fraction of more luminous objects with lower redshifts but warmer dust temperatures which makes them the main contributors to the $250\,\mu$m source counts \citep{Chapman04,Chapman10, Magnelli10,Casey09,Casey11}. Consequently, at $z\sim2$ the density of farIR selected ULIRGs is approximately 2 times higher than that of $850\,\mu$m-selected ULIRGs. This is consistent with our model: a typical ULIRG at redshift $z\sim2$ will be visible at $850\,\mu$m with a flux brighter than a few mJys and at $250\,\mu$m brighter than $\sim 100$mJy; on the other hand, roughly $\sim 20-30\%$ of objects with few mJy fluxes at $850\,\mu$m are within redshifts $z\sim1-2$ while around $\sim 50\%$ of objects with $\sim100$mJy fluxes at $250\,\mu$m are in the same redshift bin (see Figure~\ref{fig:z_dist}).

While the luminosity and colour distribution our best-fit model requires is consistent with observed $70\,\mu$m, $850\,\mu$m and $1100\,\mu$m source counts, they are not sufficient to produce enough sources at observed wavelengths in between. The inconsistency between models which successfully produce $70\,\mu$m and $850\,\mu$m counts and their results at $160\,\mu$m, is also reported in other works and has been corrected by including a population of cold galaxies at low redshifts \citep{Lagache03,Lagache04,LeBorgne09,Valiante09}. While it is important to note the under-production of  $160\,\mu$m counts in models, can be corrected equally by a modifying SED templates instead of introducing a new population (see also \citet{LeBorgne09}), there is some observational evidence for the existence of a cold population at low redshifts which is under-represented in IRAS sample and is often associated with bright spiral galaxies\citep{Stickel98,Stickel00,Chapman02,Patris03,Dennefeld05,Sajina06,Amblard10}.

Recently, flux density measurements at 250, 350 and $500\,\mu$m have become available for large samples of local galaxies from surveys with SPIRE on the Herschel Space Observatory \citep{Eales10,Oliver10,Clements10}.  In addition to problems at $160\,\mu$m, we also noticed the inconsistency between our best-fit model and the source counts provided by SPIRE observations. However, we resolved this issue by modifying our SED templates to be able to reproduce the source count simultaneously at $70\,\mu$m, $160\,\mu$m, SPIRE band, $850\,\mu$m and $1100\,\mu$m. There is also observational evidence for a population of cold galaxies residing at low redshifts (equivalent to modified SEDs) in Herschel-selected samples: using a subsample with spectroscopic or reliable photometric redshifts from Herschel ATLAS survey, \citet{Amblard10} performed isothermal graybody fits to low-redshift galaxies
detected from 70 to $500\,\mu$m, resulting in an IR luminosity-temperature relation offset to significantly lower temperatures when compared to the IRAS-based relation derived by \citet{Chapman03}. The new
relation found by \citet{Amblard10} is consistent with earlier work by
\citet{Dye09} based on BLAST data. It is also important to realize that these results
do not imply that the IRAS-based dust temperature fits are incorrect (in fact, they
are often supplemented with measurements at longer wavelengths) but they imply that an IRAS-based selection is biased towards warmer objects.
  
\subsection{Our best-fit model and previous models}
There are several phenomenological models in the literature which try to
reproduce the properties of IR galaxies at different wavelengths, with different
levels of complexity (e.g.,
\citet{Blain93,Guiderdoni97,Blain99,Chary01,Rowan-Robinson01,Dole03,Lagache04,Lewis05,LeBorgne09,Valiante09}).
In general, those models use an assumed form of luminosity evolution together with a density evolution to mimic the evolution of IR galaxy distributions. However, we have limited ourselves to a pure
  luminosity evolution with no density evolution, and as pointed out in
  Section~\ref{sec:CLFz}, the amount of density evolution allowed by the
  integrated CIB is small but does not have to be zero. On the other hand, we use evolving faint and bright ends slopes in our luminosity function. At first sight this may looks a simple substitution of free parameters; we tried to investigate this by trying to substitute the slope evolution of our model with a density evolution. However, the resulting fit with density evolution and in absence of slope evolution is significantly inferior to the fit obtained with pure luminosity evolution.
  
Another usual practice in modeling the infrared and submm source counts is to use only one or a few SED templates to represent the whole galaxy population. This approach neglects the observed colour distribution of local IR galaxies and do not leave any possibility for colour evolution. However, similar to \citet{Valiante09}, in our model we use a complete set of SED templates which are shown to be representative of local IR galaxies. This choice enabled our model to explore the evolution in colours of IR galaxies in addition to their luminosities.

Our model also differs from other works in the literature in calculating the source counts for a given evolutionary scenario: we calculate the source count for a given model by computing the probability of observing different sources
for different flux thresholds. The direct consequence of this new approach is a
fast calculation routine which enables us to calculate the source counts at very
bright observed fluxes, where Monte-Carlo based methods are very inefficient due
to the rarity of such objects, and therefore sometimes end up with very noisy results (see the bright source counts produced by \citet{Valiante09} in Figure~\ref{fig:all-wavelengths-comp}). Moreover, our fast algorithm is an important advantage when it comes to searching the parameter space for the best-fit model.

A comparison between our best-fit model and \citet{Lagache04}, as a model without colour distribution and evolution and \citet{Valiante09} as a model with colour distributions is shown in
Figure~\ref{fig:all-wavelengths}. While at $1100\,\mu$m, \citet{Valiante09} produces $\sim 2$ times more visible objects than what our model produces, all models do reasonably well in 
accounting for the comulative $850\,\mu$m number counts. The results from \citet{Valiante09} and our model are very close at SPIRE wavelengths but the \citet{Lagache04} model overpredicts the bright counts at $250\,\mu$m. At $160\,\mu$m, the \citet{Valiante09} model over-produces the faint counts and under-produces the bright objects. However, \citet{Lagache04} fit the data better, while slightly over-produces the counts for intermediate to bright flux thresholds. Finally, at $70\,\mu$m, where all the models are expected to fit the data since they use it as a starting point, the \citet{Valiante09} source counts deviate from observations by over-producing the faint counts in expense of producing too few bright objects (probably because of a too extreme modification in colour distributions which is required in their model to compensate for a factor of $\sim 5$ under-production of $160\,\mu$m sources).
  
\section{Conclusions}
\label{conclusion}

We have described a backward evolution model for the IR galaxy population, with
a small number of free parameters, emphasizing which parameters are constrained
by which observations.  We also introduced a new algorithm for calculating source counts for a given evolutionary model by direct integration of probability distributions which is faster than using Monte-Carlo sampling. This is an important advantage for searching large volumes of parameter space for the best-fit model.

While most of the earlier works used only one or a handful of SED templates to
represent the whole population of IR objects, we used a library of IR SEDs which
are able to match the IR properties of the large variety of observed
star-forming objects. This approach is necessary in order to model the colour evolution of IR galaxies in addition to produce simultaneously the counts and the redshift distributions at wavelengths shorter
than $850\,\mu$m.  

Contrary to some other models, we assumed a negligible
contribution from AGN in our SED templates, noting the inclusion of AGN is only
necessary for reproducing the properties of IR galaxies at very short IR
wavelengths\footnote{For instance at $24\,\mu$m where our model under-produces the counts at flux thresholds $0.1<S_{\rm{24}} < 10{\rm{mJy}}$ by a factor of $\sim 1.5-2$ but matches the observed data at fainter and brighter fluxes} which could also be sensitive to other modeling difficulties such as the PAH contribution to the SEDs.
  
We used available $850\,\mu$m source counts together with the redshift distribution of submm galaxies to constrain our best-fit model. At $850\,\mu$m, due to the K-correction, the source count is sensitive to the evolution of IR galaxies in a wide redshift range and out to very high redshifts. We showed that there is a degeneracy between the rate by which the characteristic luminosity of IR galaxies should increase to reproduce the source count and the maximum redshift out to which this increase should be continued; we resolve this degeneracy by requiring the model to reproduce the observed redshift distribution of submm galaxies. Moreover, we showed that our model requires a colour evolution towards cooler typical dust temperatures at higher redshifts. The employed colour evolution is similar to that used by
\citet{Valiante09} however, our best-fit model predicts a somewhat stronger colour evolution than that proposed by these authors.

Another important feature of our model is that the best-fit is obtained using
pure luminosity evolution but mildly evolving high-luminosity and low-luminosity
slopes in the LF\null. Since high-luminosity sources are rare, the evolution in
the high-luminosity slope is of little consequence. However, the evolution of
the low-lumuninosity slope affects large numbers of galaxies and if confirmed,
this effect must have a physical origin, which can be addressed using numerical
simulations of the evolution of the galaxy population, as well as with a
combination of deep Herschel and optical imaging.

The $850\,\mu$m-constrained best-fit model is consistent with observed $1100\,\mu$m
and $70\,\mu$m source counts, which confirm the consistency of the implied colour-lumionosity-redshift distribution at both low and high redshifts. However, this model under-produces the observed source counts at intermediate wavelengths, namely at $160\,\mu$m and SPIRE bands. To resolve this issue, we used the observed data at different wavelengths to find best-fit models which can reproduce their observed source counts. While the best-fit models constrained at $70\,\mu$m and $850\,\mu$m are consistent with each other and also $1100\,\mu$m, the implied evolutions for models capable of reproducing observed counts at other wavelengths are too diverse to be reconciled in a single model; specifically they need too strong colour evolutions which contradicts $850\,\mu$m observations. While the inconsistency at $160\,\mu$m has been reported in earlier works \citep{LeBorgne09,Valiante09}, we are the first to report it for $250\,\mu$m, $350\,\mu$m and $500\,\mu$m. We showed that the source counts at these wavelengths can be reproduced consistently, by adopting the best-fit model which produces correct $70\,\mu$m, $850\,\mu$m and $1100\,\mu$m source counts, together with a modification in SED templates which is equivalent to the existence of a cold population of dusty galaxies at low to intermediate redshifts which are under-represented in IRAS data. Besides the fact that there is some observational evidence for the existence of such galaxies \citep{Stickel98,Stickel00,Chapman02,Patris03,Dennefeld05,Sajina06,Amblard10}, this assumption is similar to what other models had to assume in order to reproduce adequate $160\,\mu$m sources \citep{Lagache03,Lagache04,Valiante09}.

It is important to keep in mind that phenomenological models like what we described in this paper, are mainly simple mathematical forms which relate different observations consistently rather than being physical models with explanatory power. However, their performance at different wavelengths and the distribution of sources they require for different redshifts can be used as their main predictions which also could be used to test their validity. While we used the redshift distribution of submm galaxies to constrain our model, we note that the observed redshift distribution of other wavelengths, if available, are in agreement with our best-fit model predictions \citep{Jacobs11,Berta11}.  

Additional information including tabulated data for differential and cumulative source counts at different wavelengths and their redshift distributions is available at $\mathtt{http://www.strw.leidenuniv.nl/genesis/}$

\section*{Acknowledgments}
We thank the anonymous referee for valuable comments which improved the original version of this paper. AR thanks D. Dale, H. Rottgering, J. Schaye and M. Shirazi for useful discussions. During the early stages of this work, AR was supported by a Huygens Fellowship awarded by the Dutch Ministry of Culture, Education and Science.

\section*{Appendix A: Some numerical details}
\label{sec:modeled}

For each specific evolution model, the source count at a given flux threshold
and wavelength can be calculated based on equation (\ref{eq:N-general}), where
the integration should be performed over all possible luminosities, colours and
redshifts. As mentioned in Section~\ref{sec:algorithm}, we do this by
splitting possible colour, luminosity and redshift ranges into very small bins,
assuming that in each bin the related variable is not changing significantly.\\

The finite number of SED models we are using automatically splits the colour
range into 64 bins between $0.29\leq R(60,100)\leq 1.64$ (see Section~\ref{sec:SED}). We also use logarithmically spaced bins to split the possible
luminosity range of $10^9 L_{\odot} \leq L \leq 10^{14}L_{\odot}$ into
100 bins in our calculation. This logarithmic scale which makes the
integration roughly insensitive to the number of luminosity bins, is chosen
to cope with the exponential nature of luminosity function where faint objects
are much more numerous than luminous ones. It is also worth mentioning that
the source count calculation is not sensitive to the minimum or maximum
luminosity which is used in integration, if the used luminosity range covers
the important $10^{10}-10^{13} L_{\odot}$ luminosity range; for instance,
using $L_{\rm{min}}=10^7 L_{\odot}$ instead of $L_{\rm{min}}=10^9 L_{\odot}$
as the minimum possible luminosity, does not change any of the source count
calculations we are presenting in this paper.\\

As discussed in Section~\ref{sec:algorithm}, the uniform distribution of
galaxies in redshift space is assured by the algorithm we are using,
independent of the size of redshift bins. But, for the precise calculation of
the K-correction and evolution functions, we split the redshift range of
$0 \leq z \leq 8$ using bin sizes equal to $\Delta z = 0.01$. However we noted
it is possible to use even bigger redshift bins (e.g. $\Delta z = 0.1$)
without any significant change in the results.

\end{document}